\documentclass[aps,prb,onecolumn,12pt, altaffillletter,superscriptaddress,showkeys]{revtex4-1}
\usepackage{graphicx,color}
\usepackage{amsmath}
\usepackage{txfonts}
\usepackage{siunitx}
\usepackage{xspace}
\usepackage{tabularx,threeparttable}
\usepackage{setspace}
\usepackage{bm}
\usepackage{url}
\usepackage{hyperref,hypernat}

\newcommand{\kCl}{\textit{$\kappa$}-\textrm{Cl}\xspace}
\newcommand{\esqh}{{e^2/h}\xspace}
\newcommand{\Vg}{$V_\mathrm{g}$\xspace}
\newcommand{\mathVg}{V_\mathrm{g}}
\newcommand{\mathVmic}{V_\mathrm{MIC}}
\newcommand{\mathVc}{V_\mathrm{c}}
\newcommand{\mathymic}{\sigma_\mathrm{MIC}}

\newcommand{\mathGap}{\Delta_\mathrm{c}}
\newcommand{\bvec}[1]{\mbox{\boldmath $#1$}}
\begin{document}

\title{Critical Behavior in Doping-Driven Metal--Insulator Transition on 
Single-Crystalline Organic Mott-FET}

\author{Yoshiaki Sato}
\email{yoshiaki.sato@riken.jp}
\affiliation{RIKEN, 2-1 Hirosawa, Wako, Saitama 351-0198, Japan}
\author{Yoshitaka Kawasugi}
\email{kawasugi@riken.jp}
\affiliation{RIKEN, 2-1 Hirosawa, Wako, Saitama 351-0198, Japan}
\author{Masayuki Suda}
\affiliation{RIKEN, 2-1 Hirosawa, Wako, Saitama 351-0198, Japan}
\affiliation{Research Center of Integrative Molecular Systems (CIMoS), %
Institute for Molecular Science, 38 Nishigo-Naka, Myodaiji, Okazaki 444-8585, Japan}
\author{Hiroshi M. Yamamoto}
\email{yhiroshi@ims.ac.jp}
\affiliation{RIKEN, 2-1 Hirosawa, Wako, Saitama 351-0198, Japan}
\affiliation{Research Center of Integrative Molecular Systems (CIMoS), %
Institute for Molecular Science, 38 Nishigo-Naka, Myodaiji, Okazaki 444-8585, Japan}
\author{Reizo Kato}
\affiliation{RIKEN, 2-1 Hirosawa, Wako, Saitama 351-0198, Japan}

\sloppy 

\begin{abstract}
\vspace*{1em}
We present the carrier transport properties in the vicinity of a doping-driven Mott transition observed at a field-effect transistor (FET) channel using a single crystal of the typical two-dimensional organic Mott insulator \textit{$  \kappa  $}-(BEDT-TTF)$_2$CuN(CN)$_2$Cl (\kCl). The FET shows a continuous metal--insulator transition (MIT) as electrostatic doping proceeds. The phase transition appears to involve two-step crossovers, one in Hall measurement and the other in conductivity measurement. The crossover in conductivity occurs around the conductance quantum $\esqh $, and hence is not associated with ``bad metal'' behavior, which is in stark 
contrast to the MIT in half-filled organic Mott insulators or that in doped inorganic Mott insulators. Through in-depth scaling analysis of the conductivity, it is found that the above carrier transport properties in the vicinity of the MIT can be described by a high-temperature Mott quantum critical crossover, which is theoretically argued to be a ubiquitous feature of various types of Mott transitions.  
[This document is the unedited Author’s version of a Submitted Work that was subsequently accepted for publication in Nano Letters, copyright \copyright American Chemical Society after peer review. To access the final edited and published work see http://dx.doi.org/10.1021/acs.nanolett.6b03817]
\end{abstract}

\maketitle

%%%% MAIN TEXT %%%%

Strongly correlated electrons confined in two-dimensional materials exhibit a rich variety of anomalous phases, such as high-$T_\mathrm{c}$ superconductivity, Mott insulators, and quantum spin liquids. Enormous effort has been devoted to understanding the mechanisms of the phase transitions between them,\cite{Cui2015, Tsen2015b} which are driven by the complex interplay between many-body interactions and charge--spin fluctuations. Mott insulators are particularly remarkable in viewpoints of nanoscience as well as fundamental physics since they are not only mother materials for a diverse range of strongly correlated systems, but also serve as a molecular or atomic limit for coherently coupled arrays of single electron transistors (SETs).\cite{Stafford1994, Byrnes2008} In Mott insulating state, strong Coulomb repulsion between electrons in half-filled band leads to carrier localization, which can be viewed as a situation that the collective Coulomb blockade confines an electron to the individual ‘subnano-sized SET’. The electronic/magnetic properties of Mott insulators are governed by quantum mechanics, and indeed various types of unconventional phase transitions occur due to imbalance in the particle--wave duality of electrons which is triggered by charge doping,\cite{Lee2006b} a magnetic/electric field\cite{Cario2010, Zhu2016a} pressure\cite{Sipos2008,Powell2011a} or introduction of disorders.\cite{Sasaki2012, Lahoud2014} An important issue is the possibility of quantum phase transitions in Mott insulators,\cite{Imada1998, Vojta2003} where quantum fluctuations play an essential role in the transitions with the emergence of quantum critical points at 0~K.\cite{Lohneysen2007, Millis1993, Gegenwart2008, Sachdev2009, Sebastian2010} This would hold true for a coherently coupled array of SETs, although it has not yet been recognized.

Even at finite temperatures, quantum fluctuations with an energy scale overwhelming the temperature $ k_\mathrm{B}T $ give rise to a critical crossover feature observed as a power-law singularity in physical quantities in the vicinity of the quantum critical point.\cite{Vojta2003,Sachdev2011} It has been argued that the unconventional non-Fermi liquid behavior commonly observed near the Mott transition should pertain to such quantum criticality, although there is no general consensus on its origin.

To properly investigate the doping-driven Mott transition, it is required to dope carriers in a clean manner without introducing impurities since the modification of the local electronic environment by such disorders should destroy the intrinsic many-body states. Electrostatic doping techniques using the electric double-layer transistor (EDLT) or field-effect transistor (FET) configurations are promising for such investigation. Combined with recent advances in fabrication technology for crystalline nanosheets\cite{Cui2015} and interfaces,\cite{Caviglia2008, Bollinger2011, Hwang2012} these techniques not only enable fine control of the charge density as an independent tunable parameter but also extend the possibility of novel device applications that utilize abrupt changes in physical properties triggered by a phase transition.

\begin{figure}
%\advance\leftskip 0.86in
\includegraphics[width = \textwidth]{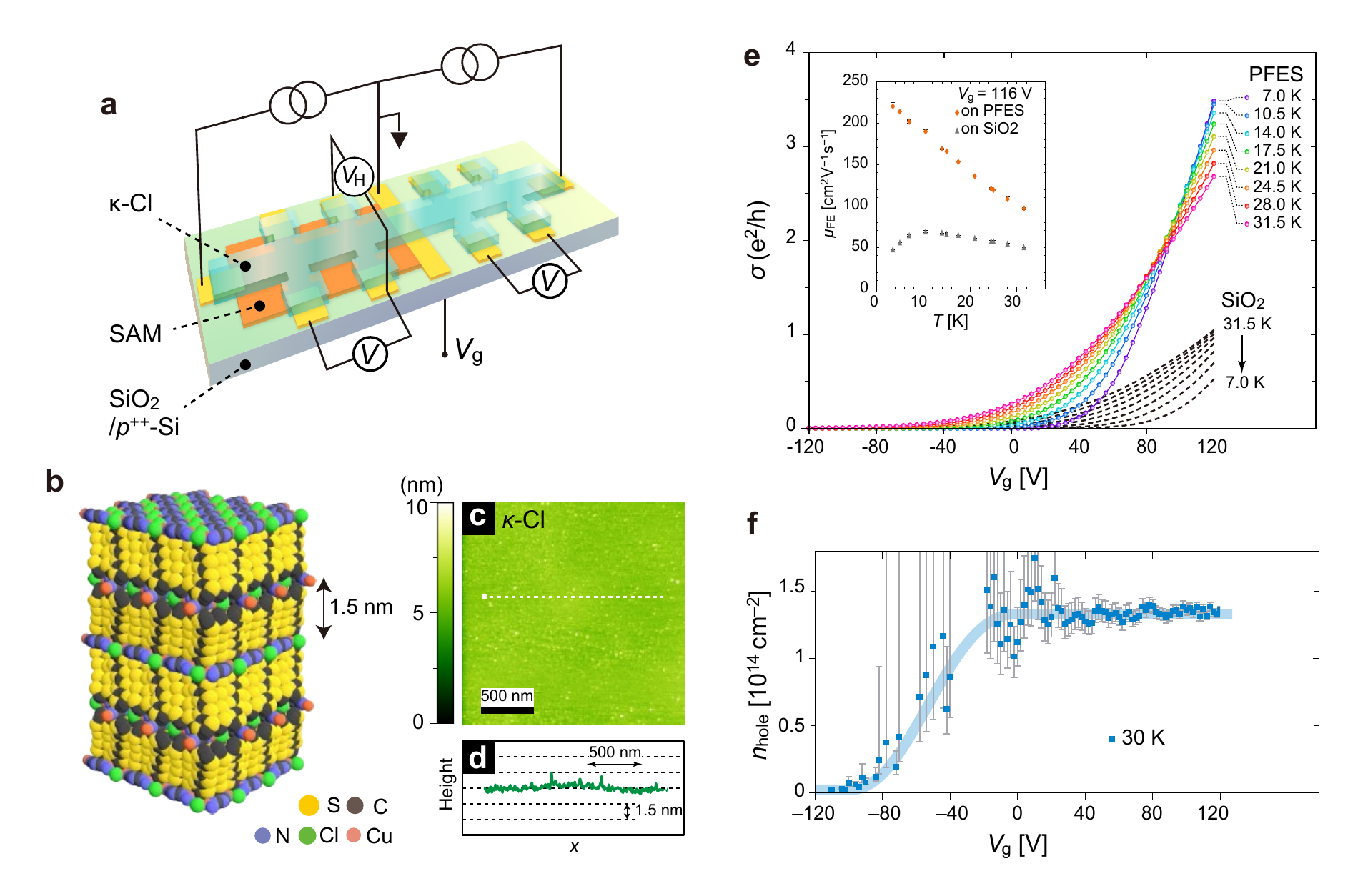}

\caption{\label{fig_Schematic} Schematic of the Mott-FET device and the doping-driven metal--insulator transition of \kCl. (a) Schematic of FET device with the micropatterned SAM interface. (b) Crystal structure of \kCl with alternately stacked cationic dimer [(BEDT-TTF)$_2$]$^+$ layers and anionic insulator layers. (c, d) AFM topographical image measured on the surface of the \kCl crystal. (e) Transfer curves for various \textit{T }(device F\#1) on a PFES-covered substrate (colored circles) and on bare SiO$_2$ (gray squares). Inset: device mobility $  \mu _\mathrm{FE}= ( 1/c_{\mathrm{g}} )\, \mathrm{d} \sigma /\mathrm{d}\mathVg $ ($ c_\mathrm{g}=7.0 \times 10^{14}$~\si{cm^{-2}V^{-1}}; areal capacitance of back gate) as a function of $ T $ in the highly doped regime ($ \mathVg =116$~V). (f) Gate voltage dependence of carrier density $ n_\mathrm{hole} $ (device F\#1) for PFES device area determined from Hall coefficients ($B= 8$~T). In the highly doped regime where the MIC occurs, $ n_\mathrm{hole} $ is almost independent of $ \mathVg $ with $ n_\mathrm{hole} \sim 1.3 \times 10^{14}$~\si{cm^{-2}}, corresponding to the cross-sectional area of  $  \sim  $ 75\%  of the first Brillouin zone. Error bars were calculated from the standard deviation of the Hall angle. The solid line (light blue) is a guide to the eye.
 }
\end{figure}

The organic Mott insulator \textit{ $  \kappa  $ }-(BEDT-TTF)$_2$CuN(CN)$_2$Cl, referred to as \kCl, is one of the two-dimensional materials suitable for the channel of electrostatic doping devices. The crystal structure of \kCl consists of alternately stacked  $  \pi  $-conducting [(BEDT-TTF)$_2$]$^+$ cation layers and atomically thin insulating polyanion layers (Figure \ref{fig_Schematic}b), showing excellent crystallinity.\cite{Furukawa2015a, Kagawa2005} An effectively half-filled band is formed by a carrier density of one hole per BEDT-TTF dimer. As the areal carrier density ($ \sim 10^{14}$~\si{\cm^{-2}}) is almost an order of magnitude lower than those of inorganic Mott insulators, a large region of the phase diagram can be investigated by electrostatic doping. Indeed, we have demonstrated that by charge doping up to a density of $ \sim 10^{14}$~\si{\cm^{-2}} using an EDLT, both electron- and hole-doping-driven Mott transitions were successfully realized on a single \kCl crystal.\cite{Kawasugi2016} Nonetheless, the EDLT technique is inconvenient for fine-tuning of the band filling at low \textit{T }owing to freezing of the ionic liquid.\cite{Hayes2015} On the other hand, a solid-gate FET is highly advantageous in terms of the applicability of various micrometer-scale interface engineering methods as well as the high controllability of the doping level in real time at low $T$.

In this Letter, we report a doping-driven Mott MIT on a solid-gate FET using a single crystal with a highly flat surface as well as a gate dielectric covered with self-assembled monolayers (SAMs). We selected hydrophobic 1H,1H,2H,2H-perfluorodecyltriethoxysilane (PFES) and octyltriethoxysilane (OTS) as SAM reagents, which are widely used to improve device performance and are also expected to eliminate the adsorbed solvents on the surface.\cite{Kobayashi2004, Wang2011} By micropatterning these SAMs,\cite{Yokota2014a, Najmaei2014} we successfully induced a continuous MIT on a \kCl single-crystal channel. Owing to the high resistivity of the bulk channel below $ T \sim 50$~K, the observed MIT was governed by carriers confined in two dimensions at the interface, exhibiting sheet conductivity equal to  $ \esqh $ ($\esqh=3.874 \times 10^{-5}~$\si{\ohm^{-1}}) on the metal--insulator crossover (MIC) line. Through in-depth analysis of the conductivity on the verge of the MIT, we demonstrated the quantum critical feature of the doping-driven Mott transition in the intermediate-$T$ crossover region. This is distinct from the quantum criticality in the low-$T$ regime for high-${T}_\mathrm{c}$ cuprates,\cite{Bollinger2011, Leng2011, Garcia-Barriocanal2013} and follows the ‘high-$T$ Mott quantum criticality’ expected to be universal to various types of Mott transition.\cite{Vucicevic2015, Terletska2011} On the basis of the scaling plot for the quantum criticality, we propose a candidate crossover line that delineates the boundary between metallic and insulating regions.

The FET device was fabricated on a SiO$_2$ (300~nm)/$p^{+}$-Si substrate with a micropatterned SAM (area: $ 50 \times 50$~\si{{\micro\meter}^2}), as schematically shown in Figure \ref{fig_Schematic}a. We fabricated four devices: one with gate surface modification by PFES (device F\#1) and the other three with gate surface modification by OTS (devices C\#1--\#3), details of which are shown in Table \ref{tbl_Device}. We found that the difference in SAM species did not noticeably influence the overall MIT behaviors, as discussed in Supporting Information \ref{sec_SAMdep}. The patterning of the SAMs and the attachment of electrodes were performed using a conventional photolithography technique (Supporting Information \ref{sec_Substrate}). Then thin crystals of \kCl (thickness: 40~nm for device F\#1 and 80~nm for device C\#2) were laminated on top of the substrate. Through the improvement of the electrochemical crystal growth process (Supporting Information \ref{sec_Crystal}), we successfully obtained thin crystals of \kCl with a flat surface. Figure \ref{fig_Schematic}c shows an atomic force microscopy (AFM) image of the surface of the typical \kCl crystal laminated on the substrate. Over micrometer-scale areas, the roughness of the surface is below 1.5~nm, or the thickness of one BEDT-TTF molecular layer (Figure \ref{fig_Schematic}b). For comparison, shown in Figure S3 is an AFM image of a typical thin crystal of the sister material $  \kappa  $-(BEDT-TTF)$_2$CuN(CN)$_2$Br ($  \kappa  $-Br), which was synthesized in accordance with a previous method reported elsewhere.\cite{Kawasugi2009a} One can recognize the steps of the BEDT-TTF molecular layers and many small vacancies for $  \kappa  $-Br. The availability of the single crystal with perfect surface flatness for \kCl is a major advantage for precisely investigating the MIT without a significant effect of disorder.

Figure \ref{fig_Schematic}e shows the transfer curves of device F\#1, which has a partly covered PFES layer and provides both \kCl/PFES/SiO$_2$ and \kCl/SiO$_2$ interfaces on an identical single crystal. Both on the SiO$_2$ and PFES regions, the sheet conductivity ($  \sigma  $) at low $T$ is suppressed to below $ 10^{-4} \esqh $ by applying a negative gate voltage (\Vg), whereas a positive \Vg induces a continuous increase in $  \sigma  $.
This ``\textit{n}-type polarity'' stems from the natural \textit{n}-type doping for \kCl surface (typically density of $ \sim10^{12}-10^{13}$~\si{cm^{-2}})\cite{Kawasugi2016} and implies that electron doping breaks the strong carrier localization of Mott insulator state.\cite{Kawasugi2009a} This description is also supported by the carrier density estimated from Hall effect measurements (Figure \ref{fig_Schematic}f): $ n= \sigma B/e\tan \theta _\mathrm{H} $, where $ B $ and $  \theta _\mathrm{H} $ are the magnetic field and Hall angle, respectively. The carrier density is nearly zero at negative \Vg, corresponding to the gapped incompressible state (small $  \left\vert \mathrm{d}n/\mathrm{d} \epsilon  \right\vert  $, with $  \epsilon  $ being the chemical potential) characterizing Mott insulators.\cite{Imada1998, Sordi2011} Upon \textit{electron} doping by as small amount as $ \sim10^{12}$~\si{cm^{-2}}, however, \textit{hole} carriers with the density of $ 1.3 \times 10^{14}$~\si{cm^{-2}} (corresponding to the cross-sectional area of 75\%  of the first Brillouin zone) rapidly recovered, evidencing a compressible state with a large Fermi surface (large $  \left\vert  \mathrm{d}n/\mathrm{d} \epsilon \right \vert  $). This change in the Hall carrier density indicates an abrupt increase in the mobile carrier spectral weight accompanied by closing of the Mott gap upon doping.\cite{Kawasugi2016, Cooper1990} Plus, continuous increase in $  \sigma  $ after establishment of large Fermi surface state with almost a constant hole carrier density is also a unique feature of nonrigid bands of strongly correlated electrons, on which doping enhances carrier coherence through modification of the charge mass or lifetime.

\begin{figure}
\centering
\includegraphics[width= 0.6\textwidth]{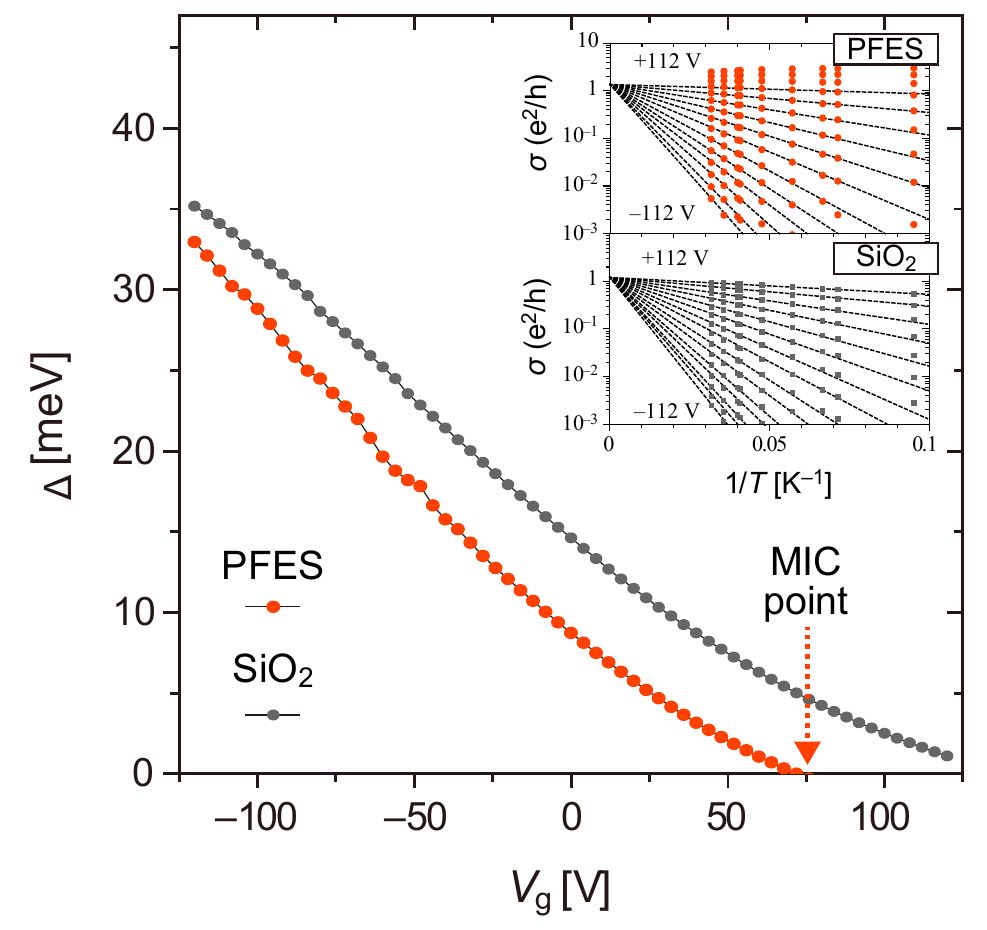}

\caption{\label{fig_Energy}Characteristics of activated transport in insulating regime (device F\#1). (Main panel) Gate voltage dependence of the transport gap $  \Delta  $ in the insulating regime for the PFES-covered (red) and bare SiO$_2$ substrate (gray) areas. (Inset) Arrhenius plots of sheet conductivity for different gate voltages on the PFES-covered and bare substrate. Symbols are experimental results; dotted lines are fits using $  \sigma _\mathrm{ins} \left( T \right) = \sigma _{ \infty}\exp ( - \Delta /2k_\mathrm{B}T )  $
. Each step is $  \Delta\mathVg=8$~V.
}
\end{figure}

A stark difference between the PFES and SiO$_2$ substrate areas manifests itself at a highly positive \Vg. On the PFES region, $  \sigma  $ exceeds $ 3 \esqh $ and its behavior becomes metallic (i.e., increasing $  \sigma  $ with decreasing $ T $). Importantly, an MIC in the conductivity occurs at $ \mathVg =75 $~V (the point of zero activation energy in Figure \ref{fig_Energy}), whereas the crossover between an incompressible Mott insulator state and a compressible state with a large Fermi surface occurs at $ \mathVg<0 $ (the jump of the $ n_\mathrm{hole} ( \mathVg )  $ curve in Figure \ref{fig_Schematic}f). This is in contrast to the MIC of other two-dimensional electron systems, where continuous changes in the carrier density and metallic transport (i.e., $ \mathrm{d} \sigma /\mathrm{d}T <0 $) take place at the same time.\cite{Radisavljevic2013, Knyazev2008}  The conductivity at the MIC is $  \sigma =1.4 \esqh $, which agrees with the Mott--Ioffe--Regel (MIR) limit in two dimensions (minimum conductivity anticipated within semiclassical Boltzmann transport theory). The device mobility $  \mu _\mathrm{FE}= ( 1/c_\mathrm{g} )\, \mathrm{d} \sigma /\mathrm{d}\mathVg $ ($ c_\mathrm{g}=7 \times 10^{10}$~\si{cm^{-2}V^{-1}}: areal capacitance of back gate) also becomes very high, reaching 220~\si{cm^{2} V^{-1} s^{-1}}  at 5~K. On the SiO$_2$ region, however, $  \sigma  $ remains insulating with a value below $ \esqh $, and the maximum mobility is 70 ~\si{cm^{2} V^{-1} s^{-1}}, much lower than that on the PFES region (Figure \ref{fig_Schematic}e inset).

On the other hand, focusing on the insulating regime the conductivity is well expressed by activated-transport behavior modeled with
$  \sigma _\mathrm{ins} ( T, \mathVg) = \sigma _{ \infty} \exp \left( -\Delta  ( \mathVg) /2k_\mathrm{B}T \right)  $  
 irrespective of SAM functionalization (Figure \ref{fig_Energy}, inset), where $  \Delta  $ 
is the transport gap and $  \sigma _{ \infty}  $ ($  \approx 1.4\esqh $
) denotes the high temperature limit of activated-transport conductivity. In the insulating limit ($ \mathVg \sim -120$~V, Figure \ref{fig_Energy}), $  \Delta  $ ($\approx$~35~meV) is comparable to that of bulk \kCl (Supporting Information \ref{sec_highT}), which confirms that the original electronic properties of \kCl are retained through the device fabrication processes. On the other hand, electron doping continuously reduces $  \Delta  $ toward zero, which occurs more rapidly on the PFES-patterned area than on the bare SiO$_2$ area (Figure \ref{fig_Energy}). The difference in $ \mathrm{d} \Delta /\mathrm{d}\mathVg $ cannot be explained solely by the doping effect from the SAM,\cite{Kobayashi2004, Suda2015} which suggests that the Mott transition is sensitively affected by interfacial disorder conditions such as impurities, the contamination by the solvent, and charge traps.

\begin{figure}
%\advance\leftskip 1.36in
\includegraphics[width=0.7\textwidth]{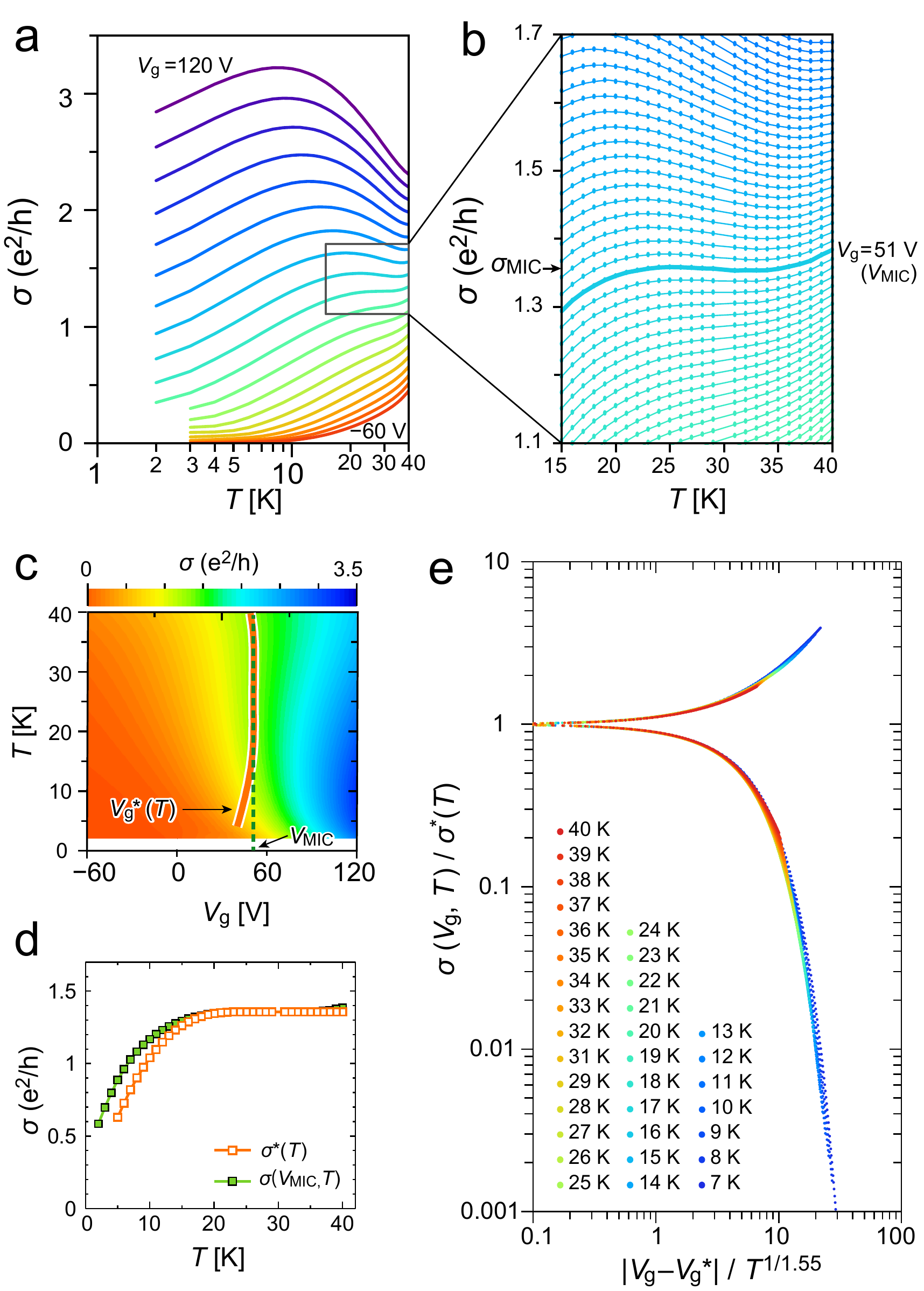}

\caption{\label{fig_Critical}Criticality in the vicinity of the MIC point (device C\#2, on OTS-covered area). (a) $  \sigma -T $ plot for various \Vg ($ =120, 112, \cdots , -54, -60\ \mathrm{V} $). For $ 21\ \mathrm{K}<T<36\ \mathrm{K} $, there is a distinct MIC point at $ \mathVmic =51$~ V where the sheet conductivity becomes $T$-independent, namely $  \sigma  \left( \mathVmic T \right) = \mathymic =1.35\esqh $ (thick line in panel b). On the metallic side ($ \mathVg > \mathVmic $), $  \sigma  \left( T \right)  $ curves exhibit maxima at certain temperatures, below which $  \sigma  \left( T \right)  $ decreases in a linear-in-$ \log T $ manner. On the other hand, for $ \mathVg <  \mathVmic $ $  \sigma  $ monotonically decreases with decreasing $T$. (b) Magnified view of $  \sigma -T $ plot for various \Vg (in increments of 1~V) in the vicinity of the MIC. Dots represent measurement points. (c) Color-coded plot of conductivity with the trajectory of the optimized crossover line (orange solid line) under the assumption of high-$T$ Mott quantum criticality (candidate QWL, see text for details). For comparison, $ \mathVg = \mathVmic $ is marked with a dashed line (green). (d) Conductivities on the optimized crossover line (orange) and the $ \mathVg =\mathVmic $ line (green). (e) Quantum critical scaling plot of $  \sigma  $ normalized by $  \sigma ^{*} ( T )  $. For $ T>7$~K, all of the conductivity data on the insulating and metallic sides collapse onto two different branches of the universal scaling function (eq. (1)). The critical exponent is $  \nu z=1.55 $.
}
\end{figure}

Figure \ref{fig_Critical}a,b displays the detailed $T$-dependence of the conductivity near the MIC, which is measured on the SAM-patterned area of device C\#2. Similar $  \sigma  ( T )  $ behaviors were observed for all the devices on the SAM-patterned areas, and hereafter we concentrate on device C\# 2 (for devices C\#1, C\#3, and F\#1, see Figure S8). The conductivity curve for each \Vg is continuous without jumps or kinks, and reproducible without hysteresis even after sweeping many times, indicating that no structural transition occurs. For 22~K$\lesssim T \lesssim$ 36~K, a well-defined MIC behavior is observed; a crossover line exists with $T$-independent conductivity $  \sigma ^{*}= \mathymic=1.35\esqh $ at $ \mathVg^{*}=\mathVmic= 51$~V, which separates the metallic and insulating regions in the $  \sigma  $ versus \Vg plane. (Hereafter, variables with asterisks represent those on the crossover line.) In the lightly doped region ($ \mathVg<\mathVmic $), all the $  \sigma  (\mathVg ,T)  $ curves gradually decrease with decreasing $T$. On the other hand, the heavily doped region ($ \mathVg>\mathVmic $) shows nonmonotonic behavior with conductivity maxima, below which the transport becomes insulating. The low-$T$ insulating transport for $  \sigma \gtrsim 0.5\esqh $ can be scaled linearly with $ \log T $. Negative magnetoresistance is also observed in the highly doped regime below  $  \sim  $ 7~K, which indicates that the linear-in-$ \log T $ insulating behavior is explained by weak localization correction.\cite{Hikami1980} Interestingly, $ \mathVmic $ is a device-dependent value, whereas $  \mathymic $ is almost identical for each device (see Figures \ref{fig_Schematic} and \ref{fig_ScalingOthers}). This means that the conductivity in the vicinity of the MIC is not affected by the variation in the thickness of \kCl crystals, and thus the MIC is clearly caused by carriers confined in two dimensions at the interface. Notably, $  \mathymic $  corresponds to the minimum metallic conductivity, and hence one cannot identify a bad-metal (BM) feature, that is, metallic transport that violates the MIR limit over the entire $T$ range. (Note that above 40~K, it is difficult to accurately extract the carrier transport on the interface from the observed conductivity because of the dominant contribution of the undoped bulk channel. Nevertheless, one cannot clearly observe BM behavior at high $T$. See Supporting Information \ref{sec_highT}.) This is in contrast to Mott transitions in bulk \kCl at half-filling\cite{Furukawa2015a, Limelette2003a} or those in inorganic Mott insulators.\cite{Emery1995, Takagi1992, Si2016}

Recently, theoretical studies based on dynamical mean-field theory (DMFT)\cite{Vucicevic2015, Terletska2011, Vucicevic2013} have advocated that the high-$T$ MICs associated with Mott transitions bear a type of quantum criticality. This high-$T$ quantum criticality is expected to be observed on any Mott systems because it is governed by high-energy-scale quantum fluctuations due to the competing electron--electron interaction and Fermi energy on Mott insulators. Although DMFT is exact in the limit of infinite dimensions, it becomes a reasonable approximation at high $T$ even for low-dimensional system such as \kCl. Indeed, it has been experimentally demonstrated that various half-filled organic Mott insulators including \kCl show quantum critical nature in the high-$T$ crossover region for a \textit{pressure-driven} Mott transition.\cite{Furukawa2015a} The experimental and theoretical phase diagrams around the quantum critical Mott crossovers are summarized in Figure S6. In brief, despite the complexity of the low-$T$ material-specific phases, the half-filled organic Mott insulators show material-independent rapid crossovers at high $T$ with the carrier transport continuously changing from insulating to BM behavior. At these crossovers, quantum criticality is commonly observed with the crossover conductivity much smaller than $ \esqh $ (i.e., BM behavior). These behaviors are completely consistent with the quantum criticality suggested by DMFT\cite{Terletska2011, Vucicevic2013} Therefore, the high-$T$ Mott quantum critical crossover scenario is appropriate for describing the continuous MIC in the conductivity, at least for a pressure-driven Mott transition. The following question arises: does this hold true for a \textit{doping-driven} Mott transition? According to the theory,33 the quantum criticality of a doping-driven Mott transition should have notable features that have not been observed in a pressure-driven Mott transition: (1) A continuous crossover is sustained even in the low-$T$ region (well below 1\%  of the half bandwidth $ D $ with $ D\sim 0.25$~eV for $  \kappa  $-type BEDT-TTF compounds\cite{Komatsu1996}). (2) The crossover conductivity coincides with the MIR limit at high $T$, whereas it may deviate at low $T$. (3) The MIT can show two-step crossovers at finite $T$; one is a rapid change in charge compressibility $  \vert \mathrm{d}n/\mathrm{d} \epsilon  \vert  $ and the other is a sign change in $ \mathrm{d} \sigma /\mathrm{d}T $ (Supplemental Material in ref. \onlinecite{Vucicevic2015}). These three points appear to be consistent with the present experimental results, and the third point can be observed in the discrepancy between the jump of the Hall carrier density and the MIC in the conductivity (Figure \ref{fig_Schematic}e,f). Thus, the high-$T$ quantum critical crossover in the doping-driven Mott transition is a reasonable scenario that explains the observed MIC.

According to DMFT studies,\cite{Vucicevic2015, Terletska2011, Vucicevic2013} the MIC of a Mott transition can be well described within the framework of finite-size quantum critical scaling theory.\cite{Sachdev2011, Phillips2012, Gantmakher2010, Sondhi1997} This theory predicts that the conductivity is amenable to a scaling law written as\cite{Dobrosavljevic1997}
\begin{equation}
  \sigma  \left( T, \delta  \right) 
  	= \sigma ^{*} \mathcal{F}_{ \pm } \left( \frac{ \vert  \delta  \vert }{T^{1/ \nu z}} \right) , 
  \label{eq_QCS} 
\end{equation}
where $  \delta  $ is the distance from the crossover line and the power exponents $  \nu  $ and $ z $ are called the correlation exponent and dynamical exponent, respectively. In contrast to magnetic/charge ordering transitions and superconducting phase transitions, the MIT generally has no obvious order parameter. Yet, in the present doping-driven Mott transition, it is natural to regard \Vg as the scaling variable, that is, $  \delta =\mathVg-\mathVg^{*} $. $ \mathcal{F}_{+} ( y )  $ and $ \mathcal{F}_{-} ( y )  $ are the universal scaling functions for the metallic and insulating regimes, respectively, satisfying $ \mathcal{F}_{ \pm } ( y )  \rightarrow 1 $ and having mirror symmetry $ \mathcal{F}_{+} ( y ) =1/\mathcal{F}_{-} ( y )  $ for $ y \rightarrow 0 $. Importantly, the crossover line in eq. (1) should be provided by the ``quantum Widom line (QWL)'',\cite{Vucicevic2015, Vucicevic2013} on which the system becomes the least stable. For scaling analysis, we attempt to propose a possible crossover line to find the QWL in the following.  \\

In analogy with the Widom line\cite{Simeoni2010} in the supercritical region of the classical gas–liquid-type phase transition, the location of the QWL in the $T$ versus \Vg plane is generally dependent on temperature.\cite{Vucicevic2015} Thus, the crossover line $ \mathVg^{*} ( T )  $ should have a complex form and deviate from the constant-\Vg crossover line widely adopted for conventional doping-driven MITs.\cite{Abrahams2001} Because it is fundamentally difficult to identify the location of the QWL from the transport properties, we instead derive a candidate crossover line through a simple optimization calculation (for details, see Supporting Information \ref{sec_Scaling}). In short, the crossover line is taken as $ \mathVg^{*}=\mathVmic $ for $ T>25$~K to satisfy the theoretically suggested high-$T$ feature of the QWL, 
$  \sigma _{\mathrm{high}-T}^{*} =\esqh$ (constant). Then it is smoothly extended to the low-$T$ region so that the renormalized conductivity data $ \sigma  ( T,\mathVg) / \sigma ^{*} ( T )  $ converge onto the universal scaling curves (eq.~(1)) with the highest scaling quality. If the resultant crossover line is appropriate for the candidate QWL, the quantum critical scaling along the crossover line should hold for sufficiently low $T$. \\

Figure \ref{fig_Critical}e depicts the results of the scaling analysis based on the high-$T$ Mott quantum critical crossover scenario, which is obtained with $  \nu z=1.55 $. It is found that the quantum critical scaling law persists over three decades of $  \sigma / \sigma ^{*} $ and that the critical region extends down to $ T \sim 7$~K. As mapped in Figure \ref{fig_Critical}c, the optimized crossover line exhibits a bow-shaped trajectory in the $T$ vs \Vg plane and deviates from the conventional critical crossover line with $ \mathVg=\mathVmic $ at low $T$. The crossover conductivity $  \sigma ^{*} $ falls from $ \mathymic $ with decreasing $T$(Figure \ref{fig_Critical}d), which is consistent with the theory.\cite{Vucicevic2015} In the lower-\textit{T }regime of $ T<7$~K, it turns out that the data for $  \sigma / \sigma ^{*} $ are no longer scaled by the universal curves. This can be understood from two possible points of view. First, the quantum critical description of the continuous MIT is obscured because of the weak localization in the low-$T$ region. Second, since long wavelength (low energy) fluctuations should come to play a significant role in low dimensional systems at a sufficiently low $T$, the high-$T$ Mott critical crossover scenario derived from a local mean field approximation such as DMFT may be no longer appropriate. We emphasize that the quantum critical crossover with similar critical exponents down to $ T \sim 10$~K was also confirmed in devices C\#1, C\#3 and F\#1 (Figure S8), even though details of the carrier transport characteristics varied from device to device. These findings manifest the essence of the present phase transition that the critical behavior is independent of specific details of the system. Furthermore, we also attempted the scaling analysis using conventional crossover lines (i.e., constant-$ \mathVg^{*} $ lines), as displayed in Figures and \ref{fig_midTscaling} and \ref{fig_conventional}. Our attempt to fit eq. (1) with a constant $ \mathVg^{*} $ was only successful in a limited $T$ range, which indicates that the crossover line in Figure \ref{fig_Critical}d is appropriate for describing the observed quantum criticality. Taken together, the proposed crossover line meets the requirements of the QWL, and the quantum critical crossover region spreads over a wide range in the $ T $ vs \Vg plane.

One might suspect the possibility of other types of MICs associated with, for example, a percolation transition,\cite{Dobrosavljevic2012} or a Mott MIC with classical criticality.\cite{Kagawa2005, Abdel-Jawad2015} However, the continuous MIC in the present study is inconsistent with criticalities derived from these mechanisms, indicting the uniqueness of the quantum critical scaling of the charge transport in the intermediate-$T$ crossover region of the doping-induced Mott transition. Applicability for classical criticality is discussed in detail in Supporting Information \ref{sec_Classical}.

Finally, we discuss the lack of BM behavior in the observed doping-driven Mott transition. In a wide variety of high-$T_\mathrm{c}$ superconductors as well as heavy Fermion systems, metallic transport without a coherent Drude peak and with high resistivity exceeding the MIR limit is observed in the crossover region between the localized and delocalized states without showing the saturation observed for conventional metals.\cite{Imada1998, Gunnarsson2003} This BM behavior is intuitively understood as collective carrier transport retaining the strongly correlated feature of the localized state (e.g., Mott insulator), whereas its origin remains to be fully understood. Early studies based on a quantum critical model\cite{Phillips2005} did not appear to explain the BM behavior, promoting various theoretical considerations such as those based on marginal Fermi liquid theory\cite{Varma1989} and holographic duality.\cite{Hartnoll2014, Davison2014} A DMFT study33 pointed out that the BM behavior is not contradictory to quantum criticality; the linear-in-$T$ resistivity beyond the MIR limit can result from the inclined quantum critical region in the $T$ versus doping density plane. In other words, the emergence of BM behavior depends on the relationship between the characteristics of the QWL and the material-specific band dispersion. Taking into account the remarkable similarity between the high-$T$ criticality in the present experiment and the DMFT study, we speculate that the lack of BM behavior in the present experiment is due to the disordered environment and the two-dimensional band structure that can modify the QWL. So far, BM behavior has been believed to be an inherent property of Mott insulators. However, our results provide experimental counterevidence, which indicates that the non-Fermi-liquid feature in the crossover region should be understood in terms of the high-$T$ Mott quantum criticality. Optical conductivity measurements may reveal a clearer picture of the quantum critical region of the doping-driven Mott transition in future.

In summary, we successfully obtained single-crystal Mott-FETs with high performance, the device mobility exceeding  $  \sim  $ 200~\si{cm^2 V^{-1} s^{-1}} at low temperatures by a combination of thin \kCl crystals with a flat surface and interfacial modification of the device using SAMs. Fine control of the charge doping via gate voltage revealed two-step metal--insulator crossovers in the intermediate-temperature region: a rapid increase in the hole carrier density, followed by inversion of the sign of the temperature coefficient of the conductivity at $ \sim \esqh $
 without the occurrence of BM behavior. The high device quality also provided an opportunity to accurately examine the transport properties in the vicinity of the MIC, which revealed quantum criticality in the conductivity. The quantum criticality was retained down to $ T \sim 7$~K, whereas at lower temperatures metallic transport became insulating owing to a disorder-induced interference effect (weak localization). A bow-shaped crossover line of the MIT with the high temperature limit of $  \sigma _{\mathrm{high}-T}^{*} \sim \esqh $
 was successfully proposed, which implies that the origin of the observed criticality was the high-$T$ Mott quantum critical crossover, a ubiquitous feature for various types of Mott transition suggested on the basis of DMFT. This work provides a key insight into the criticality in the doping-driven Mott transition, as well as its relationship with the pressure-driven Mott transition at half-filling and the interaction-driven MIT in two-dimensional electron gas systems.

\begin{acknowledgements}
We thank K. Yokota, Y. Yokota and T. Sato for useful discussions. Financial support of this work was provided by Grants-in-Aid for Scientific Research (S) (JSPS KAKENHI Grant Number JP22224006, JP16H06346), RIKEN Molecular System, JST ERATO, and Nanotechnology Platform Program (Molecule and Material Synthesis) from the Ministry of Education, Culture, Sports, Science and Technology of Japan. 
\end{acknowledgements}

\vfill

\pagebreak
\appendix

%%%%%%%%%% Merge with Supporting Information %%%%%%%%%%
%--- Prefix a "S" to all equations, figures, tables and reset the counter------
\setcounter{equation}{0}
\setcounter{figure}{0}
\setcounter{table}{0}
\makeatletter
\renewcommand{\theequation}{S\arabic{equation}}
\renewcommand{\thefigure}{S\arabic{figure}}
\renewcommand{\thetable}{S\arabic{table}}
%\renewcommand{\bibnumfmt}[1]{[S#1]}
%\renewcommand{\citenumfont}[1]{S#1}
%%%%%%%%%%%%%%%%%%%%%%%%%%%%%%%%%%%%%%%%%%%%%%%%%%%

%% ======================================
%%  Below: SAM substrate preparation
%% ======================================

\section{Preparation of SAM-patterned substrate}
\label{sec_Substrate}

\begin{figure*}[!bt]
\centering
\includegraphics[width=\textwidth]{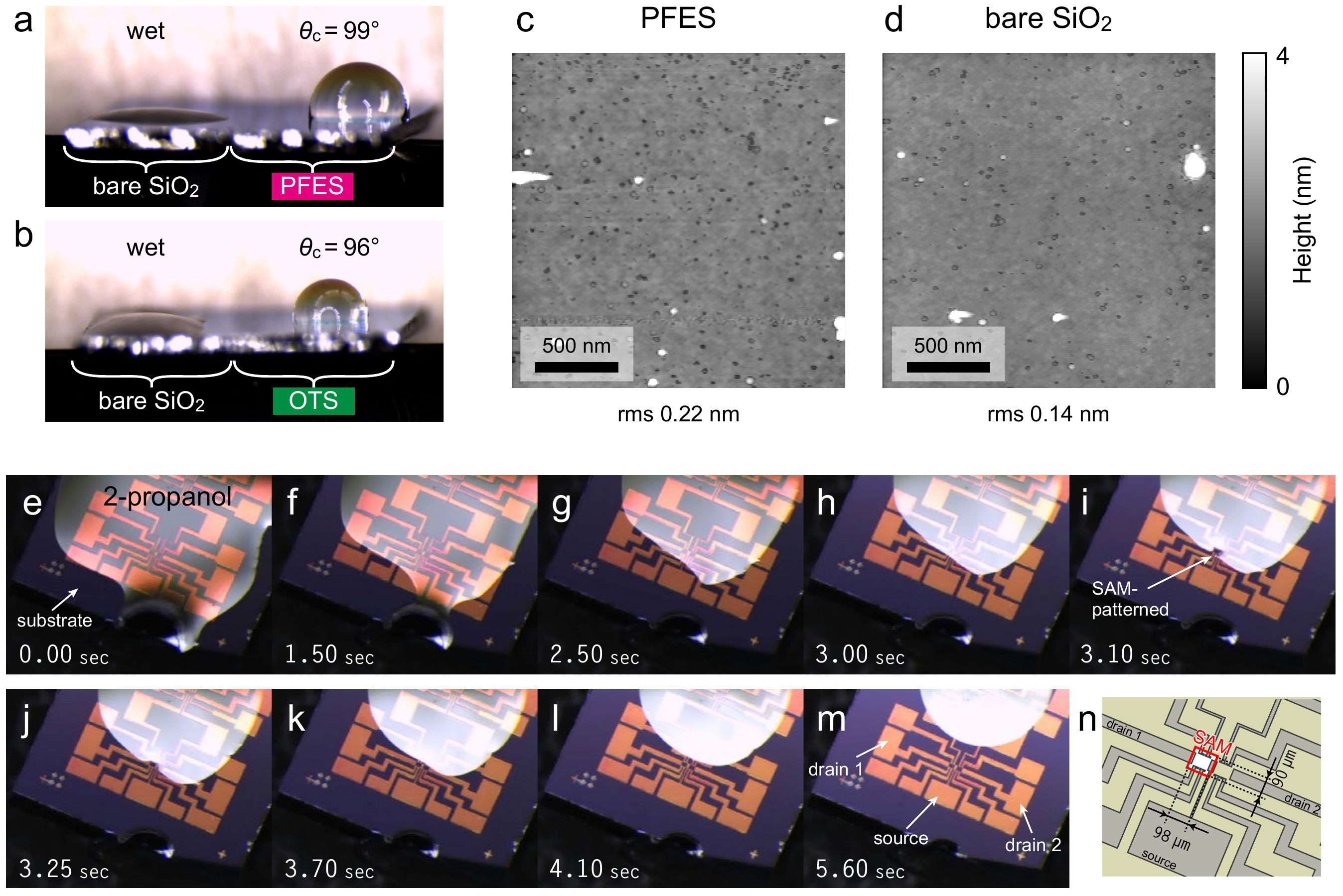}
\caption{\label{fig_SAM_estimation}
\textbf{Estimation for SAM-patterned substrate surfaces.}
(a, b) Wettability test by water droplets (side view). 
The right halves of the $\mathrm{SiO_2}$/Si substrate (6~mm $\times$ 6~mm $\times$ 0.3~mm) 
are functionalized by PFES (a) or OTS (b) through the same process 
as the SAM patterning of the channel areas. 
The $\theta_\mathrm{c}$ denotes the contact angle. 
(c, d) AFM topographic images for the substrate surface measured 
on the PFES-patterned area (c) and the bare $\mathrm{SiO_2}$ area (d). 
(e--m) Pre-lamination testing for wettability of the SAM-patterned substrate. 
Shown is the same substrate as is designed for conductivity measurements, 
on which Au/Ti microelectrodes (orange colored) and a SAM-functionalized area 
(PFES; 90~\si{\micro\meter} $\times$ 98~\si{\micro\meter} in dimension, 
as shown in panel (n)) are patterned. 
After dipping the substrate into the lamination solvent (2-propanol), 
the droplet of 2-propanol is left on the substrate ((e), bright part). 
During several seconds, 2-propanol is vaporized and the droplet gradually shrinks. 
When the edge of the droplet goes through the channel area, 
2-propanol on the SAM-patterned area is immediately repelled, 
which can be observed as deformation of the droplet edges (i, j). 
(n) Magnified view of electrode/SAM-pattern design around the channel area. 
SAM pattern is located between the electrodes labeled as drain 1 and source 
(highlighted by a red square).
}
\end{figure*}

Highly \textit{p}-doped Si wafers with a 300 nm thick thermally grown 
SiO$_{2}$ layer were obtained from a commercial source and used as 
the gate substrate. First, the Ti (20 nm)/Au (180 nm) electrodes were 
formed on top of the SiO$_{2}$ wafers by photolithography (NanoSystem 
Solutions, DL-1000RS) and thermal deposition (Katagiri engineering). The 
wafers were cleaned by the oxygen plasma and UV/ozone treatments, 
followed by a second photolithographic process to make micropatterns 
for subsequent SAM formation. After short time immersion in alkaline 
hydrogen peroxide (2:1:1 solution of 30\% w/w aqueous H$_{2}$O$_{2}$, 
alkaline photoresist developer (AZ developer, Microchemicals), and pure 
water), the wafers were immersed for 2--3 days in 2\% toluene/hexane 
solution of silylating agents: OTS (octyltriethoxysilane, TCI) or PFES 
(1H,1H,2H,2H-perfluorodecyltriethoxysilane, TCI). The wafers were 
sequentially rinsed by \textit{o}-xylene, 2-propanol/acetone, and 
alkaline hydrogen peroxide, followed by sonication in 2-propanol/acetone 
for 2 minutes. The dried substrates were stored in N$_{2}$. The 
completion of SAM functionalization was confirmed by wetness against 
pure water. 
Contact angles were 99\si{\degree} and 96\si{\degree} for the PFES
and OTS patterned areas, respectively (Figure \ref{fig_SAM_estimation}a,b).
Non-contact mode atomic force microscopy (AFM, 
Nanonavi II, SII) revealed that residues of the photography process 
were hardly seen on the substrate surface
(Figure \ref{fig_SAM_estimation}c,d).

%% ======================================
%%  Below: Crystal synthesis & lamination
%% ======================================

\section{Crystal growth and lamination process}
\label{sec_Crystal}

Growth processes of the \textit{$\kappa$}-(BEDT-TTF)$_{2}$Cu$[$N(CN)$_{
2}$$]$Cl (\kCl) crystal were performed in a similar manner 
described in the previous reports\cite{Kawasugi2008,Suda2015} 
although slight modifications were included to 
efficiently obtain a quantity of thin crystals with good surface 
flatness. The mixture of bis-(ethylenedithio)tetrathiafulvalene 
(BEDT-TTF; 24 mg), CuCl (52 mg), TTP\textperiodcentered$[$N(CN)$_{2}$$]$ (TPP = 
tetraphenylphosphonium ; 200 mg), and TPP\textperiodcentered Cl (90 mg) was dissolved in 
1,1,2-trichloroethane (10\% v/v ethanol) solution (50 mL), and 
transferred to the specially-designed electrochemical cell after the 
residues were filtered. Electrochemical oxidation of BEDT-TTF was 
performed in Ar by applying galvalnostatic current
(5~\si{\micro\ampere}) for 120 hours 
at 30~\si{\degreeCelsius}. Then the electrochemical cell was 
opened to air and kept at 20~\si{\degreeCelsius}
 for several hours until a large number of rhombic thin-layer
crystals were precipitated. Particularly thin \kCl crystals 
(30--150 nm thick) with a well-defined shape were selected using optical 
microscopy and pipetted into a petri dish filled with pure 2-propanol 
for rinsing. After the single crystal was gently transferred on an 
electrode-patterned substrate in 2-propanol, the substrate was carefully dried 
to complete lamination of \kCl. In order to confirm that the 
SAM-patterned area is steadily formed on the substrate, 
we performed a simple wettability test using 2-propanol 
\textit{just before} transferring \kCl onto it (Figure \ref{fig_SAM_estimation}e--m).
Solvents and reagents were 
used as received from commercial sources.

\begin{figure*}[tb]
\centering
\includegraphics[width=\textwidth]{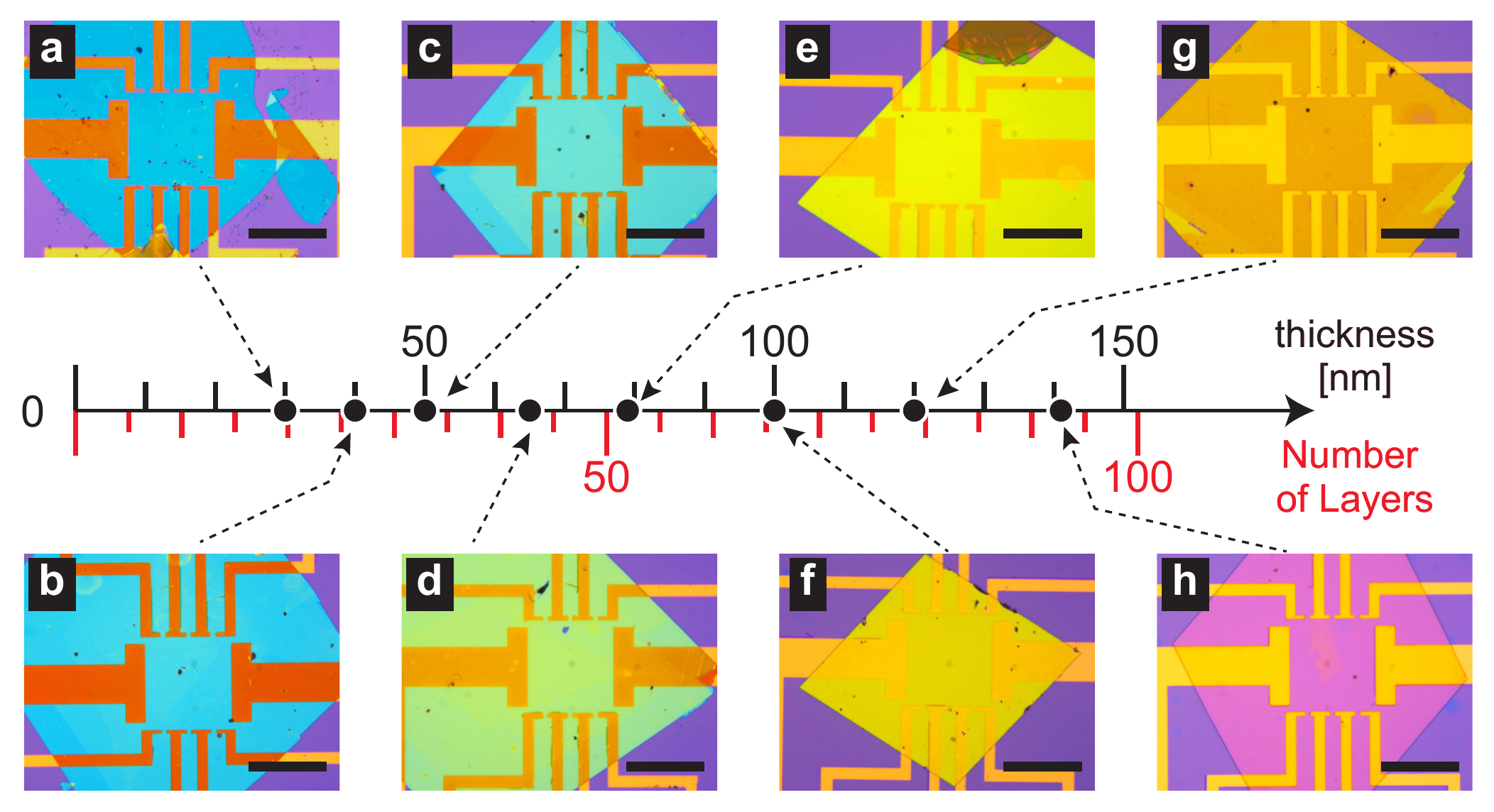}
\caption{\label{fig_photo_kCl}
\textbf{Optical photo images of Mott-FETs for various 
thicknesses of \kCl single crystals.} 
Thickness: 30~nm (a), 40~nm (b), 50~nm (c), 65~nm (d), 79~nm (e), 100~nm (f), 120~nm (g), 
and 141~nm (h), corresponding to 20, 27, 33, 43, 53, 67, 80, and 94 
layers of BEDT-TTF, respectively. Apparent colors of
\kCl crystals are dependent on their thicknesses, which is likely to be due to 
the optical interference at \kCl/(SAM/)SiO${}_{2}$/%
$p^{+}$-Si interfaces. Scale bar: 100~\si{\micro\meter}.
}
\end{figure*}

%% ======================================
%%  Below: Crystal Assessment
%% ======================================

\section{Assessment of thin-layer crystal channel}

\begin{figure*}[tp]
\centering
\includegraphics[width=0.8\textwidth]{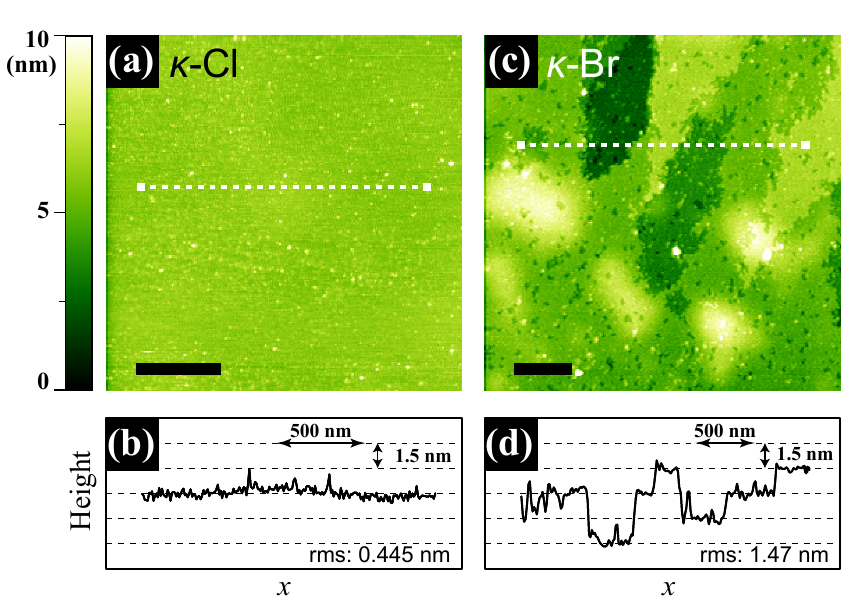}
\caption{\label{fig_AFM_kCl}
\textbf{AFM images of the surface for \kCl and 
\textit{$\kappa$}-Br.} (a,b) Topography for the surface of typical \kCl
crystals used in the present study (the same images as those shown in 
Figure \ref{fig_Schematic}c,d in the main part). Scale bar: 500~nm. (c,d) Topography for the surface of 
\textit{$\kappa$}-Br crystals, which are fabricated in accordance with the 
method previously reported.\cite{Kawasugi2008} As contrasted with \kCl, 
the surface of \textit{$\kappa$}-Br has many pits and molecular steps, the 
height of which corresponds to the thickness of one layer (1.5~nm) or 
two layers (3 nm) of BEDT-TTF.
}
\end{figure*}

Thickness of the laminated \kCl crystal was estimated by a surface 
profiler (P-15, KLA-Tencor). Figure \ref{fig_photo_kCl} displays microscope images of 
the laminated thin-layer \kCl crystals with various
thicknesses. Isolated crystals became almost colorless and 
transparent when their thicknesses are below $\sim$100~nm. However, by 
lamination on the (SAM/)SiO$_{2}$/\textit{p}-doped Si substrate, 
the thin-layer \kCl crystals became clearly visible and 
differently colored depending on their thickness, probably due to an optical
interference effect\cite{Schlich2013, Blake2007} (note that the SAM functionalization of the interface 
did not alter apparent colors of the \kCl crystals). Thus, one 
can estimate the thickness as well as the cleanness of the crystal by 
optical microscopy, which enables one to immediately screen out samples 
unsuitable for subsequent measurements. The selected devices were 
further analyzed by using AFM, and we eventually obtained 
\kCl single crystals without any grains on the surface and with the 
flatness of the single BEDT-TTF molecular layer level (1.5 nm; see 
Figure \ref{fig_AFM_kCl}). After forming a Hall-bar geometry 
by the laser ablation of the \kCl crystal, carrier transport 
measurements were performed.

\begin{table*}
\caption{\label{tbl_Device}Device characteristics.}
\begin{ruledtabular}
\begin{tabular}{lcccc}
Device & F\#1 & C\#1 & C\#2 & C\#3 \\\hline
SAM species & PFES & OTS & OTS & OTS \\
Crystal thickness of \kCl [nm] & 40 & 140 & 80 & 120 \\
$\mathVmic$ on SAM-covered area [V] & 76 & 76 & 51 & 48 \\
\end{tabular}
\end{ruledtabular}
\end{table*}

\vfill
\phantom{0}

\clearpage

%% ======================================
%%  Below: SAM dependence
%% ======================================

\section{SAM species dependence on transport properties}
\label{sec_SAMdep}

In the main part, we demonstrated that the metal--insulator transition (MIT) 
in \kCl FET is facilitated by chemical functionalization of 
the channel/gate dielectric interfaces using hydrophobic SAMs. 
To elucidate the interfacial SAM functionalization effect in depth, 
we present dependences of carrier transport properties on the SAM species in this section.

\begin{figure*}[tb]
\centering
\includegraphics[width=\textwidth]{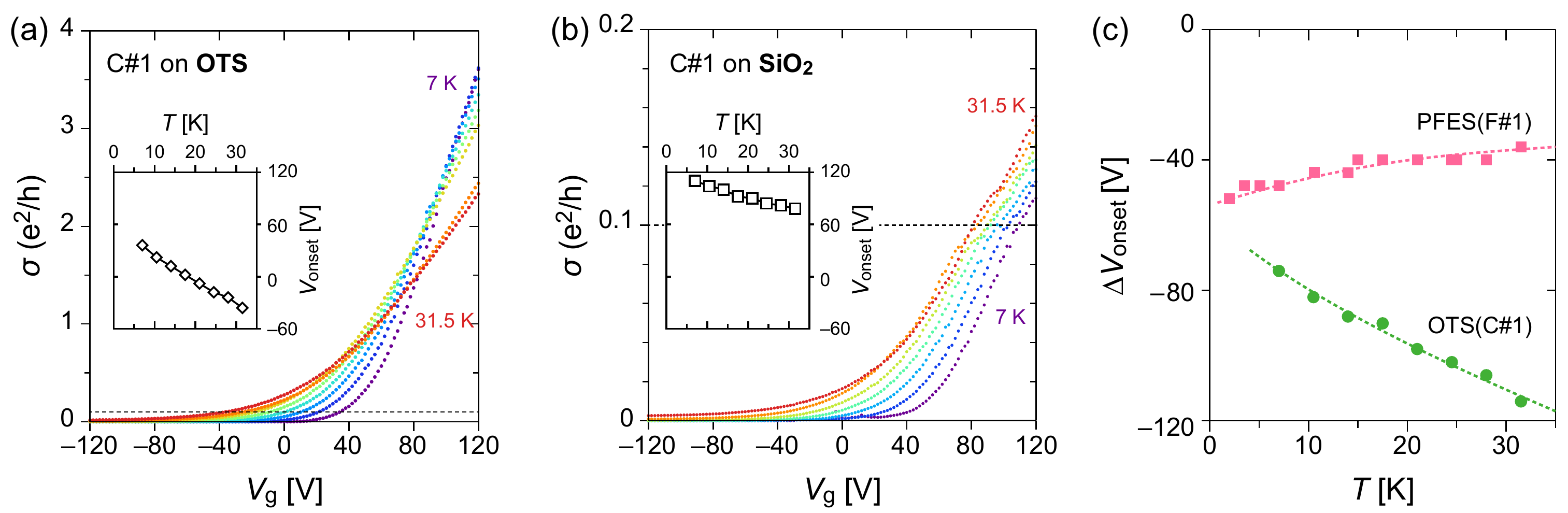}
\caption{\label{fig_SAM_dependence}
\textbf{OTS-SAM functionalization effect on carrier transport properties 
of a \kCl Mott-FET and comparison with the PFES-SAM functionalization effect.}
Transfer curves of device C\#1 for various $T$ on a OTS-covered substrate (a)
and on bare $\mathrm{SiO_2}$ (b). Temperatures are 7 to 35~K by 3.5~K steps. 
Dashed lines mark $\sigma=0.1\esqh$. 
The insets of (a) and (b) indicate the $T$-dependence of the onset voltage $V_\mathrm{onset}$, 
where the transfer curves crosses the $\sigma=0.1\esqh$ line. 
(c) Onset voltage shift due to SAM functionalization, $\Delta V_\mathrm{onset}$,
as a function of $T$ for devices F\#1 and C\#1. Dotted lines are guides to the eye.
}
\end{figure*}

Figure \ref{fig_SAM_dependence} shows the transfer curves for device C\#1, 
which has the same device geometries as device F\#1 (Figure \ref{fig_Schematic}a in the main part) 
but OTS is used as SAM instead of PFES. Since device C\#1 provides both \kCl/OTS/$\mathrm{SiO_2}$  and
\kCl/$\mathrm{SiO_2}$ interfaces on an identical single crystal \kCl, 
one can extract effects of the OTS functionalization on transport properties. 
For the OTS-covered area (Figure \ref{fig_SAM_dependence}a), a clear MIT is observed, 
where $  \sigma  $  increases (decreases) with decreasing $T$ on the highly (lightly) doped regime. 
One can recognize that the series of transfer curves are similar to that 
for the PFES-functionalized area on device F\#1 (Figure \ref{fig_Schematic}e in the main part); 
indeed, gate voltages of the crossover point ($\mathVmic $) are
identical (Table \ref{tbl_Device}), and the device mobility for device C\#1
 ($  \sim 180\ \si{cm^{2}.V^{-1}.s^{-1}}$   for $\mathVg =120$~V at 7 K)
is comparable to that for F\#1
($  \sim 200 \ \si{cm^{2}.V^{-1}.s^{-1}} $ for $ \mathVg =120$~V at 7 K). 
The bare $\mathrm{SiO_2}$ area exhibits insulating behaviors 
with \textit{n}-type FET characteristics in the whole \Vg\ range (Figure  \ref{fig_SAM_dependence}b),
which is qualitatively similar to the bare  $\mathrm{SiO_2}$ area on device F\#1. 
However, the magnitude of conductivity on bare $\mathrm{SiO_2}$ area of 
device C\#1 is suppressed to lower than that of device F\#1; 
for example, $  \sigma  \left(\mathVg =120\ \mathrm{V} \right)  $ at 31.5~K is $ 1.0\esqh $
for device F\#1, but only $ 0.16 \esqh $ for device C\#1. 
This discrepancy implies that there is a difference in the \textit{intrinsic} doping levels 
of \kCl between devices C\#1 and F\#1; that is, \kCl on device 
C\#1 should be at deeper insulating (less doped) region in the phase diagram 
in the absence of the extrinsic doping by the gate electric field or SAMs. 
The origin of the device-dependent intrinsic doping remains unclear, 
yet it is possibly due to charge imbalance at the crystal surface, 
which is commonly observed in crystalline materials consisting of cation/anion
stacking layers.\cite{Nakagawa2006}

In order to investigate the effect induced by SAMs, 
let us define the onset voltage $ V_\mathrm{onset} $
 of transfer curves as the gate voltage at which $  \sigma  $
 exceeds $ 0.1\esqh $. (Note that the magnitude of 
the threshold conductivity is arbitrary, and choice of conductivity 
other than $ 0.1\esqh $ does not change the results qualitatively.) 
The $ V_\mathrm{onset} $ shown in the insets of Figure
\ref{fig_SAM_dependence}a,b 
serves as an indicator of the boundary between the strongly localized regime 
and the relatively conductive regime. By taking the difference 
of $ V_\mathrm{onset} $ between the SAM-covered region and the bare $\mathrm{SiO_2}$ region
in an identical single-crystal device, $  \Delta V_\mathrm{onset}=V_\mathrm{onset} 
\left( \mathrm{SAM} \right) -V_\mathrm{onset} \left( \mathrm{SiO_2} \right)  $, 
one can cancel out intrinsic doping effect of \kCl and 
extract the SAM functionalization effect. 
Figure \ref{fig_SAM_dependence}c compares $  \Delta V_\mathrm{onset} $
for devices C\#1 and F\#1. Over the entire $T$ range, 
device C\#1 (OTS) shows more negative $  \Delta V_\mathrm{onset} $
than device F\#1 (PFES). The tendency of $  \Delta V_\mathrm{onset} 
\left( \mathrm{PFES} \right) > \Delta V_\mathrm{onset} \left( \mathrm{OTS} \right)  $
is consistent with the polarities of SAMs revealed in the
previous semiconductor-based device experiments;\cite{Kobayashi2004, Wang2011, YokotaK2011} 
that is, the PFES SAM induces positive charge on the FET channel 
due to the electric dipole moment of PFES molecules, 
whereas the OTS SAM serves as a neutral one. 
On the other hand, large negative offsets for $  \Delta V_\mathrm{onset} $
on both devices imply other dominant effects by SAM functionalization 
than the simple doping in the present Mott FETs. 
At the present stage it is difficult to determine the origin of the dominant factors 
due to lack of experimental methods that can access the buried interface 
underneath the several tens of nm thick \kCl crystal. 
Nevertheless, we can speculate that 
\textit{reduction of interfacial contaminants by hydrophobic SAMs} 
will bring about apparent negative $  \Delta V_\mathrm{onset} $. 
It is widely known that the SAM functionalization at the interface
effectively minimizes effects of charge traps, 
impurity-induced potential fluctuations, and atmospheric doping (e.g., oxygen or moisture) 
on the $\mathrm{SiO_2}$ substrate surface.\cite{Wang2011,Liu2011e} 
Furthermore, poor wetness of the SAM-covered region is favorable to 
reduce contamination during the \kCl lamination process. 
In the lamination process (see Supporting Information \ref{sec_Crystal}), trace amounts of ingredients 
used for electrochemical synthesis of \kCl such as neutral BEDT-TTF molecules, 
$\mathrm{Cl^{-}}$ and TTP$^{+}$ are unavoidably contained 
in the droplet of 2-propanol solvent even after careful rinsing.
Such ingredients are to be concentrated during evaporation of the solvent, 
and eventually left as residues in the interface. 
This type of contamination would be avoided on hydrophobic SAMs 
because their low wettability to 2-propanol results in quick elimination 
of solvent before the ingredients are fully concentrated (Figure \ref{fig_SAM_estimation}). 
Because the Mott transition is particularly sensitive to disorder, 
cleaner interfaces expected on the hydrophobic-SAM-covered area 
facilitates the MIT with low-level electron doping, 
leading to significant negative shift of $ V_\mathrm{onset} $. 
As a consequence, the charge doping by SAMs, the most prominent 
effect of SAM functionalization on conventional semiconductor devices, 
is considered to be merely a minor effect on the present Mott FET.

%% ======================================
%%  Below: high-T transport
%% ======================================
\section{High-\textit{T} transport}
\label{sec_highT}

In the main part, scaling analysis was performed for $T<40$ K. This is 
due to the fact that the charge transport above $T>50$ K no more exhibits 
the quantum critical behavior described by eq.~(1). In this section, we 
briefly consider the cause for such deviation.

\begin{figure*}
\centering
\includegraphics[width=0.9\textwidth]{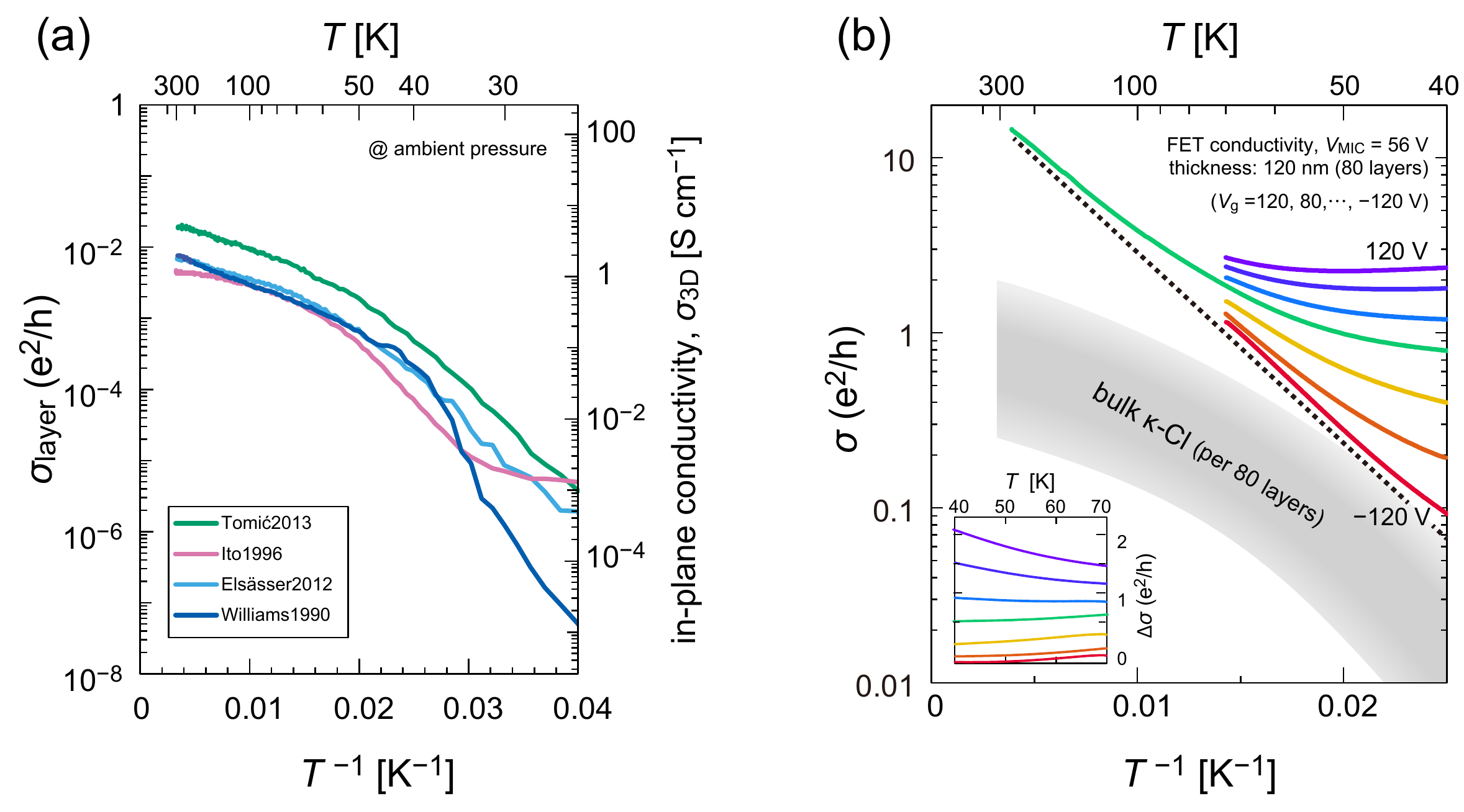}

\caption{\label{fig_highT}
\textbf{Carrier transport in the bulk channel of \kCl single crystal and 
its contribution to the FET conductivity at high \textit{T} ($\boldsymbol{T>}$ 50~K). }%
(a) Temperature dependence of the in-plane conductivity in bulk \kCl single
crystals found in literature. Data are adapted from 
Refs.~\onlinecite{Tomic2013,Ito1996,Elsasser2012,Williams1990}.
Left axis ($\sigma _\mathrm{layer}$) indicates the conductivity
per one BEDT-TTF conducting layer (thickness: 1.5~nm), 
which is calculated from three dimensional conductivity (right axis).
(b) Sheet conductivity of the FET device at high $T$
plotted for various gate voltages (from $\mathVg=-120$~V to 120~V 
with the step of 40~V, colored solid lines). 
The MIC point in the intermediate-\textit{T} region ($T=25-40$~K)
was $\mathVmic =+56$~V ($\mathymic =1.4 \esqh$), 
and the crystal thickness was 120~nm (80 layers). 
Dashed line provides \Vg-independent part of sheet conductivity. 
For comparison, shaded area shows the expected sheet conductivity 
of the bulk \kCl crystal, which is obtained by conversion of the three 
dimensional conductivities (panel a) into the values for the 
thickness of 120~nm (i.e., $\sigma=80\times\sigma_\mathrm{layer}$). 
Inset: \Vg-dependent part of the high-$T$ sheet conductivity
as a function of $ T $ for varying \Vg.
}
\end{figure*}

The electronic and magnetic properties of bulk \kCl have 
been intensively studied including the pressure effects.
\cite{Limelette2003a, Kagawa2009,Kanoda2006,Lefebvre2000}
At ambient pressure, bulk \kCl is a strong insulator showing 
basically activation-type dc-transport characteristics. In decreasing 
$T$, the system undergoes three transport regimes: (i) 
\textit{the high-T phase} above $T\sim 50$ K, with relatively small 
transport gap, (ii) \textit{the paramagnetic Mott insulator phase} 
(PMI, $T>25$~K), in which the electron correlation becomes significant and $
\mathrm{d}\sigma /\mathrm{d}T$ exhibits a larger transport gap, (iii)\textit{ 
the antiferromagnetic Mott insulator phase} (AFI) at low \textit{T}. 
Figure \ref{fig_highT}a compares the in-plane conductivity of \kCl at 
ambient pressure, which were reported in literature.
\cite{Tomic2013,Ito1996,Elsasser2012,Williams1990}
Although the absolute values of the conductivity are different, semiconductor--PMI 
crossovers are commonly observed near $T\sim 50$ K. 
The transport gap $\Delta$, which is related to the conductivity by 
$\sigma \propto \exp(\Delta /2k_\mathrm{B}T)$,
 is $\sim 20$ meV in the semiconductor regime, whereas that in the PMI 
regime depends on samples taking the values within 50--100 meV. 

We can estimate the sheet conductivity $\sigma _\mathrm{layer}$ per layer 
from the in-plane conductivity ($\sigma _\mathrm{3D}=L/WDR$, where $R$ is 
resistance and $L$, $W$, $D$ are the channel length, width, and 
thickness, respectively) and the thickness of one layer ($1.5$~nm), 
as indicated in Figure \ref{fig_highT}a. As is mentioned previously, the typical 
thicknesses of thin-layer \kCl crystals used in our FET 
devices are in the range of 40--140~nm, corresponding to 26--93 BEDT-TTF 
layers. Since the observed $\sigma $ is sum of conductances of the 
internal and surface channels, the contribution from the non-doped internal 
channel becomes significant in the carrier transport on FET in the high-$T$
regime.\cite{Kawasugi2009a} On the other hand, the rapid decrease of
$\sigma_\mathrm{layer}$ from $\sim 10^{-3}\,\esqh$ with lowering \textit{T} is 
favorable to precisely evaluate the critical phenomena around the MIC on the 
surface channel occurring at $\sigma \sim \lbrack 10^{-2}-10^{0}\rbrack 
\esqh$. 

Figure \ref{fig_highT}b represents the high-\textit{T} sheet conductivity of the \kCl
FET (thickness of \kCl: 120~nm) for various gate 
voltages. Note that here the thin-layer \kCl crystal is 
laminated on the OTS/SiO$_{2}$/doped-Si substrate and is subject to tensile 
strain upon cooling.\cite{Yamamoto2013} A metal--insulator crossover (MIC) is 
observed at $\mathVmic=56$~V in the intermediate-$T$ region 
around 30 K (not shown). In increasing $T$, \Vg
-dependence of $\sigma $ becomes smaller and $\sigma (T,\mathVg)$ 
curves converge to the insulating one for $\mathVg=0$ V. To roughly 
estimate the surface-channel contribution on $\sigma $, we plot the 
\Vg-dependent part of $\sigma (T,\mathVg)$ ($\Delta \sigma $, 
difference between the observed $\sigma $ and 
the \Vg-independent part of conductivity indicated by dashed line in 
Figure~\ref{fig_highT}) 
for varying \Vg. The metallic transport ($\mathrm{d}\Delta\sigma 
/\mathrm{d}T<0$) is limited to $\Delta\sigma \gtrsim \esqh$; therefore no hallmarks 
of bad-metal (BM) behavior were found up to 70~K. The apparent transport gap of 
the FET in the high-\textit{T} limit is $\sim$50~meV, which is more than 
two times higher than that for bulk \kCl crystal. This 
could be due to the fact that, for the FETs, the thin-layer 
\kCl crystal laminated on the silicon substrate is subject to tensile 
strain (effectively \textit{negative} in-plane pressure) by decreasing 
\textit{T} because of the large difference in the thermal compressibility 
between the soft \kCl crystal and the incompressible 
substrate.\cite{Kawasugi2009a,Yamamoto2013,Suda2014} 
Such an effective pressure is expected to reduce 
the transfer integrals between BEDT-TTF dimers, enhancing the insulating 
character for the internal channel of the laminated \kCl of 
FET than the purely bulk crystal. On the other hand, the sheet 
conductivity of FET exceeds $10\esqh$ at 250~K, which 
corresponds to $0.2\esqh$ per one-layer BEDT-TTF channel, 
and is nearly an order of magnitude higher than the 
conductivity expected for a bulk crystal. It is unclear what makes this discrepancy; 
one possibility is the difference in the measuring geometry (the aspect ratio of the
channel and the wiring geometries) between the bulk and FET samples.

%% ======================================
%%  Below: 3D phase diagram
%% ======================================

\section{Phase diagrams of doping/pressure-driven Mott transitions}
\label{sec_PhaseDiagram}

\begin{figure*}[!bth]
\centering
\includegraphics[width=\textwidth]{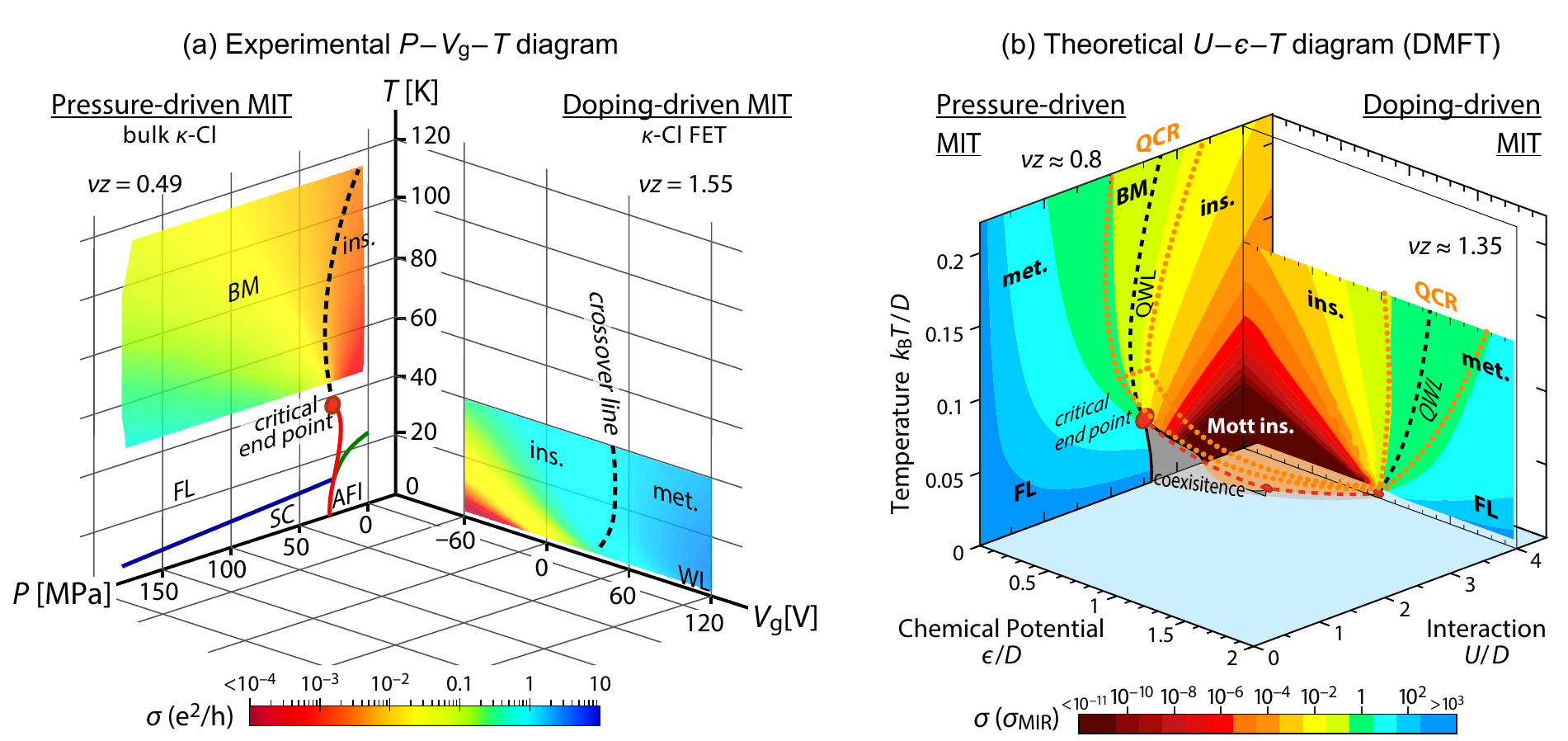}

\caption{\label{fig_3Dmap}
\textbf{Phase diagrams for doping-driven and pressure-driven 
metal--insulator transitions around a Mott insulator.} (a) 
Experimentally determined pressure--gate-voltage--temperature($
P-\mathVg -T$) phase diagram for \kCl with contour maps of conductivity 
in units of $\esqh$. The phase diagram for a pressure-driven MIT at 
half-filling (left half of the panel) is determined from measurement for 
a bulk \kCl single crystal (adapted from ref.~\onlinecite{Furukawa2015a}). 
(b) Interaction--chemical-potential--temperature ($U-\epsilon -T$)
phase diagram based on DMFT solutions of the maximally frustrated Hubbard 
model, which is adapted from refs.~\onlinecite{Vucicevic2015,Vucicevic2013}. $U$ and $D$ 
represent on-site Coulomb repulsion energy in the Hubbard Hamiltonian and 
the half bandwidth, respectively, and $\epsilon $ is the chemical 
potential measured from the half-filled level. 
}
\end{figure*}

Here we map out the 3D phase diagrams around a Mott insulator based on 
both experiments and theories. Figure \ref{fig_3Dmap}a shows the experimental mapping of 
the phases and the crossover lines. The phase diagram for a doping-driven 
metal--insulator transition (MIT), shown in the $T-\mathVg$ plane,
is determined from the present experiment on a
single-crystal \kCl FET (device C\#2). The conductivity of bulk 
\kCl at half-filling (the $T-P$ plane)\cite{Furukawa2015a} is converted to that per 
layer ($\sigma _\mathrm{layer}$, defined in Supporting 
Information \ref{sec_highT}) for comparison with the sheet conductivity obtained from 
measurement on the FET. The MIC lines are denoted by 
dashed lines. Taking into account the \textit{n}-type characteristics 
and the tensile effect for the \kCl FET fabricated onto the SiO${}_2$/Si
substrate,\cite{Yamamoto2013} the coordinate origin for the doping-driven MIT is 
shifted toward the negative $P$ and positive $\mathVg$ directions.
The quantum critical scaling exponents for the crossover regions 
are $\nu z=0.49$ (ref.~\onlinecite{Furukawa2015a}) and 1.55 for the pressure-driven and 
doping-driven transitions, respectively.

Figure \ref{fig_3Dmap}b shows the theoretical phase diagram calculated by a DMFT 
method.\cite{Vucicevic2015,Vucicevic2013} The conductivity is represented in units of 
the Mott--Ioffe--Regel limit $\sigma _\mathrm{MIR}$
($\approx \sqrt[]{2\pi}\esqh$ for quasi-2D systems\cite{Gunnarsson2003}), 
which is obtained by the condition of $l\approx a$ 
($l$: mean free path of carrier, $a$: lattice spacing).\cite{Hussey2004}
The low-\textit{T} coexisistence domes reside 
at the boundary of the Mott insulator and Fermi liquid (FL) 
phases, which accompany high-\textit{T} quantum critical regions 
(QCR, the areas enclosed by the orange dotted lines) along the quantum Widom 
lines (QWL). Those phases continuously connect in the $U-\epsilon -T $ 
space,\cite{Vucicevic2015,Sordi2012} yielding similarity between the phase diagrams for 
the pressure-driven and doping-driven Mott transitions. However, the nature 
of the QCR is significantly different between the two types of Mott 
transitions; the QCR for the doping-driven transition extends to low \textit{T
} ($\nu z\approx 1.35$, ref.~\onlinecite{Vucicevic2015}) and the critical conductivity is $
\sim \sigma _\mathrm{MIR}$, whereas that for the pressure-driven transition at 
half-filling is observed above a high critical-end-point temperature ($
\nu z\approx 0.8$, ref.~\onlinecite{Vucicevic2013}) and the critical conductivity is suppressed to 
below $\sim 0.1 \sigma _\mathrm{MIR}$ (BM behavior). The experimental phase 
diagram is consistent with the theoretical one, although the weak 
localization (WL) is observed on the actual \kCl FET instead 
of FL behavior possibly due to disorders at the FET interface. (AFI: 
antiferromagnetic Mott insulator phase, SC: superconducting phase.)

%% ======================================
%%  Below: Critical Scaling analysis with QUANTUM THOERY
%% ======================================

\section{Scaling analysis for quantum criticality}
\label{sec_Scaling}

As described in the main text, the quantum phase transition (QPT) is a
phase transition between competing ground states,
which can be tuned by a non-thermal parameter of the quantum systems. 
In contrast to classical phase transitions driven by 
the thermal fluctuation, the QPT is driven by the quantum fluctuation stemming from 
Heisenberg's uncertainty principle (or, more specifically, non-commutativity 
between the kinetic and potential operators in the Hamiltonian). Therefore, 
the QPT is intrinsically characterized by coupling of the spatial (static) 
and temporal (dynamical) fluctuations. Mathematically, these 
fluctuations are formulated as the correlation length $\xi $ in the 
spatial direction and the dynamical correlation length $\xi _{\tau }$ 
in the (imaginary) time direction, respectively. As the system 
approaches the critical point by tuning external parameters (e.g., 
magnetic/electric field, pressure, etc.), the two characteristic lengths 
diverge in coupled power-law forms of $\xi \propto \vert \delta \vert ^{-\nu 
}$ and $\xi _{\tau }\propto\xi ^{z}$, where $\delta $ is the distance 
from the critical point, and the power exponents $\nu $ and $z$ are 
the correlation exponent and the dynamical exponent, respectively. At finite 
\textit{T}, $\xi _{\tau }$ is cutoff by $\hbar /k_\mathrm{B}T$, so that the 
characteristics of the system are described by one parameter $\vert 
\delta \vert T^{-1/{\nu z}}$ in the QCR 
(i.e., $T\gtrsim c\vert \delta \vert ^{\nu z}$). Applying this finite-size 
QPT theory to the MIT, one can obtain the 
expression of eq.~(\ref{eq_QCS}) in the main text.

\begin{figure}
\centering
\includegraphics[width=0.5\textwidth]{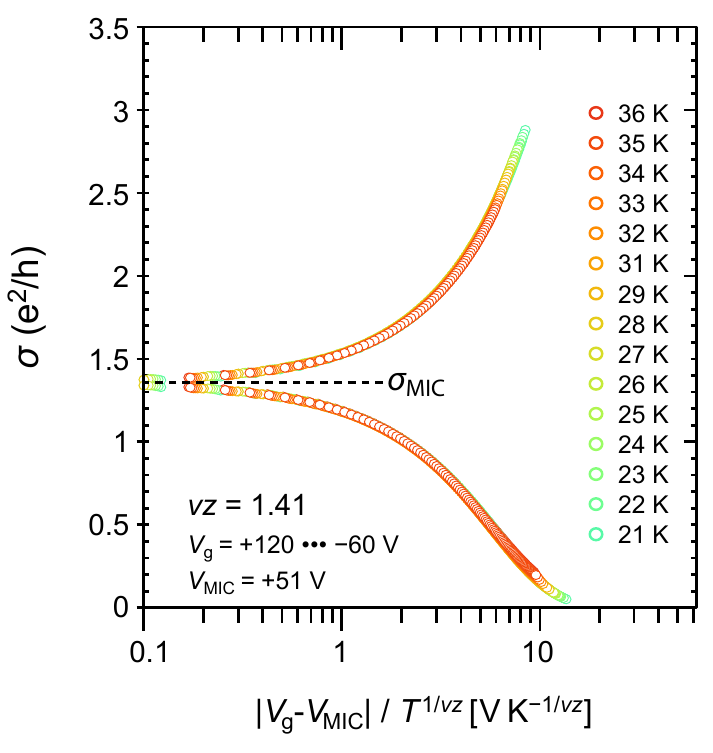}

\caption{\label{fig_midTscaling}
\textbf{One-parameter scaling at intermediate \textit{T} for device C\#2.}
Family of conductivity data plotted as a function of
$\vert \mathVg-\mathVmic \vert T^{-1/\nu z}$. 
Conductivity at the point of  $\vert \mathrm{d}\sigma / \mathrm{d}T
\vert_{\mathVg}=0$ (i.e., $\mathymic$) is marked by a dashed line. $\nu z=1.41$ is 
estimated from a least-square analysis.
}
\end{figure}

In the simplest one-parametric scaling on two-dimensional MITs, 
the conductivity of the crossover line, $\sigma ^{*}$
(corresponding to the `separatrix' in the 
nomenclature of renormalization group theory), is a 
constant of $\sigma _\mathrm{MIC}\sim \esqh$, and the critical point ($
\delta =0$) is defined on the constant scaling variable line in eq. 
(1), which results in
\begin{equation}
\sigma =
\sigma _\mathrm{MIC}\mathcal{F}_{\pm }
	\left(\frac{\vert \mathVg-\mathVmic\vert }{T^{1/\nu z}}\right),
\label{eq_midTscaling}
\end{equation}
for the MIC induced by gate-electric-field doping. In fact, eq.~(\ref{eq_midTscaling})
\textit{does} describe the conductivity for a narrow \textit{T}-range 
of 22 K $\lesssim T\lesssim$ 35 K, at which one can see an obvious crossover from 
insulating to metallic regimes with a conductivity plateau 
$\vert {\mathrm{d}\sigma }/{\mathrm{d}T}\vert _{\mathVg=\mathVmic}=0$
in between (Figure~\ref{fig_Critical}b in the main part). In Figure \ref{fig_midTscaling}, 
family of $\sigma (\mathVg,T)$ for device 
C\#2 are re-plotted with respect to $\vert \mathVg - \mathVmic\vert T^{-1/\nu 
z}$. All the points including both the metallic (${\sigma }/{\mathymic}>1$)
and insulating (${\sigma }/{\mathymic}<1$) 
sides collapse onto two branches by providing an exponent as $\nu z=1.4\pm 
0.2$. In terms of the form of the  universal scaling function, the metallic side 
can be approximated by a linear function: $\mathcal{F}_{+}(y)=1+Ay$ with $
A=0.14$, that is, the metallic conductivity is given by the sum of 
a constant and a term proportional to $T^{-1/\nu z}$, which 
represents the non-Fermi liquid behavior in the vicinity of the MIC. For the 
the insulating side, the universal scaling function has a mirror symmetry of
$\mathcal{F}_{-}(y)={1}/{\mathcal{F}_{+}(y)}$ for $\sigma \sim \esqh$, yet in the deeper 
insulating regime, it has an exponential asymptotic form of $
\mathcal{F}_{-}(y)\simeq \exp (-By^{C})$ with $B=0.12$ and $C=1.3$. Using $\nu 
z=1.4$, one obtains $\sigma _\mathrm{ins}\simeq \mathymic\exp\lbrack -({\Delta 
}/2T)^{0.92}\rbrack $, which can be approximated by the activated-transport
behavior $\ln \sigma \propto 1/T$ above the intermediate-\textit{T} regime 
(Figure \ref{fig_Energy} in the main part).

In the framework of renormalization group theory, however, $\sigma 
^{*}(T)$ is not necessarily identical to the na{\"i}ve definition of the MIT, 
$\sigma ^{*}=\mathymic$. That is, metal and insulator are 
distinguished by behaviors of the dimensionless conductance on the scale 
transformation of the length scale and the imaginary time scale (or 
temperature); the conductance goes to zero in the limit of $L\to \infty 
$ or $T\to 0$ through renormalization procedures for insulator, whereas 
not for metal. The 'separatrix' is the line which delimits the boundary 
between metal and insulator regimes in the flow diagram of 
renormalization, and often has a \textit{T}-dependent form of $
\sigma ^{*}$ in practice. Therefore, it is essential for a scaling 
analysis to adequately identify the location of the 'separatrix' or 
the crossover line in the phase diagram depending on the system under 
study.

\vspace{1em}

\textit{High-T Mott quantum critical scaling }(\textit{HT-MQCS}).
---As discussed in the main text, it is generally expected 
for Mott transitions that the MIC in the conductivity behavior exhibits quantum 
criticality along a crossover line referred to as `quantum Widom line 
(QWL)'. The QWL may show a bow-shaped trajectory in the phase diagram, and 
should smoothly converge to a critical end point at a low $T$ (if 
present). For the pressure-driven Mott transition at half-filling, where the 
almost perfect mirror symmetry $\mathcal{F}_{+}(y)=1/\mathcal{F}_{-}(y)$ holds for the wide 
conductivity range in the crossover region, one can identify the QWL simply 
by tracing the inflection points of conductivity, $\mathrm{d}^{2}\log \sigma /\mathrm{d}P^{2}
$ ($P$: pressure).\cite{Furukawa2015a} For the doping-driven Mott transition in 
the present study, however, such a straightforward identification of the QWL 
was found to be impossible due to various factors (i.e., poor mirror 
symmetry, weak localization at low $ T $, and lack of the 
critical end point in the range of observations). Therefore, we figured 
out a possible candidate for the QWL by an optimization calculation with 
the following procedure:

\newcounter{numberedCntG}
\begin{enumerate}
\item Setting series of initial $\lbrace \sigma ^{}(T_{i})\rbrace $ for 
each temperature point $T_{i}$.

\item Calculating $\lbrace \mathVg^{}(T_{i})\rbrace $ by linear 
interpolation from $\sigma ^{}(T_{i})$ and the adjacent experimental data 
points for $\sigma (\mathVg,T_{i})$.

\item Obtaining the above-mentioned approximate/asymptotic forms for 
the universal scaling function $\mathcal{F}_{\pm }({\vert \mathVg-\mathVg^{*}\vert 
}/{T^{1/\nu z}})$.

\item Executing a least squares calculation under a constraint 
condition (see below) so that the family of conductivity data are 
converged to $\sigma ^{*}(T)\times \mathcal{F}_{\pm }({\vert 
\mathVg-\mathVg^{*}\vert }/{T^{1/\nu z}})$, with a single critical exponent 
$\nu z$. Here, the fitting parameters are $\nu z$, series of $
\lbrace \sigma ^{*}(T_{i})\rbrace $, and the variables included in 
the approximate/asymptotic forms of $\mathcal{F}_{\pm }$. The series of $\lbrace 
\mathVg^{*}(T_{i})\rbrace $ are updated (procedure 2) for every 
optimization calculation loop. 
\setcounter{numberedCntG}{\theenumi}
\end{enumerate}
\begin{quotation}
\noindent
\textit{\underline{CONSTRAINT CONDITION}: }$\sigma ^{*}(T)=\mathymic$
for $T>23$~K (for definition of $\mathymic$, see Figure \ref{fig_midTscaling} 
and the main text). This condition is introduced for consistency with 
the theoretically suggested QWL for a doping-driven Mott transition;\cite{Vucicevic2015} 
that is, at high \textit{T}, the conductivity on the QWL becomes
independent of \textit{T} and coincides with the Mott--Ioffe--Regel limit.
\end{quotation}
Consequently, one can obtain $\nu $z giving the best scaling, as well 
as the \textit{optimized} crossover line $\mathVg^{*}(T)$.

\begin{figure*}[tbh]
\centering
\includegraphics[width=0.9\textwidth]{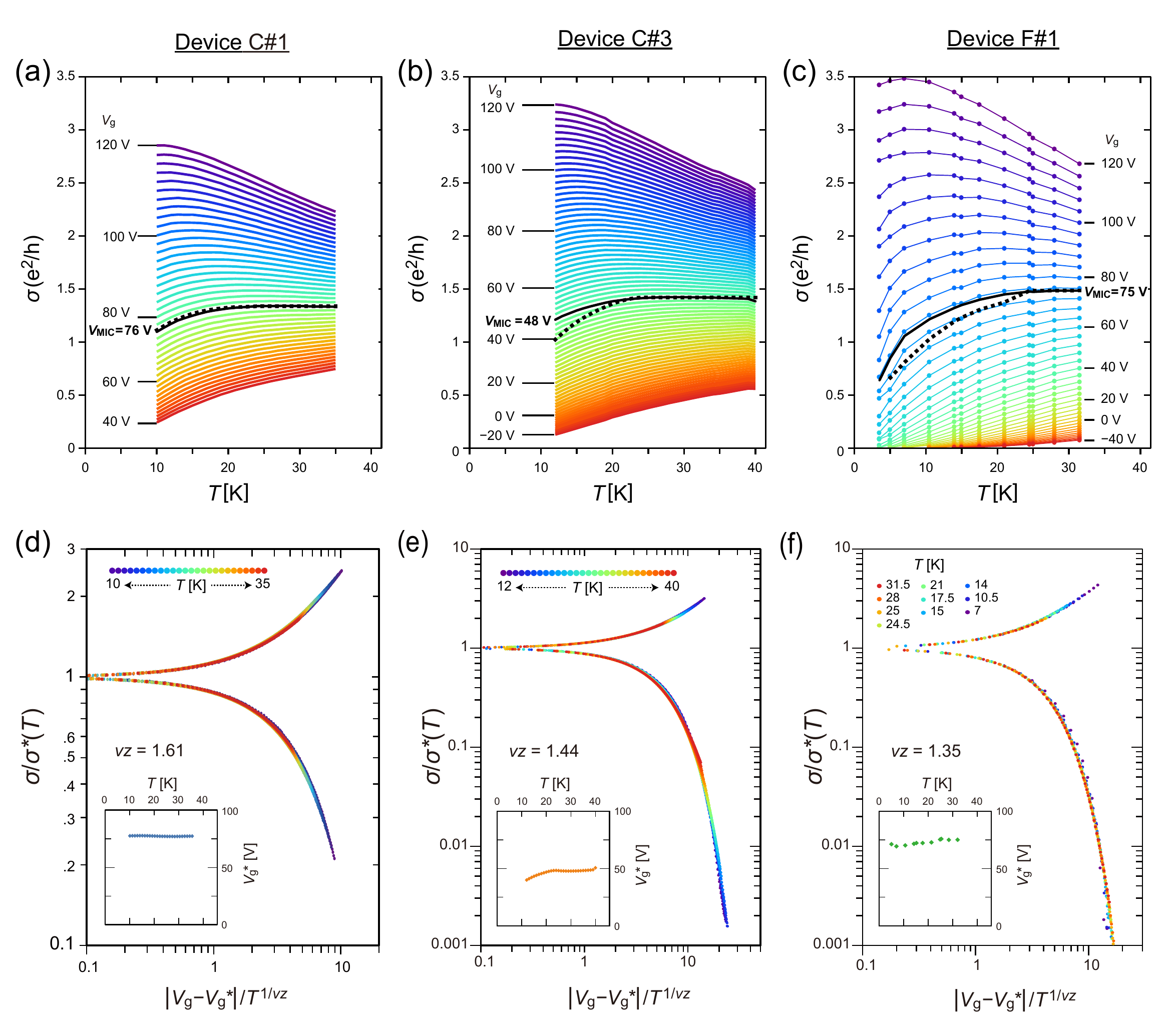}

\caption{\label{fig_ScalingOthers}
\textbf{Transport characteristics in the vicinity of 
the doping-driven metal--insulator transition for devices 
C\#1, C\#3, and F\#1.}
 (a-c) Temperature dependence 
of sheet conductivity as a function of \Vg\ for devices 
C\#1 (a), C\#3 (b), and F\#1 (c). Single crystals of \kCl are 
laminated on the OTS/SiO${}_2$ (300~nm)/doped-Si substrates
(C\#1 and C\#3) or the PFES/SiO${}_2$ (300~nm)/doped-Si substrates (F\#1).
Solid black lines represent $\mathVg = \mathVmic$ 
lines giving $\vert \mathrm{d}\sigma /\mathrm{d}T
\vert_{\mathVg=\mathVmic}=0$ in the intermediate-\textit{T} regime
($T\sim 25-35$~K). Dashed lines highlight the crossover lines 
$\sigma^{*}$ between the metallic and insulating regimes (possible QWL),
by which the quantum critical scaling (eq.~(\ref{eq_QCS}) in the main text) is retained
in the widest \textit{T} range with $\mathrm{d}\sigma^{*}/\mathrm{d}T=0$
at intermediate \textit{T}. (d-f) Scaling plots of conductivity 
normalized by $\sigma^{*} (T)$ for devices C\#1 (d), C\#3 (e), and F\#1 (f),
yielding the critical exponents $\nu z=$ 1.61, 1.44, and 1.35, respectively. 
Insets: Trajectories of $\sigma^{*}$ on the \Vg\ versus \textit{T} plane.
}
\end{figure*}

Figure \ref{fig_ScalingOthers} shows the result for analysis of HT-MQCS for devices C\#1, 
C\#3, and F\#1, following the above procedure. As the result for device C\#2 
shown in the main part, one can recognize good convergence of family of 
conductivity data on the two branches of the universal scaling function for a
wide range of conductivity. The critical exponents $\nu z$ are 1.66, 
1.41, and 1.35 for devices C\#1, C\#3, and F\#1, respectively, being 
close to that for device C\#2, $\nu z=1.55$ (Figure \ref{fig_Critical} in the main 
part). For all devices, the crossover conductivities $\sigma ^{*}(T)$ are 
increasing functions with \textit{T} at low \textit{T}, which is 
consistent with DMFT.\cite{Vucicevic2015}

\clearpage

\textit{Quantum critical scaling with a conventional crossover line.}%
---To validate the HT-MQCS using an optimization calculation described above, 
let us consider a conventional approach to experimentally identify the 
crossover line of the MIC within the one-parametric scaling. The crossover line 
is expected to fulfill two conditions as follows:\cite{Dobrosavljevic2012}

\newcounter{numberedCntD}
\begin{enumerate}

\item
$\sigma ^{*}=\sigma (\mathVg=\mathVc,T)$. Namely, the crossover line is 
defined on a constant doping-density (critical density) line, $
\mathVg^{*}=\mathVc$ ($\mathVc$: critical gate voltage). 
It is implicitly postulated that the gate voltage (or 
the doping density) should be a good scaling variable.

\item
$\sigma ^{*}\sim T^{p}$. The power-law in the crossover conductivity 
reflects the self-similarity nature of underlying physics at the critical 
point. The specific case of $p=0$ corresponds to the na{\"i}ve definition of the
MIC with ${\mathrm{d}\sigma ^{*}}/{\mathrm{d}T}=0$, and is adopted in some systems 
exhibiting doping-driven metal--insulator (or superconductor--insulator) 
transitions\cite{Gantmakher2010}. Nonzero $p$ is also, indeed, applicable to the 
case of low-density two-dimensional electron gas systems.
\cite{Knyazev2008,Pradhan2015}
\setcounter{numberedCntD}{\theenumi}
\end{enumerate}

\begin{figure*}[tbh]
\centering
\includegraphics[width=0.95\textwidth]{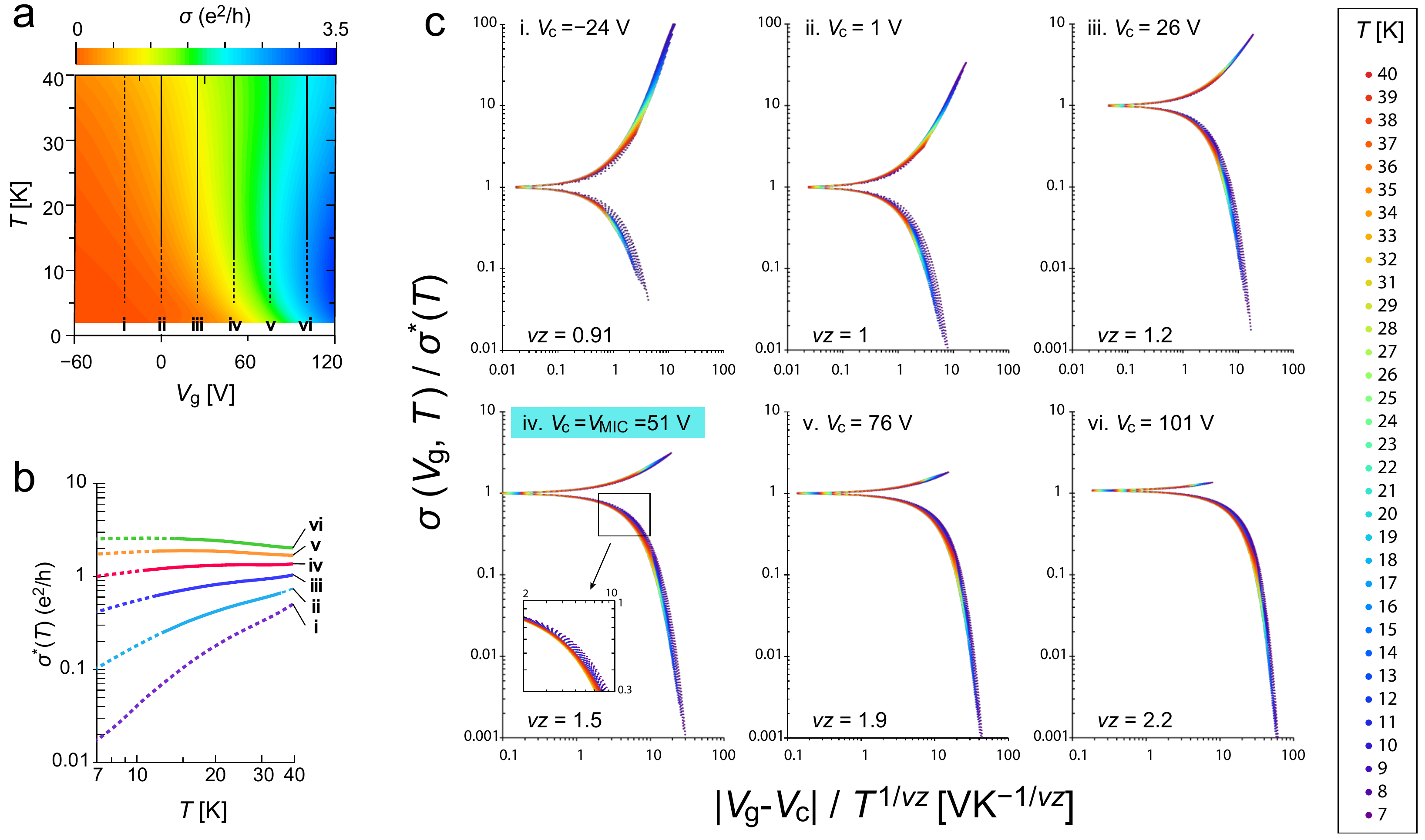}

\caption{\label{fig_conventional}
\textbf{Quantum critical scaling along the crossover lines 
with constant critical gate voltage,} $\mathbf{\bvec{V}_{g}^{*}%
=\bvec{V}_{c}}$. Scaling 
analyses were performed for varying $\mathVc$: i)~$-$24~V, ii)~1~V, 
iii)~26~V, iv)~51~V($=\mathVmic$), v)~76~V, and vi)~101~V. (a) Location of 
the crossover lines in the $\mathVg-T$ plane. (b) Crossover 
conductivities $\sigma ^{*}$ as a function of temperature in the 
log-log plot. For (a) and (b), solid (dashed) lines represent the 
\textit{T} regions where the scaling of the conductivity data is 
successful (failed) for each $\mathVc$. (c) Scaling plots in the 
\textit{T} region where the HT-MQCS is successful (Figure \ref{fig_Critical} in the main 
part). The inset of panel (c-iv) highlights the poor scaling quality in 
the insulating side of ${\sigma }/{\sigma ^{*}}\lesssim1$ at low 
\textit{T}. All data were collected for device C\#2.
}
\end{figure*}

Figure \ref{fig_conventional} displays the results of the quantum critical scaling under various 
$\mathVc$ ranging from $-24$~V (insulating regime) to 101 V (metallic 
regime at intermediate \textit{T}). The characteristics of scaling 
sensitively change depending on the choice of $\mathVc$ as follows. 
First, the critical exponent $\nu z$ for $\mathVc=\mathVmic$ is 1.5, 
being close to the value under the HT-MQCS, but it becomes smaller (larger) 
for higher (lower) $\mathVc$, ranging from 0.9 to 2.2 (Figure~\ref{fig_conventional}c). 
Second, the quality of the conductivity scaling also depends on $\mathVc$; 
the scaling is successful in the widest \textit{T} range at $
\mathVc \sim \mathVmic$ (solid lines in Figure
\ref{fig_conventional}a,b), whereas the 
conductivity data convergence becomes worse as $\mathVc$ is away from $
\mathVmic$. Even for $\mathVc \sim \mathVmic$, poor data convergence below 
$T\sim 12$ K becomes prominent particularly for the insulating regime 
(inset in Figure \ref{fig_conventional}c-iv), which is in contrast to the HT-MQCS yielding 
successful scaling for $T>7$ K. The \textit{T}-dependence of 
the crossover conductivity deviates from the power-law form irrespective of $
\mathVc$ (Figure \ref{fig_conventional}b), which means that no crossover lines meet the 
condition 2 above mentioned. Indeed, the accurate low-$ T $ behavior 
of the crossover line remains unclear in doping-driven Mott transitions, 
and awaits further theoretical investigation beyond DMFT. Nevertheless, 
these tendencies imply that HT-MQCS is essential for the doping-driven MIC 
in Mott insulators.

\clearpage
%% ======================================
%%  Below: Critical Scaling analysis with CLASSICAL THEORY
%% ======================================

\section{Classical critical scaling}
\label{sec_Classical}

A number of previous studies\cite{Limelette2003, Kagawa2005, Abdel-Jawad2015} 
for Mott insulators have revealed that the pressure-driven Mott transitions at half-filled 
show phase diagrams analogous to the liquid--gas transition in a classical fluid.
Around a critical end point (CEP) at a finite $T$, 
physical quantities such as electrical conductivity were recognized 
to exhibit classical critical behaviors driven by thermal fluctuations, 
whereas transient from classical to quantum critical behaviors occur at a higher $T$.
In this section, we verify the applicability of the classical phase transition for the present system. 
First, we present the formalism we use for the classical critical scaling analysis. 
Starting from a fundamental homogenous equation describing the classical criticality, 
two different interpretations for conductivity scaling are introduced. 
Then we compare the derived theoretical scaling formula with the experimentally 
observed conductivity behavior around the doping-driven metal--insulator transition in the present study.

\begin{figure*}[b]
\centering
\includegraphics[width=0.5\textwidth ]{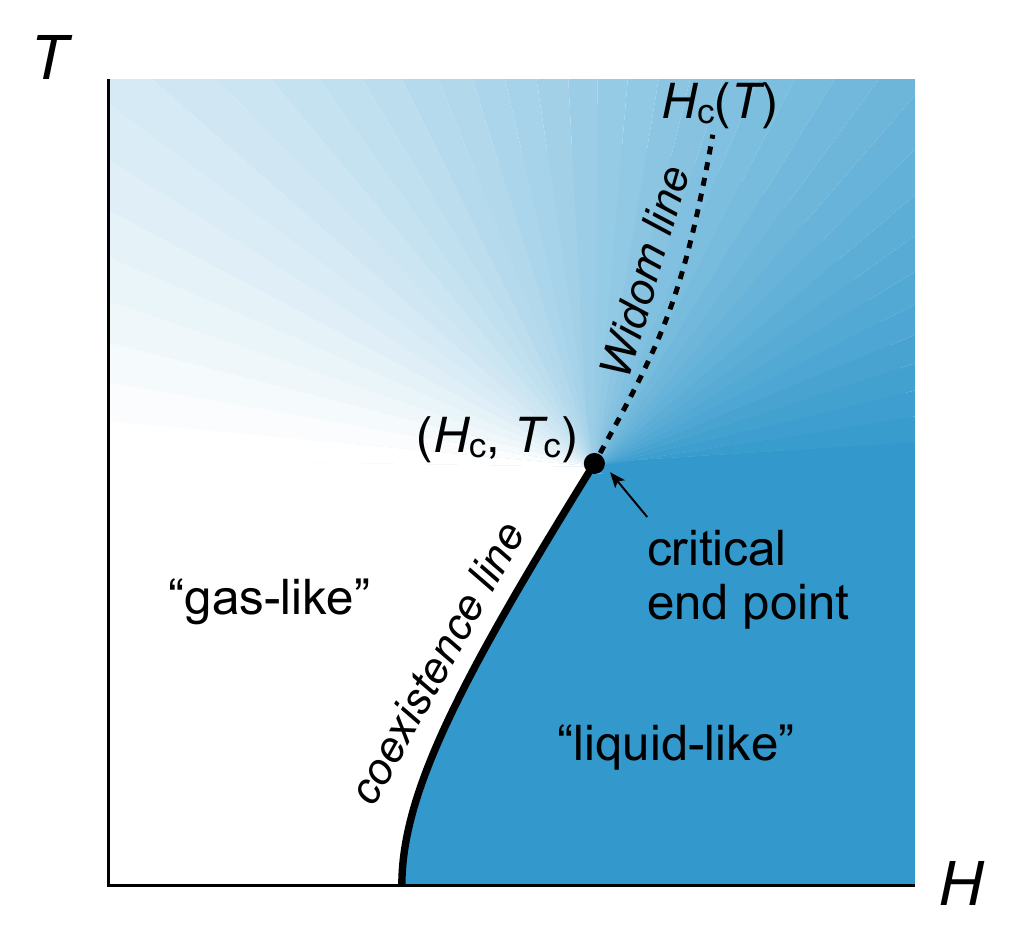}

\caption{\label{fig_generalClassical}
\textbf{General phase diagram for liquid--gas transition of classical fluids. }
There is a critical end point at $(H_\mathrm{c},T_\mathrm{c} )$, 
below which the liquid-like and gas-like phases are separated
by a coexistence line. Above $T_\mathrm{c}$, the liquid--gas crossover 
is characterized by the Widom line. 
Here, these two lines are collectively expressed as $H_\mathrm{c} (T)$.
}
\end{figure*}

Figure \ref{fig_generalClassical} shows the general phase diagram of the gas--liquid-type
phase transition.\cite{Pelissetto2002} The transition is controlled
not only by $T$ but also by $H$, the field conjugate to the order parameter $\phi$  in the Hamiltonian.
At a low $T$, the dense ``liquid'' and dilute ``gas'' phases are separated by a coexistence line. 
In going across the coexistence line, the thermal fluctuation drives the phase transition 
with a discontinuous jump of $\phi$. 
The coexistence line terminates at a CEP $\left( H, T \right) = \left( H_{\mathrm{c}},T_{\mathrm{c}} \right) $.
Above $ T_{\mathrm{c}} $, it is possible that the system undergoes a continuous crossover between 
the liquid-like phase to the gas-like phase, which are delineated 
by a crossover line referred to as Widom line\cite{Simeoni2010}. 
With regard to the difference from the quantum phase transition, 
the most prominent feature in the classical phase transition is the singularity 
of the thermodynamic quantities when the system approaches to the finite-$T$ CEP:
 $  \left( H,T \right)  \rightarrow  \left( H_{\mathrm{c}},T_{\mathrm{c}} \right)  $.
The critical behaviors can be described by scaling laws, which are derived as a consequence of 
a renormalization group analysis around the CEP.

In the scaling theory, it is assumed that the singular part of the free energy density $ f_\mathrm{s} $
is characterized by the correlation length $  \xi  $
in the vicinity of the CEP. Given the distances from the coexistence line ($ T<T_\mathrm{c} $)
or the crossover line ($ T>T_\mathrm{c} $) as 
$ h= \left( H-H_\mathrm{c} \left( T \right)  \right) /H_\mathrm{c} \left( T \right)  $
and $ t= \left( T-T_\mathrm{c} \right) /T_\mathrm{c} $, the renormalization group theory states that $ f_\mathrm{s} $
and $ \xi $  are formulated as homogeneous forms using a rescaling factor
$ b \left(>1 \right): $\cite{Pelissetto2002, Simons1997}
\begin{subequations}
	\begin{align}
		L^d f_\mathrm{s} \left( h, t \right) &= \left(L/b \right)^d  f_\mathrm{s} \left( b^{s_h}h, b^{s_t}t \right),
		\label{eq_CHomog:a} \\
		\xi  \left( h, t \right) &= b \xi  \left( b^{s_h}h, b^{s_t}t \right), 
		\label{eq_CHomog:b}
	\end{align}
\end{subequations}
where $ L $ and $ d $  are the system size and the dimension, respectively. 
These equations represent the relation when the system is seen from two different length scales, 
i.e., the original scale $\bm{x} $ and the $ b $-fold coarser scale $ \bm{x}^{\prime}=\bm{x}/b $.
Integrating out the short-range fluctuation contribution in the latter scale 
leads to renormalization for the parameters as $ h^{\prime}=b^{s_{h}}h $ and $ t^{\prime}=b^{s_{t}}t $,
by which the free energy density can be expressed by the \textit{same} function 
as that in the original scale, $ f_\mathrm{s} \left( h^{\prime},t^{\prime} \right)  $.
Since the scale transformation also reduces the localization length and 
the system volume to $  \xi^{\prime}= \xi /b $
and $ \left(L^{\prime}\right) ^{d}= \left( L/b \right) ^{d} $, respectively, the above two equations are yielded. 
Note that for the classical liquid--gas type transition, both $ h $  and $ t $
are relevant parameters, and thus the exponents $ s_{h} $ and $ s_{t} $
are positive. Choosing $ b^{s_{h}}=1/ \vert h \vert  $
in eqs. (\ref{eq_CHomog:a}) and (\ref{eq_CHomog:b}), 
one can obtain the following two scaling relations:\cite{Simons1997}
\begin{subequations}
	\begin{equation}
		f_\mathrm{s} = \left | h \right |^{d/{s_h}} f_\mathrm{s} \left(\pm 1, t/{\left | h \right |^{s_t/s_h}} \right)
					\equiv \left | h \right |^{d/{s_h}} g_{f1, h\pm, t\pm}
					\left( \frac{ h }{ |t|^{\mathGap}} \right),
		\label{eq_Cf-h} 
	\end{equation}
	\begin{equation}
		\xi  \left( h, t \right) =\left | h \right |^{-\nu_c} g_{\xi 1, h\pm, t\pm} \left( \frac{ h }{ |t|^{\mathGap}} \right), 
		\label{eq_Cxi-h}
	\end{equation}
\end{subequations}
where $  \mathGap =s_{h}/s_{t} $  is called as gap exponent and $  \nu _{c}=1/s_{h} $
is the critical exponent for correlation length. Similarly, choice of $ b^{s_{t}}=1/ \vert t \vert  $
in eqs. (\ref{eq_CHomog:a}) and (\ref{eq_CHomog:b}) leads to
\begin{subequations}
	\begin{equation}
		f_\mathrm{s} = \left | t \right |^{d/{s_t}} f_\mathrm{s} \left(h/{\left | t \right |^{s_t/s_h}} ,\pm 1 \right)
					\equiv \left | t \right |^{d/{s_t}} g_{f2, h\pm, t\pm}
					\left( \frac{ h }{ |t|^{\mathGap}} \right),
		\label{eq_Cf-t} 
	\end{equation}
	\begin{equation}
		\xi  \left( h, t \right) =\left | t \right |^{-\nu} g_{\xi 2, h\pm, t\pm} \left( \frac{ h }{ |t|^{\mathGap}} \right), 
		\label{eq_Cxi-t}
	\end{equation}
\end{subequations}
with $  \nu =1/s_{t} $. In each equation, the single-parameter function $ g_{x,h \pm ,t \pm } $
is a universal scaling function, where the pair of symbols $  \pm  $
denotes four branches defined on the sectors with different signs of $ t $ and $ h $.

Scaling relations for physical quantities can also be derived from the homogeneous formula
for the free energy density (eq. (\ref{eq_CHomog:a})). For example, the thermally averaged 
order parameter $ m= \partial f_\mathrm{s}/ \partial h $
 (corresponding to magnetization in a magnetic phase transition) satisfies
\begin{equation}
	m\left( h, t \right) = b^{s_h-d} m\left( b^{s_h} h, b^{s_t} t \right).
\end{equation}
With the proper choice of $ b $ as above, one can obtain a pair of scaling relations:
\begin{subequations}
	\begin{align}
		m \left( h, t \right) &=\left | h \right |^{\frac{d}{s_h} -1} g_{m1, h\pm, t\pm} \left( \frac{ h }{ |t|^{\mathGap}} \right),
		\label{eq_Cm-h} 
		\\
		m \left( h, t \right) &=\left | t \right |^{\frac{d}{s_t} -\mathGap} g_{m2, h\pm, t\pm} \left( \frac{ h }{ |t|^{\mathGap}} \right), 
		\label{eq_Cm-t}
	\end{align}
\end{subequations}
which are akin to the scaling relations for the free energy density except that the exponents of $  \vert h \vert  $
and $  \vert t \vert  $  are different. In general, an arbitrary physical quantity should be 
amenable to a similar scaling relation as a function of $ {h}/{ \vert t \vert ^{\mathGap}} $.\cite{Simons1997}

% here table for classical scaling exponents is to be placed

\begin{table}
\centering
\begin{threeparttable}[h]
	\caption{\label{tbl_clsExponent} Classical scaling exponents.}
	\begin{ruledtabular}
	\setstretch{1.5}
	\begin{tabularx}{0.8\textwidth}{ccX}
	{\footnotesize exponent} & {\footnotesize  representation} & {\footnotesize  description} \\\hline
		$\alpha$ & 
		$2-\frac{ d }{ s_t }$ &
		 {\small $C\propto |t|^{-\alpha}$ for $h=0$ ($C$: specific heat) }\\
		$\beta$ & 
		$\frac{ d }{ s_t }-\mathGap$ &
		  {\small $m \propto |t|^{\beta}$ for $h=0$ and $t<0$} \\
		$\gamma$ & 
		$2\mathGap-\frac{ d }{ s_t }$ &
		  {\small $\chi \propto |t|^{-\gamma}$ for $h=0$ and $t>0$
			($\chi=\partial m/\partial h$: susceptibility)} \\
		$\delta$ & 
		$\frac{ s_h }{ d - s_h }$ &
		  {\small $m \propto |h|^{1/\delta}$ for $t=0$} \\
		$\nu$ & 
		$\frac{ 1 }{ s_t }$ &
		  {\small $\xi \propto |t|^{-\nu}$ for $h=0$} \\
		$\nu_\mathrm{c}$ & 
		$\frac{ 1 }{ s_h }$ &
		  {\small $\xi \propto |h|^{-\nu_\mathrm{c}}$ for $t=0$} \\
	\end{tabularx}
	\begin{tablenotes}
	\footnotesize
	\setstretch{0.8}
		\item[a]{
			The exponent identities are deduced from above representations:
			$\alpha+2\beta+\gamma=2$ (Rushbrooke's identiy), 
			$\beta (\delta-1)=\gamma$ (Widom's identity), 
			and  $\alpha = 2-\nu d$ (Josephson's indenty).
			}
		\item[b]{
			$\mathGap = s_h / s_t = \beta \delta = \gamma \delta / (\delta -1 )$.	}
	\end{tablenotes}
	
	\end{ruledtabular}
	
\end{threeparttable}
\end{table}

Critical exponents of the transition can be straightforwardly obtained by taking the limit of
 $ h \rightarrow 0 $ or $ t \rightarrow 0 $ for the above scaling relations. 
Relations between the critical exponents and the rescaling factors are summarized in Table \ref{tbl_clsExponent}.

From an experimental standpoint, the Mott critical behaviors have been investigated 
mainly by precise measurements of the dc conductivity in the vicinity
of the pressure-induced Mott transition at half-filling. 
A conductivity scaling identical to the form of eqs. (\ref{eq_Cf-h}), (\ref{eq_Cxi-h}), and (\ref{eq_Cm-h})
has commonly observed around the CEP ($ H_\mathrm{c},T_\mathrm{c} $) 
in various Mott insulators, which reads\cite{Limelette2003, Kagawa2005, Abdel-Jawad2015}
\begin{equation}
	\sigma_\mathrm{metal}\left( h, t \right)-\sigma^* (T)
		=\left | h \right |^{s_{\sigma h}} g_{\sigma h, \pm} \left( \frac{ h }{ |t|^{\mathGap}} \right),
	\label{eq_Ccond-h}
\end{equation}
where $\sigma^* (T)$ is the $T$-dependent crossover conductivity, and
pressure $  \left[ P-P_\mathrm{c} \left( T \right)  \right] /P_\mathrm{c} \left( T \right)  $
is taken as $ h $ for pressure-induced Mott transition. The $  \sigma _\mathrm{metal} $
indicates that the scaling holds in the metallic region ($ H>H_\mathrm{c} $). 
This type of conductivity scaling is also confirmed by theoretical studies based on DMFT calculation.\cite{Georges2004}
Although the conductivity scaling seems convincing, the universality class for 
the Mott transition at half-filled is still controversial. 
This is attributed to the fact that all the physical quantities exhibit mathematically
identical scaling formula as shown above, and one cannot know solely from the 
conductivity scaling which physical quantity should be 
assigned to the experimentally observed $  \sigma_\mathrm{metal} - \sigma^{*} $. 
Therefore, as one interpretation,\cite{Limelette2003, Kagawa2005} it is postulated that
 $  \sigma_\mathrm{metal} - \sigma^* $ is proportional to $ m $
 (i.e., the density of doublon--holon pairs). The scaling formula for $ m $
then can be applicable to $  \sigma_\mathrm{metal} - \sigma^{*} $, 
leading to $ s_{ \sigma h}=1/ \delta  $, $ \mathGap = \beta  \delta  $, and
\begin{equation}
	\sigma_\mathrm{metal} \left( h=0, t<0 \right) - \sigma^* (T) \sim \left | t \right |^{\beta},
	\label{eq_Ccritical_beta}
\end{equation}
\begin{equation}
	\left(\partial / \partial h \right) \left [ \sigma^* (T) - \sigma_\mathrm{metal} \left( h=0, t> 0 \right) \right ] 
		\sim\left | t \right |^{-\gamma},
	\label{eq_Ccritical_gamma}
\end{equation}
\begin{equation}
	\sigma_\mathrm{metal} \left( h, t=0 \right)- \sigma^* (T) \sim\left | h \right |^{1/\delta}.
	\label{eq_Ccritical_delta}
\end{equation}
On the basis of this analysis, (V$_{1-x}$Cr$_{x}$)$_{2}$O$_{3}$,
a three-dimensional Mott insulator, gives
$  \beta  \approx 1/2 $,  $  \delta  \approx 3 $, $  \gamma  \approx 1 $, and $  \mathGap \approx 3/2 $, 
which correspond to the mean-field values for the Ising universality class.\cite{Limelette2003} 
On the other hand, a two-dimensional organic Mott insulator \kCl provides 
$  \beta  \approx 1 $, $  \delta  \approx 2 $, $   \gamma  \approx 1 $, and $  \mathGap \approx 2 $,
which cannot be accounted for by any conventional universality classes.\cite{Kagawa2005}

On the other hand, it was recently proposed that the conductivity behavior reflects the singularity of $  \xi  $
 around the CEP rather than the order parameter itself. 
According to the experimental argument by Abdel-Jawad \textit{et al.},\cite{Abdel-Jawad2015}
$\sigma _\mathrm{metal}- \sigma^* (T) \propto 1/ \xi  $
holds at $ T=T_\mathrm{c} $. Extension for $ T \neq T_\mathrm{c} $
can be readily done by assuming
\begin{equation}
	\sigma_\mathrm{metal} - \sigma^* (T) = \frac{ 1 }{ \xi \left (h, t \right)} \vartheta\left(\frac{h}{|t|^{\mathGap}}\right),
	\label{eq_Ccond-xi-ext}
\end{equation}
where $  \vartheta  \left( y \right)  $ is a universal function satisfying $  \vartheta  \left(  \infty \right) =1 $ and $  \vartheta  \left(  0 \right) = 0 $.
Using eq. (\ref{eq_CHomog:b}), one obtains
\begin{subequations}
	\begin{align}
		\sigma_\mathrm{metal} -\sigma^* (T) &=
			\left | h \right |^{\nu_\mathrm{c}} g_{\sigma h \pm} \left( \frac{ h }{ |t|^{\mathGap}} \right),
		\label{eq_Ccond-xi-h} 
		\\
		\sigma_\mathrm{metal} -\sigma^* (T) &=
			\left | t \right |^{\nu} g_{\sigma t \pm} \left( \frac{ h }{ |t|^{\mathGap}} \right).
		\label{eq_Ccond-xi-t}
	\end{align}
\end{subequations}

The first equation has the same form as eq. (\ref{eq_Cxi-h}), and we obtain $ s_{ \sigma h}= \nu _\mathrm{c} $
and $  \mathGap ={ \nu }/{ \nu _{c}}= \beta  \delta  $. 
In this description, scaling analysis for (V$_{1-x}$Cr$_{x}$)$_{2}$O$_{3}$
yields $  \left(  \nu , \nu _\mathrm{c}, \mathGap \right)  \approx  \left(1/2, 1/3, 3/2 \right)  $, 
which again agrees with mean-field value for the Ising universality class. 
On the other hand, the critical exponents for two-dimensional organic Mott insulators
 (\kCl and $ \mathrm{EtMe_3P[Pd(dmit)_2]_2}$) are 
$  \left(  \nu , \nu _\mathrm{c}, \mathGap \right)  \approx  \left( 1, 1/2 ,2 \right)  $, 
which are close to theoretical values for \textit{two-dimensional} Ising systems, 
$  \left(  \nu , \nu _\mathrm{c}, \mathGap \right) _\mathrm{theo}= \left( 1,1/2,15/8 \right)  $.\cite{Abdel-Jawad2015}

We now examine whether the scaling theory for classical liquid--gas-type transition 
is applicable to our experimental results for doping-induced metal--insulator transition
on the two-dimensional Mott insulator \kCl. 
In analogy with paired scaling relations about the free energy density 
(eqs. (\ref{eq_Cf-h}) and (\ref{eq_Cf-t})), we consider eq. (\ref{eq_Ccond-h}) and its counterpart: 
\begin{equation}
	\sigma_\mathrm{metal}\left( h, t \right) -  \sigma^* (T)
		 =\left | t \right |^{s_{\sigma t}} g_{\sigma t, \pm} \left( \frac{ h }{ |t|^{\mathGap}} \right),
	\label{eq_Ccond-t}
\end{equation}
where $ s_{ \sigma t}=s_{ \sigma h} \mathGap $ and 
$ g_{ \sigma t, \pm } \left( 0 \right) = \left . {\mathrm{d} g_{ \sigma t, \pm } ( y )} / {\mathrm{d}y} \right \vert _{h=0}=0 $
are satisfied. In our case, we take the gate voltage as a scaling variable, 
defining $ h= \left( \mathVg -\mathVg^{*} \right) / \mathVg^{*} $. 
The most significant parts lacking for the quantum critical scaling (eq.~(\ref{eq_QCS}) in the main part)
are the prefactors $  \vert h \vert ^{s_{ \sigma h}} $ and $  \vert t \vert ^{s_{ \sigma t}} $, 
which yield non-monotonic behaviors in conductivity around the CEP.
Indeed, taking the derivative of eqs. (\ref{eq_Ccond-h}) and (\ref{eq_Ccond-t}) with respect to $ h $, 
one can immediately obtain
\begin{subequations}
	\begin{align}
		\left. \frac{ \mathrm{d} \sigma_\mathrm{metal} }{ \mathrm{d} h } \right \vert _{t=0} 
			& \sim \left | h \right |^{s_{\sigma h} -1},			
		\label{eq_Ccritical_exp_h} 
		\\
		\left. \frac{ \mathrm{d} \sigma_\mathrm{metal} }{ \mathrm{d} h } \right \vert _{h=0} 
			& \sim \left | t \right |^{s_{\sigma t} -\mathGap },	
		\label{eq_Ccritical_exp_t}
	\end{align}
\end{subequations}
which lead to warping of $  \sigma  \left( h,t \right)  $ unless $ s_{ \sigma h}=1 $. 
In our results, however, \Vg-dependence of conductivity is quite monotonic in the vicinity of crossover line
without any points showing $ {\mathrm{d} \sigma _\mathrm{metal} }/{\mathrm{d} h}=0 $
or $  {\mathrm{d} \sigma _\mathrm{metal} }/{\mathrm{d} h}\rightarrow \infty $
in the measurement $T$ range (see Figure \ref{fig_Schematic} in the main part). 
Therefore, we have only to concentrate on two possibilities for the classical transition: 
(i) the scaling exponents satisfy $ s_{ \sigma h}=1 $ and $ s_{ \sigma t}=1/ \mathGap $, or
(ii) $ T_\mathrm{c} $ is well below the measurement $T$ range, i.e.,  $ T_\mathrm{c} < 2\ \mathrm{K} $

\begin{figure*}[tb]
\centering
\includegraphics[width=0.95\textwidth]{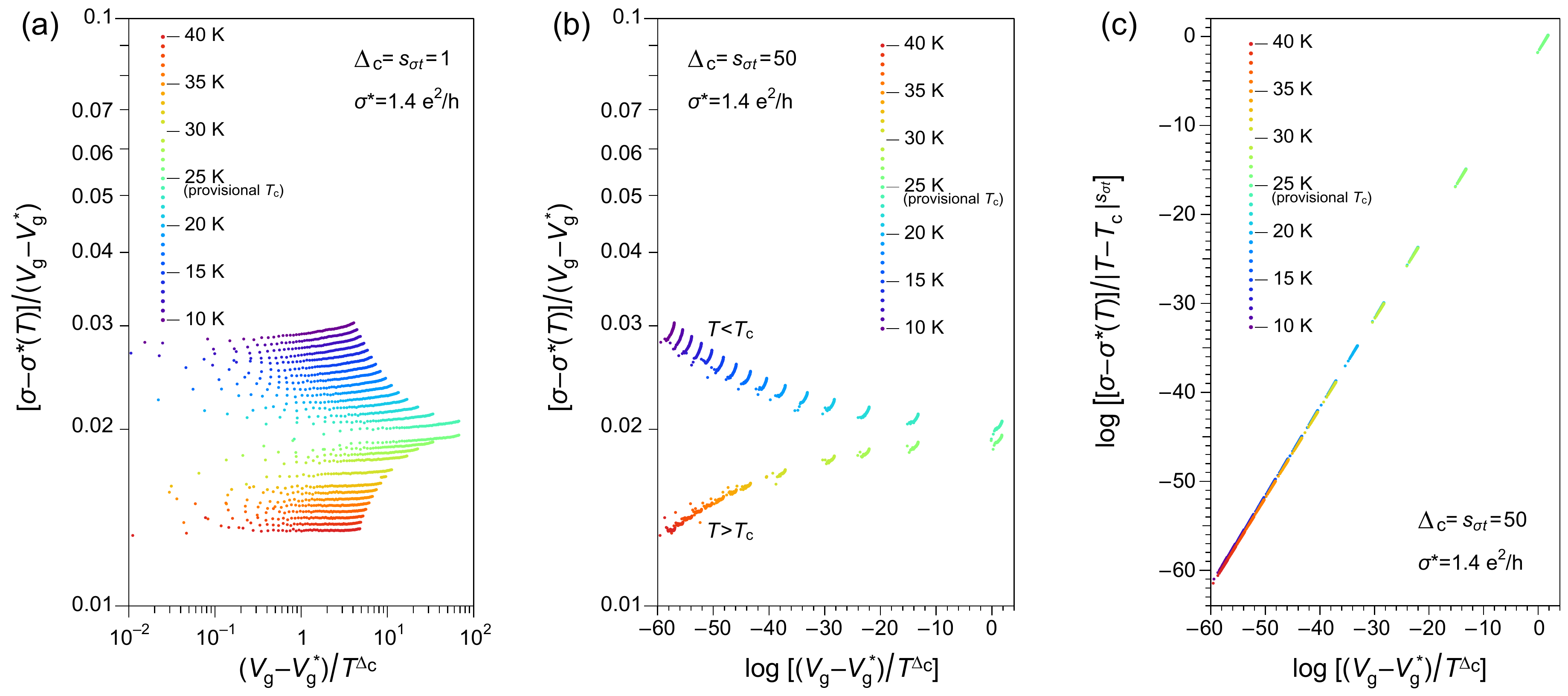}
\caption{\label{fig_classical_scaling_unity}
\textbf{Classical critical scaling for a provisional $T_\mathrm{c}=25$~K in the metallic regime.}
 (a) The plot of
 $\left (\sigma_\mathrm{metal}-\sigma^* \right) / \left(\mathVg - \mathVg^* \right) ^{s_{\sigma h} }$
with respect to $\left(\mathVg - \mathVg^* \right)/T^{\mathGap}$ for $s_{\sigma t}=\mathGap =1$. 
(b) The semi-log plot of
 $\left (\sigma_\mathrm{metal}-\sigma^* \right) / \left(\mathVg - \mathVg^* \right) ^{s_{\sigma h} }$
with respect to $\left(\mathVg - \mathVg^* \right)/T^{\mathGap}$ for  $s_{\sigma t}=\mathGap =50$ $(\gg1)$. 
(c) The log-log plot of
 $\left (\sigma_\mathrm{metal}-\sigma^* \right) / \left(T - T_\mathrm{c} \right) ^{s_{\sigma t} }$
with respect to  $\left(\mathVg - \mathVg^* \right)/T^{\mathGap}$ for $s_{\sigma t}=\mathGap =50$ $(\gg1)$. 
For all, $s_{\sigma h}=1$ is assumed.
}
\end{figure*}

As for case (i), the scaling was found to fail for any choice of $ T_\mathrm{c} $
and $  \sigma ^{*} \left(\mathVg^{*} \right)  $. 
Figure \ref{fig_classical_scaling_unity} shows examples of the scaling plots for
provisionally assumed $ T_\mathrm{c}$ ($=25 \ \mathrm{K} $). 
(Note that other choice for the provisional $T_\mathrm{c}$ 
in the intermediate $T$ range does not affect the conclusion.) 
For $ s_{ \sigma h}= \mathGap =1 $ (Figure \ref{fig_classical_scaling_unity}a), 
the data are completely dispersed, and no universal curves can be identified. 
By assuming $ s_{ \sigma h}= \mathGap \gg 1 $
(Figure \ref{fig_classical_scaling_unity}b,c),
the scaling for $ T>T_\mathrm{c} $ 
(the lower branches of Figure \ref{fig_classical_scaling_unity}b) is improved, 
whereas the data for $ T<T_\mathrm{c} $ 
(the upper branches of Figure \ref{fig_classical_scaling_unity}b) is still scattered 
from a single curve. Hence the possibility for the universality class giving $ s_{ \sigma h}=1 $
should be ruled out.

\begin{figure*}[tb]
\centering
\includegraphics[width=0.5\textwidth]{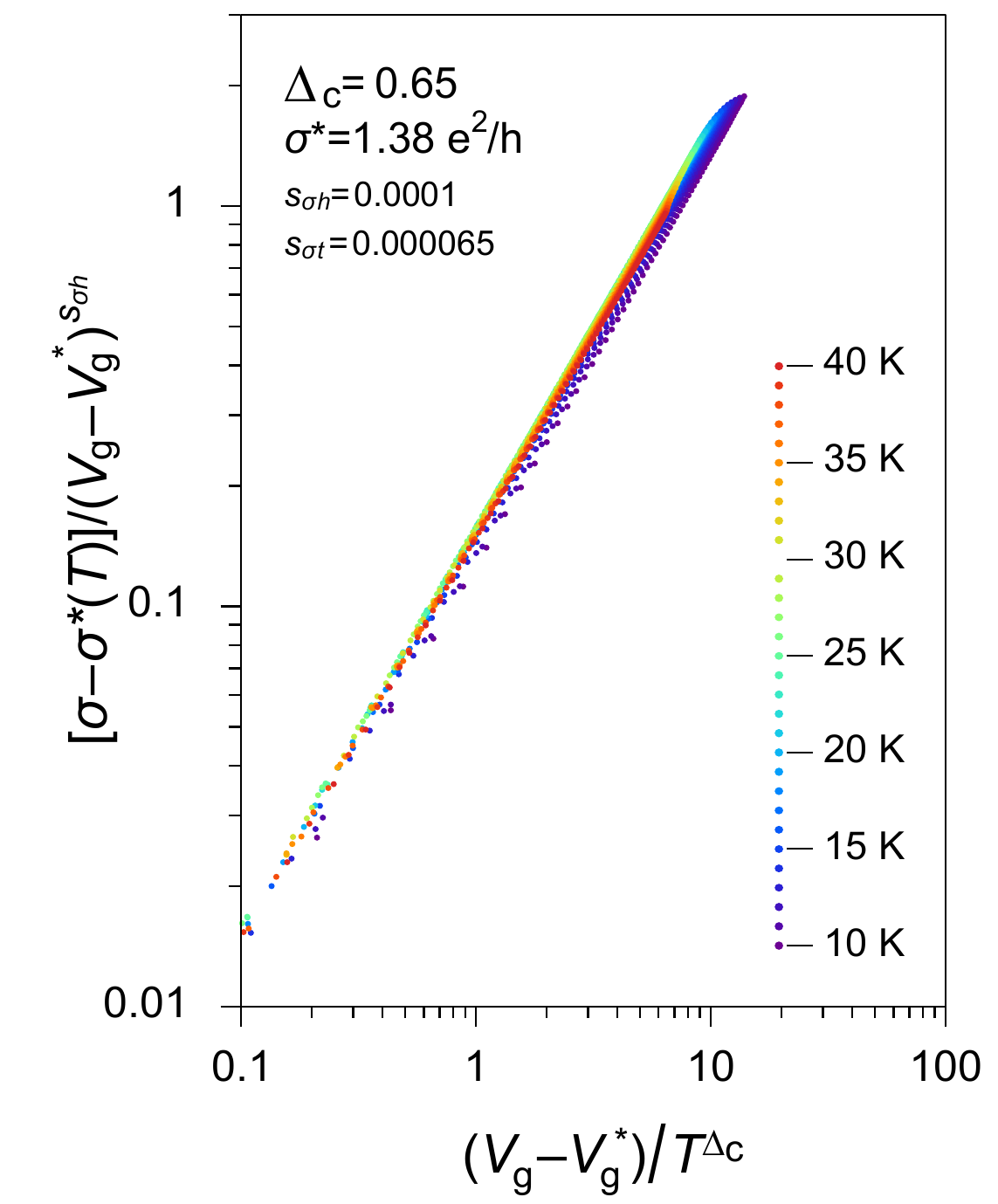}

\caption{\label{fig_classical_scaling_Tczero}
\textbf{Classical critical scaling for $T_\mathrm{c}\rightarrow 0$ in the metallic regime.} 
A constant conductivity line ($\sigma^{*} = 1.38 \esqh$) was taken as the crossover line. 
Best quality of scaling was obtained for $s_{\sigma h}, s_{\sigma_t} \rightarrow 0$
and $\mathGap=0.65$, while the scaling gradually breaks down 
in decreasing $T$ ($T<15$~K). Since the plot of 
$\left (\sigma_\mathrm{metal}-\sigma^* \right) / \left(T- T_\mathrm{c}\right) ^{s_{\sigma t} }$
 (eq. (\ref{eq_Ccond-t}))  is almost identical to that of
$\left (\sigma_\mathrm{metal}-\sigma^* \right) / \left(\mathVg - \mathVg^* \right) ^{s_{\sigma h} }$
(eq. (\ref{eq_Ccond-h})), only the latter is shown here.
}
\end{figure*}

On the other hand, under assumption of (ii) $ T_\mathrm{c}<2 \ \mathrm{K} $ 
better scaling is possible for $ T_\mathrm{c} \rightarrow 0 $ as shown in Figure \ref{fig_classical_scaling_Tczero}. 
Because the scaling quality is not sensitive to the way of choice for 
the crossover line $\mathVg^{*} \left( T \right)  $ in this case,
the constant conductivity line $  \sigma ^{*}= \sigma _\mathrm{c}=1.38 \esqh $
was taken for the crossover line for simplicity. 
The scaling analysis leads to $ s_{ \sigma h},s_{ \sigma t} \rightarrow 0 $,
and a finite gap exponent $  \mathGap = s_{\sigma}/s_{\sigma_h} = s_{h}/s_{t} \approx 0.65 $. 
Namely, the conductivity scaling formula is given by 
$ \sigma_\mathrm{metal} (h,T) - \sigma^{*}(T) = g_{\sigma t,\pm}  \left ( h / T^{\mathGap } \right ) $,
which is almost equivalent to the quantum critical scaling formula (eq.~(\ref{eq_QCS}) in the main text).
The meaning of these exponents depends on how to associate 
the conductivity to relevant physical quantities. 
If we adopted the $  \sigma - \sigma^* \sim 1/ \xi  $
correspondence (i.e., $   \nu _\mathrm{c}=s_{ \sigma h} $ and $  \nu =s_{ \sigma t} $), 
the exponents in the homogeneous scaling formula (eqs. (\ref{eq_CHomog:a}) and (\ref{eq_CHomog:b}))
would take physically inadequate values: $ s_{h}, s_{t} \rightarrow + \infty $. 
Therefore, the $  \sigma - \sigma^* \sim m $ correspondence (namely,
$ 1/ \delta =s_{ \sigma h} $  and $  \beta =s_{ \sigma t} $) 
is the only acceptable interpretation for our results. 
From eqs. (\ref{eq_Cm-h}) and (\ref{eq_Cm-t}), 
$ s_{h}=d=2 $ and $ s_{t}=d/ \mathGap \approx 3.1 $ are obtained, 
yielding the critical exponents of $  \beta =0 $, $  \gamma = \Delta _{c} \approx 0.65 $,
 $  \delta = \infty $, and $  \nu = \Delta _{c}/d \approx 0.33 $, 
which cannot be explained by any known universality classes.

To summarize, attempts for the classical critical scaling of the conductivity data 
are generally unsuccessful in our doping-driven Mott transition experiments 
since there is no singularity in the observed $\mathrm{d}\sigma/\mathrm{d}\mathVg$
that implies the existence for a finite-$T$ CEP. 
In the particular case of $T_\mathrm{c}\rightarrow 0$,
the classical critical scaling is possible 
although the scaling quality is worse than that for the quantum critical scaling. 
It is noteworthy that the classical scaling on the Mott transition has been found 
to hold within a narrow $T$ range around the CEP, 
empirically $ T\lesssim 2T_\mathrm{c} $.\cite{Furukawa2015a, Vucicevic2013}
Hence one should be cautious about accepting the above discussion 
in which metallic conductivity data are scaled for a wide $T$ range of
$ 10 \ \mathrm{K} < T < 40 \ \mathrm{K} $ in spite of the very low $ T_\mathrm{c} $.
We emphasize that the observed $ T_\mathrm{c} \rightarrow 0 $ behavior
with wide-$T$-range scaling rather implies the zero-temperature critical point
for quantum phase transitions driven by the high-energy scale fluctuations. 
Reminding the fact that in the quantum critical description the insulating regimes 
can also be scaled with an identical critical exponent (Figure~\ref{fig_Critical} in the main part), 
it is more plausible to consider that the continuous conductivity crossover 
at an intermediate  $T$  is brought about by a quantum phase transition. 

\vspace{10mm}

\section*{Cover Image}
\begin{center}
\includegraphics[width=0.8\textwidth]{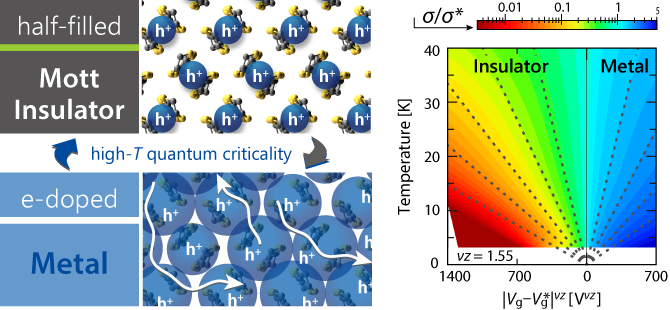}
\end{center}

%%%%%%%%%%   REFERENCE %%%%%%%%%%%
\pagebreak
%\bibliography{Ref}

\begin{thebibliography}{82}%
\makeatletter
\providecommand \@ifxundefined [1]{%
 \@ifx{#1\undefined}
}%
\providecommand \@ifnum [1]{%
 \ifnum #1\expandafter \@firstoftwo
 \else \expandafter \@secondoftwo
 \fi
}%
\providecommand \@ifx [1]{%
 \ifx #1\expandafter \@firstoftwo
 \else \expandafter \@secondoftwo
 \fi
}%
\providecommand \natexlab [1]{#1}%
\providecommand \enquote  [1]{``#1''}%
\providecommand \bibnamefont  [1]{#1}%
\providecommand \bibfnamefont [1]{#1}%
\providecommand \citenamefont [1]{#1}%
\providecommand \href@noop [0]{\@secondoftwo}%
\providecommand \href [0]{\begingroup \@sanitize@url \@href}%
\providecommand \@href[1]{\@@startlink{#1}\@@href}%
\providecommand \@@href[1]{\endgroup#1\@@endlink}%
\providecommand \@sanitize@url [0]{\catcode `\\12\catcode `\$12\catcode
  `\&12\catcode `\#12\catcode `\^12\catcode `\_12\catcode `\%12\relax}%
\providecommand \@@startlink[1]{}%
\providecommand \@@endlink[0]{}%
\providecommand \url  [0]{\begingroup\@sanitize@url \@url }%
\providecommand \@url [1]{\endgroup\@href {#1}{\urlprefix }}%
\providecommand \urlprefix  [0]{URL }%
\providecommand \Eprint [0]{\href }%
\providecommand \doibase [0]{http://dx.doi.org/}%
\providecommand \selectlanguage [0]{\@gobble}%
\providecommand \bibinfo  [0]{\@secondoftwo}%
\providecommand \bibfield  [0]{\@secondoftwo}%
\providecommand \translation [1]{[#1]}%
\providecommand \BibitemOpen [0]{}%
\providecommand \bibitemStop [0]{}%
\providecommand \bibitemNoStop [0]{.\EOS\space}%
\providecommand \EOS [0]{\spacefactor3000\relax}%
\providecommand \BibitemShut  [1]{\csname bibitem#1\endcsname}%
\let\auto@bib@innerbib\@empty
%</preamble>
\bibitem [{\citenamefont {Cui}\ \emph {et~al.}(2015)\citenamefont {Cui},
  \citenamefont {Lee}, \citenamefont {Kim}, \citenamefont {Arefe},
  \citenamefont {Huang}, \citenamefont {Lee}, \citenamefont {Chenet},
  \citenamefont {Zhang}, \citenamefont {Wang}, \citenamefont {Ye},
  \citenamefont {Pizzocchero}, \citenamefont {Jessen}, \citenamefont
  {Watanabe}, \citenamefont {Taniguchi}, \citenamefont {Muller}, \citenamefont
  {Low}, \citenamefont {Kim},\ and\ \citenamefont {Hone}}]{Cui2015}%
  \BibitemOpen
  \bibfield  {author} {\bibinfo {author} {\bibfnamefont {X.}~\bibnamefont
  {Cui}}, \bibinfo {author} {\bibfnamefont {G.-H.}\ \bibnamefont {Lee}},
  \bibinfo {author} {\bibfnamefont {Y.~D.}\ \bibnamefont {Kim}}, \bibinfo
  {author} {\bibfnamefont {G.}~\bibnamefont {Arefe}}, \bibinfo {author}
  {\bibfnamefont {P.~Y.}\ \bibnamefont {Huang}}, \bibinfo {author}
  {\bibfnamefont {C.-H.}\ \bibnamefont {Lee}}, \bibinfo {author} {\bibfnamefont
  {D.~A.}\ \bibnamefont {Chenet}}, \bibinfo {author} {\bibfnamefont
  {X.}~\bibnamefont {Zhang}}, \bibinfo {author} {\bibfnamefont
  {L.}~\bibnamefont {Wang}}, \bibinfo {author} {\bibfnamefont {F.}~\bibnamefont
  {Ye}}, \bibinfo {author} {\bibfnamefont {F.}~\bibnamefont {Pizzocchero}},
  \bibinfo {author} {\bibfnamefont {B.~S.}\ \bibnamefont {Jessen}}, \bibinfo
  {author} {\bibfnamefont {K.}~\bibnamefont {Watanabe}}, \bibinfo {author}
  {\bibfnamefont {T.}~\bibnamefont {Taniguchi}}, \bibinfo {author}
  {\bibfnamefont {D.~A.}\ \bibnamefont {Muller}}, \bibinfo {author}
  {\bibfnamefont {T.}~\bibnamefont {Low}}, \bibinfo {author} {\bibfnamefont
  {P.}~\bibnamefont {Kim}}, \ and\ \bibinfo {author} {\bibfnamefont
  {J.}~\bibnamefont {Hone}},\ }\href {\doibase 10.1038/nnano.2015.70}
  {\bibfield  {journal} {\bibinfo  {journal} {Nat. Nanotechnol.}\ }\textbf
  {\bibinfo {volume} {10}},\ \bibinfo {pages} {534} (\bibinfo {year}
  {2015})}\BibitemShut {NoStop}%
\bibitem [{\citenamefont {Tsen}\ \emph {et~al.}(2015)\citenamefont {Tsen},
  \citenamefont {Hunt}, \citenamefont {Kim}, \citenamefont {Yuan},
  \citenamefont {Jia}, \citenamefont {Cava}, \citenamefont {Hone},
  \citenamefont {Kim}, \citenamefont {Dean},\ and\ \citenamefont
  {Pasupathy}}]{Tsen2015b}%
  \BibitemOpen
  \bibfield  {author} {\bibinfo {author} {\bibfnamefont {A.~W.}\ \bibnamefont
  {Tsen}}, \bibinfo {author} {\bibfnamefont {B.}~\bibnamefont {Hunt}}, \bibinfo
  {author} {\bibfnamefont {Y.~D.}\ \bibnamefont {Kim}}, \bibinfo {author}
  {\bibfnamefont {Z.~J.}\ \bibnamefont {Yuan}}, \bibinfo {author}
  {\bibfnamefont {S.}~\bibnamefont {Jia}}, \bibinfo {author} {\bibfnamefont
  {R.~J.}\ \bibnamefont {Cava}}, \bibinfo {author} {\bibfnamefont
  {J.}~\bibnamefont {Hone}}, \bibinfo {author} {\bibfnamefont {P.}~\bibnamefont
  {Kim}}, \bibinfo {author} {\bibfnamefont {C.~R.}\ \bibnamefont {Dean}}, \
  and\ \bibinfo {author} {\bibfnamefont {A.~N.}\ \bibnamefont {Pasupathy}},\
  }\href {\doibase 10.1038/nphys3579} {\bibfield  {journal} {\bibinfo
  {journal} {Nat. Phys.}\ }\textbf {\bibinfo {volume} {12}},\ \bibinfo {pages}
  {208} (\bibinfo {year} {2015})}\BibitemShut {NoStop}%
\bibitem [{\citenamefont {Stafford}\ and\ \citenamefont {{Das
  Sarma}}(1994)}]{Stafford1994}%
  \BibitemOpen
  \bibfield  {author} {\bibinfo {author} {\bibfnamefont {C.~A.}\ \bibnamefont
  {Stafford}}\ and\ \bibinfo {author} {\bibfnamefont {S.}~\bibnamefont {{Das
  Sarma}}},\ }\href {\doibase 10.1103/PhysRevLett.72.3590} {\bibfield
  {journal} {\bibinfo  {journal} {Phys. Rev. Lett.}\ }\textbf {\bibinfo
  {volume} {72}},\ \bibinfo {pages} {3590} (\bibinfo {year}
  {1994})}\BibitemShut {NoStop}%
\bibitem [{\citenamefont {Byrnes}\ \emph {et~al.}(2008)\citenamefont {Byrnes},
  \citenamefont {Kim}, \citenamefont {Kusudo},\ and\ \citenamefont
  {Yamamoto}}]{Byrnes2008}%
  \BibitemOpen
  \bibfield  {author} {\bibinfo {author} {\bibfnamefont {T.}~\bibnamefont
  {Byrnes}}, \bibinfo {author} {\bibfnamefont {N.~Y.}\ \bibnamefont {Kim}},
  \bibinfo {author} {\bibfnamefont {K.}~\bibnamefont {Kusudo}}, \ and\ \bibinfo
  {author} {\bibfnamefont {Y.}~\bibnamefont {Yamamoto}},\ }\href {\doibase
  10.1103/PhysRevB.78.075320} {\bibfield  {journal} {\bibinfo  {journal} {Phys.
  Rev. B}\ }\textbf {\bibinfo {volume} {78}},\ \bibinfo {pages} {075320}
  (\bibinfo {year} {2008})}\BibitemShut {NoStop}%
\bibitem [{\citenamefont {Lee}\ \emph {et~al.}(2006)\citenamefont {Lee},
  \citenamefont {Nagaosa},\ and\ \citenamefont {Wen}}]{Lee2006b}%
  \BibitemOpen
  \bibfield  {author} {\bibinfo {author} {\bibfnamefont {P.~A.}\ \bibnamefont
  {Lee}}, \bibinfo {author} {\bibfnamefont {N.}~\bibnamefont {Nagaosa}}, \ and\
  \bibinfo {author} {\bibfnamefont {X.-G.}\ \bibnamefont {Wen}},\ }\href
  {\doibase 10.1103/RevModPhys.78.17} {\bibfield  {journal} {\bibinfo
  {journal} {Rev. Mod. Phys.}\ }\textbf {\bibinfo {volume} {78}},\ \bibinfo
  {pages} {17} (\bibinfo {year} {2006})}\BibitemShut {NoStop}%
\bibitem [{\citenamefont {Cario}\ \emph {et~al.}(2010)\citenamefont {Cario},
  \citenamefont {Vaju}, \citenamefont {Corraze}, \citenamefont {Guiot},\ and\
  \citenamefont {Janod}}]{Cario2010}%
  \BibitemOpen
  \bibfield  {author} {\bibinfo {author} {\bibfnamefont {L.}~\bibnamefont
  {Cario}}, \bibinfo {author} {\bibfnamefont {C.}~\bibnamefont {Vaju}},
  \bibinfo {author} {\bibfnamefont {B.}~\bibnamefont {Corraze}}, \bibinfo
  {author} {\bibfnamefont {V.}~\bibnamefont {Guiot}}, \ and\ \bibinfo {author}
  {\bibfnamefont {E.}~\bibnamefont {Janod}},\ }\href {\doibase
  10.1002/adma.201002521} {\bibfield  {journal} {\bibinfo  {journal} {Adv.
  Mater.}\ }\textbf {\bibinfo {volume} {22}},\ \bibinfo {pages} {5193}
  (\bibinfo {year} {2010})}\BibitemShut {NoStop}%
\bibitem [{\citenamefont {Zhu}\ \emph {et~al.}(2016)\citenamefont {Zhu},
  \citenamefont {Peng}, \citenamefont {Zou}, \citenamefont {Prokes},
  \citenamefont {Mahanti}, \citenamefont {Hong}, \citenamefont {Mao},
  \citenamefont {Liu},\ and\ \citenamefont {Ke}}]{Zhu2016a}%
  \BibitemOpen
  \bibfield  {author} {\bibinfo {author} {\bibfnamefont {M.}~\bibnamefont
  {Zhu}}, \bibinfo {author} {\bibfnamefont {J.}~\bibnamefont {Peng}}, \bibinfo
  {author} {\bibfnamefont {T.}~\bibnamefont {Zou}}, \bibinfo {author}
  {\bibfnamefont {K.}~\bibnamefont {Prokes}}, \bibinfo {author} {\bibfnamefont
  {S.~D.}\ \bibnamefont {Mahanti}}, \bibinfo {author} {\bibfnamefont
  {T.}~\bibnamefont {Hong}}, \bibinfo {author} {\bibfnamefont {Z.~Q.}\
  \bibnamefont {Mao}}, \bibinfo {author} {\bibfnamefont {G.~Q.}\ \bibnamefont
  {Liu}}, \ and\ \bibinfo {author} {\bibfnamefont {X.}~\bibnamefont {Ke}},\
  }\href {\doibase 10.1103/PhysRevLett.116.216401} {\bibfield  {journal}
  {\bibinfo  {journal} {Phys. Rev. Leet.}\ }\textbf {\bibinfo {volume} {116}},\
  \bibinfo {pages} {216401} (\bibinfo {year} {2016})}\BibitemShut {NoStop}%
\bibitem [{\citenamefont {Sipos}\ \emph {et~al.}(2008)\citenamefont {Sipos},
  \citenamefont {Kusmartseva}, \citenamefont {Akrap}, \citenamefont {Berger},
  \citenamefont {Forr{\'{o}}},\ and\ \citenamefont {Tutis}}]{Sipos2008}%
  \BibitemOpen
  \bibfield  {author} {\bibinfo {author} {\bibfnamefont {B.}~\bibnamefont
  {Sipos}}, \bibinfo {author} {\bibfnamefont {A.~F.}\ \bibnamefont
  {Kusmartseva}}, \bibinfo {author} {\bibfnamefont {A.}~\bibnamefont {Akrap}},
  \bibinfo {author} {\bibfnamefont {H.}~\bibnamefont {Berger}}, \bibinfo
  {author} {\bibfnamefont {L.}~\bibnamefont {Forr{\'{o}}}}, \ and\ \bibinfo
  {author} {\bibfnamefont {E.}~\bibnamefont {Tutis}},\ }\href {\doibase
  10.1038/nmat2318} {\bibfield  {journal} {\bibinfo  {journal} {Nat. Mater.}\
  }\textbf {\bibinfo {volume} {7}},\ \bibinfo {pages} {960} (\bibinfo {year}
  {2008})}\BibitemShut {NoStop}%
\bibitem [{\citenamefont {Powell}\ and\ \citenamefont
  {McKenzie}(2011)}]{Powell2011a}%
  \BibitemOpen
  \bibfield  {author} {\bibinfo {author} {\bibfnamefont {B.~J.}\ \bibnamefont
  {Powell}}\ and\ \bibinfo {author} {\bibfnamefont {R.~H.}\ \bibnamefont
  {McKenzie}},\ }\href {\doibase 10.1088/0034-4885/74/5/056501} {\bibfield
  {journal} {\bibinfo  {journal} {Rep. Prog. Phys.}\ }\textbf {\bibinfo
  {volume} {74}},\ \bibinfo {pages} {056501} (\bibinfo {year}
  {2011})}\BibitemShut {NoStop}%
\bibitem [{\citenamefont {Sasaki}(2012)}]{Sasaki2012}%
  \BibitemOpen
  \bibfield  {author} {\bibinfo {author} {\bibfnamefont {T.}~\bibnamefont
  {Sasaki}},\ }\href {\doibase 10.3390/cryst2020374} {\bibfield  {journal}
  {\bibinfo  {journal} {Crystals}\ }\textbf {\bibinfo {volume} {2}},\ \bibinfo
  {pages} {374} (\bibinfo {year} {2012})}\BibitemShut {NoStop}%
\bibitem [{\citenamefont {Lahoud}\ \emph {et~al.}(2014)\citenamefont {Lahoud},
  \citenamefont {Meetei}, \citenamefont {Chaska}, \citenamefont {Kanigel},\
  and\ \citenamefont {Trivedi}}]{Lahoud2014}%
  \BibitemOpen
  \bibfield  {author} {\bibinfo {author} {\bibfnamefont {E.}~\bibnamefont
  {Lahoud}}, \bibinfo {author} {\bibfnamefont {O.~N.}\ \bibnamefont {Meetei}},
  \bibinfo {author} {\bibfnamefont {K.}~\bibnamefont {Chaska}}, \bibinfo
  {author} {\bibfnamefont {A.}~\bibnamefont {Kanigel}}, \ and\ \bibinfo
  {author} {\bibfnamefont {N.}~\bibnamefont {Trivedi}},\ }\href {\doibase
  10.1103/PhysRevLett.112.206402} {\bibfield  {journal} {\bibinfo  {journal}
  {Phys. Rev. Lett.}\ }\textbf {\bibinfo {volume} {112}},\ \bibinfo {pages}
  {206402} (\bibinfo {year} {2014})}\BibitemShut {NoStop}%
\bibitem [{\citenamefont {Imada}\ \emph {et~al.}(1998)\citenamefont {Imada},
  \citenamefont {Fujimori},\ and\ \citenamefont {Tokura}}]{Imada1998}%
  \BibitemOpen
  \bibfield  {author} {\bibinfo {author} {\bibfnamefont {M.}~\bibnamefont
  {Imada}}, \bibinfo {author} {\bibfnamefont {A.}~\bibnamefont {Fujimori}}, \
  and\ \bibinfo {author} {\bibfnamefont {Y.}~\bibnamefont {Tokura}},\ }\href
  {\doibase 10.1103/RevModPhys.70.1039} {\bibfield  {journal} {\bibinfo
  {journal} {Rev. Mod. Phys.}\ }\textbf {\bibinfo {volume} {70}},\ \bibinfo
  {pages} {1039} (\bibinfo {year} {1998})}\BibitemShut {NoStop}%
\bibitem [{\citenamefont {Vojta}(2003)}]{Vojta2003}%
  \BibitemOpen
  \bibfield  {author} {\bibinfo {author} {\bibfnamefont {M.}~\bibnamefont
  {Vojta}},\ }\href {\doibase 10.1088/0034-4885/66/12/R01} {\bibfield
  {journal} {\bibinfo  {journal} {Rep. Prog. Phys.}\ }\textbf {\bibinfo
  {volume} {66}},\ \bibinfo {pages} {2069} (\bibinfo {year}
  {2003})}\BibitemShut {NoStop}%
\bibitem [{\citenamefont {L{\"{o}}hneysen}\ \emph {et~al.}(2007)\citenamefont
  {L{\"{o}}hneysen}, \citenamefont {Rosch}, \citenamefont {Vojta},\ and\
  \citenamefont {W{\"{o}}lfle}}]{Lohneysen2007}%
  \BibitemOpen
  \bibfield  {author} {\bibinfo {author} {\bibfnamefont {H.~v.}\ \bibnamefont
  {L{\"{o}}hneysen}}, \bibinfo {author} {\bibfnamefont {A.}~\bibnamefont
  {Rosch}}, \bibinfo {author} {\bibfnamefont {M.}~\bibnamefont {Vojta}}, \ and\
  \bibinfo {author} {\bibfnamefont {P.}~\bibnamefont {W{\"{o}}lfle}},\ }\href
  {\doibase 10.1103/RevModPhys.79.1015} {\bibfield  {journal} {\bibinfo
  {journal} {Rev. Mod. Phys.}\ }\textbf {\bibinfo {volume} {79}},\ \bibinfo
  {pages} {1015} (\bibinfo {year} {2007})}\BibitemShut {NoStop}%
\bibitem [{\citenamefont {Millis}(1993)}]{Millis1993}%
  \BibitemOpen
  \bibfield  {author} {\bibinfo {author} {\bibfnamefont {A.~J.}\ \bibnamefont
  {Millis}},\ }\href {\doibase 10.1103/PhysRevB.48.7183} {\bibfield  {journal}
  {\bibinfo  {journal} {Phys. Rev. B}\ }\textbf {\bibinfo {volume} {48}},\
  \bibinfo {pages} {7183} (\bibinfo {year} {1993})}\BibitemShut {NoStop}%
\bibitem [{\citenamefont {Gegenwart}\ \emph {et~al.}(2008)\citenamefont
  {Gegenwart}, \citenamefont {Si},\ and\ \citenamefont
  {Steglich}}]{Gegenwart2008}%
  \BibitemOpen
  \bibfield  {author} {\bibinfo {author} {\bibfnamefont {P.}~\bibnamefont
  {Gegenwart}}, \bibinfo {author} {\bibfnamefont {Q.}~\bibnamefont {Si}}, \
  and\ \bibinfo {author} {\bibfnamefont {F.}~\bibnamefont {Steglich}},\ }\href
  {\doibase 10.1038/nphys892} {\bibfield  {journal} {\bibinfo  {journal} {Nat.
  Phys.}\ }\textbf {\bibinfo {volume} {4}},\ \bibinfo {pages} {186} (\bibinfo
  {year} {2008})}\BibitemShut {NoStop}%
\bibitem [{\citenamefont {Sachdev}(2009)}]{Sachdev2009}%
  \BibitemOpen
  \bibfield  {author} {\bibinfo {author} {\bibfnamefont {S.}~\bibnamefont
  {Sachdev}},\ }\href {\doibase 10.1002/pssb.200983037} {\bibfield  {journal}
  {\bibinfo  {journal} {Phys. Stat. Solidi B}\ }\textbf {\bibinfo {volume}
  {543}},\ \bibinfo {pages} {6} (\bibinfo {year} {2009})}\BibitemShut {NoStop}%
\bibitem [{\citenamefont {Sebastian}\ \emph {et~al.}(2010)\citenamefont
  {Sebastian}, \citenamefont {Harrison}, \citenamefont {Altarawneh},
  \citenamefont {Mielke}, \citenamefont {Liang}, \citenamefont {Bonn},
  \citenamefont {Hardy},\ and\ \citenamefont {Lonzarich}}]{Sebastian2010}%
  \BibitemOpen
  \bibfield  {author} {\bibinfo {author} {\bibfnamefont {S.~E.}\ \bibnamefont
  {Sebastian}}, \bibinfo {author} {\bibfnamefont {N.}~\bibnamefont {Harrison}},
  \bibinfo {author} {\bibfnamefont {M.~M.}\ \bibnamefont {Altarawneh}},
  \bibinfo {author} {\bibfnamefont {C.~H.}\ \bibnamefont {Mielke}}, \bibinfo
  {author} {\bibfnamefont {R.}~\bibnamefont {Liang}}, \bibinfo {author}
  {\bibfnamefont {D.~A.}\ \bibnamefont {Bonn}}, \bibinfo {author}
  {\bibfnamefont {W.~N.}\ \bibnamefont {Hardy}}, \ and\ \bibinfo {author}
  {\bibfnamefont {G.~G.}\ \bibnamefont {Lonzarich}},\ }\href {\doibase
  10.1073/pnas.0913711107} {\bibfield  {journal} {\bibinfo  {journal} {Proc.
  Natl. Acad. Sci. USA}\ }\textbf {\bibinfo {volume} {107}},\ \bibinfo {pages}
  {6175} (\bibinfo {year} {2010})}\BibitemShut {NoStop}%
\bibitem [{\citenamefont {Sachdev}(2011)}]{Sachdev2011}%
  \BibitemOpen
  \bibfield  {author} {\bibinfo {author} {\bibfnamefont {S.}~\bibnamefont
  {Sachdev}},\ }\href@noop {} {\emph {\bibinfo {title} {{Quantum phase
  transition}}}},\ \bibinfo {edition} {2nd}\ ed.\ (\bibinfo  {publisher}
  {Cambridge University Press},\ \bibinfo {year} {2011})\BibitemShut {NoStop}%
\bibitem [{\citenamefont {Caviglia}\ \emph {et~al.}(2008)\citenamefont
  {Caviglia}, \citenamefont {Gariglio}, \citenamefont {Reyren}, \citenamefont
  {Jaccard}, \citenamefont {Schneider}, \citenamefont {Gabay}, \citenamefont
  {Thiel}, \citenamefont {Hammerl}, \citenamefont {Mannhart},\ and\
  \citenamefont {Triscone}}]{Caviglia2008}%
  \BibitemOpen
  \bibfield  {author} {\bibinfo {author} {\bibfnamefont {A.~D.}\ \bibnamefont
  {Caviglia}}, \bibinfo {author} {\bibfnamefont {S.}~\bibnamefont {Gariglio}},
  \bibinfo {author} {\bibfnamefont {N.}~\bibnamefont {Reyren}}, \bibinfo
  {author} {\bibfnamefont {D.}~\bibnamefont {Jaccard}}, \bibinfo {author}
  {\bibfnamefont {T.}~\bibnamefont {Schneider}}, \bibinfo {author}
  {\bibfnamefont {M.}~\bibnamefont {Gabay}}, \bibinfo {author} {\bibfnamefont
  {S.}~\bibnamefont {Thiel}}, \bibinfo {author} {\bibfnamefont
  {G.}~\bibnamefont {Hammerl}}, \bibinfo {author} {\bibfnamefont
  {J.}~\bibnamefont {Mannhart}}, \ and\ \bibinfo {author} {\bibfnamefont
  {J.-M.}\ \bibnamefont {Triscone}},\ }\href {\doibase 10.1038/nature07576}
  {\bibfield  {journal} {\bibinfo  {journal} {Nature}\ }\textbf {\bibinfo
  {volume} {456}},\ \bibinfo {pages} {624} (\bibinfo {year}
  {2008})}\BibitemShut {NoStop}%
\bibitem [{\citenamefont {Bollinger}\ \emph {et~al.}(2011)\citenamefont
  {Bollinger}, \citenamefont {Dubuis}, \citenamefont {Yoon}, \citenamefont
  {Pavuna}, \citenamefont {Misewich},\ and\ \citenamefont
  {Bo{\v{z}}ovi{\'{c}}}}]{Bollinger2011}%
  \BibitemOpen
  \bibfield  {author} {\bibinfo {author} {\bibfnamefont {A.~T.}\ \bibnamefont
  {Bollinger}}, \bibinfo {author} {\bibfnamefont {G.}~\bibnamefont {Dubuis}},
  \bibinfo {author} {\bibfnamefont {J.}~\bibnamefont {Yoon}}, \bibinfo {author}
  {\bibfnamefont {D.}~\bibnamefont {Pavuna}}, \bibinfo {author} {\bibfnamefont
  {J.}~\bibnamefont {Misewich}}, \ and\ \bibinfo {author} {\bibfnamefont
  {I.}~\bibnamefont {Bo{\v{z}}ovi{\'{c}}}},\ }\href {\doibase
  10.1038/nature09998} {\bibfield  {journal} {\bibinfo  {journal} {Nature}\
  }\textbf {\bibinfo {volume} {472}},\ \bibinfo {pages} {458} (\bibinfo {year}
  {2011})}\BibitemShut {NoStop}%
\bibitem [{\citenamefont {Hwang}\ \emph {et~al.}(2012)\citenamefont {Hwang},
  \citenamefont {Iwasa}, \citenamefont {Kawasaki}, \citenamefont {Keimer},
  \citenamefont {Nagaosa},\ and\ \citenamefont {Tokura}}]{Hwang2012}%
  \BibitemOpen
  \bibfield  {author} {\bibinfo {author} {\bibfnamefont {H.~Y.}\ \bibnamefont
  {Hwang}}, \bibinfo {author} {\bibfnamefont {Y.}~\bibnamefont {Iwasa}},
  \bibinfo {author} {\bibfnamefont {M.}~\bibnamefont {Kawasaki}}, \bibinfo
  {author} {\bibfnamefont {B.}~\bibnamefont {Keimer}}, \bibinfo {author}
  {\bibfnamefont {N.}~\bibnamefont {Nagaosa}}, \ and\ \bibinfo {author}
  {\bibfnamefont {Y.}~\bibnamefont {Tokura}},\ }\href {\doibase
  10.1038/nmat3223} {\bibfield  {journal} {\bibinfo  {journal} {Nat. Mater.}\
  }\textbf {\bibinfo {volume} {11}},\ \bibinfo {pages} {103} (\bibinfo {year}
  {2012})}\BibitemShut {NoStop}%
\bibitem [{\citenamefont {Furukawa}\ \emph {et~al.}(2015)\citenamefont
  {Furukawa}, \citenamefont {Miyagawa}, \citenamefont {Taniguchi},
  \citenamefont {Kato},\ and\ \citenamefont {Kanoda}}]{Furukawa2015a}%
  \BibitemOpen
  \bibfield  {author} {\bibinfo {author} {\bibfnamefont {T.}~\bibnamefont
  {Furukawa}}, \bibinfo {author} {\bibfnamefont {K.}~\bibnamefont {Miyagawa}},
  \bibinfo {author} {\bibfnamefont {H.}~\bibnamefont {Taniguchi}}, \bibinfo
  {author} {\bibfnamefont {R.}~\bibnamefont {Kato}}, \ and\ \bibinfo {author}
  {\bibfnamefont {K.}~\bibnamefont {Kanoda}},\ }\href {\doibase
  10.1038/nphys3235} {\bibfield  {journal} {\bibinfo  {journal} {Nat. Phys.}\
  }\textbf {\bibinfo {volume} {11}},\ \bibinfo {pages} {221} (\bibinfo {year}
  {2015})}\BibitemShut {NoStop}%
\bibitem [{\citenamefont {Kagawa}\ \emph {et~al.}(2005)\citenamefont {Kagawa},
  \citenamefont {Miyagawa},\ and\ \citenamefont {Kanoda}}]{Kagawa2005}%
  \BibitemOpen
  \bibfield  {author} {\bibinfo {author} {\bibfnamefont {F.}~\bibnamefont
  {Kagawa}}, \bibinfo {author} {\bibfnamefont {K.}~\bibnamefont {Miyagawa}}, \
  and\ \bibinfo {author} {\bibfnamefont {K.}~\bibnamefont {Kanoda}},\ }\href
  {\doibase 10.1038/nature03806} {\bibfield  {journal} {\bibinfo  {journal}
  {Nature}\ }\textbf {\bibinfo {volume} {436}},\ \bibinfo {pages} {534}
  (\bibinfo {year} {2005})}\BibitemShut {NoStop}%
\bibitem [{\citenamefont {Kawasugi}\ \emph {et~al.}(2016)\citenamefont
  {Kawasugi}, \citenamefont {Seki}, \citenamefont {Edagawa}, \citenamefont
  {Sato}, \citenamefont {Pu}, \citenamefont {Takenobu}, \citenamefont {Yunoki},
  \citenamefont {Yamamoto},\ and\ \citenamefont {Kato}}]{Kawasugi2016}%
  \BibitemOpen
  \bibfield  {author} {\bibinfo {author} {\bibfnamefont {Y.}~\bibnamefont
  {Kawasugi}}, \bibinfo {author} {\bibfnamefont {K.}~\bibnamefont {Seki}},
  \bibinfo {author} {\bibfnamefont {Y.}~\bibnamefont {Edagawa}}, \bibinfo
  {author} {\bibfnamefont {Y.}~\bibnamefont {Sato}}, \bibinfo {author}
  {\bibfnamefont {J.}~\bibnamefont {Pu}}, \bibinfo {author} {\bibfnamefont
  {T.}~\bibnamefont {Takenobu}}, \bibinfo {author} {\bibfnamefont
  {S.}~\bibnamefont {Yunoki}}, \bibinfo {author} {\bibfnamefont {H.~M.}\
  \bibnamefont {Yamamoto}}, \ and\ \bibinfo {author} {\bibfnamefont
  {R.}~\bibnamefont {Kato}},\ }\href {\doibase 10.1038/ncomms12356} {\bibfield
  {journal} {\bibinfo  {journal} {Nat. Commun.}\ }\textbf {\bibinfo {volume}
  {7}},\ \bibinfo {pages} {12356} (\bibinfo {year} {2016})}\BibitemShut
  {NoStop}%
\bibitem [{\citenamefont {Hayes}\ \emph {et~al.}(2015)\citenamefont {Hayes},
  \citenamefont {Warr},\ and\ \citenamefont {Atkin}}]{Hayes2015}%
  \BibitemOpen
  \bibfield  {author} {\bibinfo {author} {\bibfnamefont {R.}~\bibnamefont
  {Hayes}}, \bibinfo {author} {\bibfnamefont {G.~G.}\ \bibnamefont {Warr}}, \
  and\ \bibinfo {author} {\bibfnamefont {R.}~\bibnamefont {Atkin}},\ }\href
  {\doibase 10.1021/cr500411q} {\bibfield  {journal} {\bibinfo  {journal}
  {Chem. Rev.}\ }\textbf {\bibinfo {volume} {115}},\ \bibinfo {pages} {6357}
  (\bibinfo {year} {2015})}\BibitemShut {NoStop}%
\bibitem [{\citenamefont {Kobayashi}\ \emph {et~al.}(2004)\citenamefont
  {Kobayashi}, \citenamefont {Nishikawa}, \citenamefont {Takenobu},
  \citenamefont {Mori}, \citenamefont {Shimoda}, \citenamefont {Mitani},
  \citenamefont {Shimotani}, \citenamefont {Yoshimoto}, \citenamefont {Ogawa},\
  and\ \citenamefont {Iwasa}}]{Kobayashi2004}%
  \BibitemOpen
  \bibfield  {author} {\bibinfo {author} {\bibfnamefont {S.}~\bibnamefont
  {Kobayashi}}, \bibinfo {author} {\bibfnamefont {T.}~\bibnamefont
  {Nishikawa}}, \bibinfo {author} {\bibfnamefont {T.}~\bibnamefont {Takenobu}},
  \bibinfo {author} {\bibfnamefont {S.}~\bibnamefont {Mori}}, \bibinfo {author}
  {\bibfnamefont {T.}~\bibnamefont {Shimoda}}, \bibinfo {author} {\bibfnamefont
  {T.}~\bibnamefont {Mitani}}, \bibinfo {author} {\bibfnamefont
  {H.}~\bibnamefont {Shimotani}}, \bibinfo {author} {\bibfnamefont
  {N.}~\bibnamefont {Yoshimoto}}, \bibinfo {author} {\bibfnamefont
  {S.}~\bibnamefont {Ogawa}}, \ and\ \bibinfo {author} {\bibfnamefont
  {Y.}~\bibnamefont {Iwasa}},\ }\href {\doibase 10.1038/nmat1105} {\bibfield
  {journal} {\bibinfo  {journal} {Nat. Mater.}\ }\textbf {\bibinfo {volume}
  {3}},\ \bibinfo {pages} {317} (\bibinfo {year} {2004})}\BibitemShut {NoStop}%
\bibitem [{\citenamefont {Wang}\ \emph {et~al.}(2011)\citenamefont {Wang},
  \citenamefont {Xu}, \citenamefont {Wang}, \citenamefont {Du},\ and\
  \citenamefont {Xie}}]{Wang2011}%
  \BibitemOpen
  \bibfield  {author} {\bibinfo {author} {\bibfnamefont {X.}~\bibnamefont
  {Wang}}, \bibinfo {author} {\bibfnamefont {J.-B.}\ \bibnamefont {Xu}},
  \bibinfo {author} {\bibfnamefont {C.}~\bibnamefont {Wang}}, \bibinfo {author}
  {\bibfnamefont {J.}~\bibnamefont {Du}}, \ and\ \bibinfo {author}
  {\bibfnamefont {W.}~\bibnamefont {Xie}},\ }\href {\doibase
  10.1002/adma.201100476} {\bibfield  {journal} {\bibinfo  {journal} {Adv.
  Mater.}\ }\textbf {\bibinfo {volume} {24}},\ \bibinfo {pages} {2464}
  (\bibinfo {year} {2011})}\BibitemShut {NoStop}%
\bibitem [{\citenamefont {Yokota}\ \emph {et~al.}(2014)\citenamefont {Yokota},
  \citenamefont {Takai}, \citenamefont {Kudo}, \citenamefont {Sato},\ and\
  \citenamefont {Enoki}}]{Yokota2014a}%
  \BibitemOpen
  \bibfield  {author} {\bibinfo {author} {\bibfnamefont {K.}~\bibnamefont
  {Yokota}}, \bibinfo {author} {\bibfnamefont {K.}~\bibnamefont {Takai}},
  \bibinfo {author} {\bibfnamefont {Y.}~\bibnamefont {Kudo}}, \bibinfo {author}
  {\bibfnamefont {Y.}~\bibnamefont {Sato}}, \ and\ \bibinfo {author}
  {\bibfnamefont {T.}~\bibnamefont {Enoki}},\ }\href@noop {} {\bibfield
  {journal} {\bibinfo  {journal} {Phys. Chem. Chem. Phys.}\ }\textbf {\bibinfo
  {volume} {16}},\ \bibinfo {pages} {4313} (\bibinfo {year}
  {2014})}\BibitemShut {NoStop}%
\bibitem [{\citenamefont {Najmaei}\ \emph {et~al.}(2014)\citenamefont
  {Najmaei}, \citenamefont {Zou}, \citenamefont {Er}, \citenamefont {Li},
  \citenamefont {Jin}, \citenamefont {Gao}, \citenamefont {Zhang},
  \citenamefont {Park}, \citenamefont {Ge}, \citenamefont {Lei}, \citenamefont
  {Kono}, \citenamefont {Shenoy}, \citenamefont {Yakobson}, \citenamefont
  {George}, \citenamefont {Ajayan},\ and\ \citenamefont {Lou}}]{Najmaei2014}%
  \BibitemOpen
  \bibfield  {author} {\bibinfo {author} {\bibfnamefont {S.}~\bibnamefont
  {Najmaei}}, \bibinfo {author} {\bibfnamefont {X.}~\bibnamefont {Zou}},
  \bibinfo {author} {\bibfnamefont {D.}~\bibnamefont {Er}}, \bibinfo {author}
  {\bibfnamefont {J.}~\bibnamefont {Li}}, \bibinfo {author} {\bibfnamefont
  {Z.}~\bibnamefont {Jin}}, \bibinfo {author} {\bibfnamefont {W.}~\bibnamefont
  {Gao}}, \bibinfo {author} {\bibfnamefont {Q.}~\bibnamefont {Zhang}}, \bibinfo
  {author} {\bibfnamefont {S.}~\bibnamefont {Park}}, \bibinfo {author}
  {\bibfnamefont {L.}~\bibnamefont {Ge}}, \bibinfo {author} {\bibfnamefont
  {S.}~\bibnamefont {Lei}}, \bibinfo {author} {\bibfnamefont {J.}~\bibnamefont
  {Kono}}, \bibinfo {author} {\bibfnamefont {V.~B.}\ \bibnamefont {Shenoy}},
  \bibinfo {author} {\bibfnamefont {B.~I.}\ \bibnamefont {Yakobson}}, \bibinfo
  {author} {\bibfnamefont {A.}~\bibnamefont {George}}, \bibinfo {author}
  {\bibfnamefont {P.~M.}\ \bibnamefont {Ajayan}}, \ and\ \bibinfo {author}
  {\bibfnamefont {J.}~\bibnamefont {Lou}},\ }\href {\doibase 10.1021/nl404396p}
  {\bibfield  {journal} {\bibinfo  {journal} {Nano Lett.}\ }\textbf {\bibinfo
  {volume} {14}},\ \bibinfo {pages} {1354} (\bibinfo {year}
  {2014})}\BibitemShut {NoStop}%
\bibitem [{\citenamefont {Leng}\ \emph {et~al.}(2011)\citenamefont {Leng},
  \citenamefont {Garcia-Barriocanal}, \citenamefont {Bose}, \citenamefont
  {Lee},\ and\ \citenamefont {Goldman}}]{Leng2011}%
  \BibitemOpen
  \bibfield  {author} {\bibinfo {author} {\bibfnamefont {X.}~\bibnamefont
  {Leng}}, \bibinfo {author} {\bibfnamefont {J.}~\bibnamefont
  {Garcia-Barriocanal}}, \bibinfo {author} {\bibfnamefont {S.}~\bibnamefont
  {Bose}}, \bibinfo {author} {\bibfnamefont {Y.}~\bibnamefont {Lee}}, \ and\
  \bibinfo {author} {\bibfnamefont {A.~M.}\ \bibnamefont {Goldman}},\ }\href
  {\doibase 10.1103/PhysRevLett.107.027001} {\bibfield  {journal} {\bibinfo
  {journal} {Phys. Rev. Lett.}\ }\textbf {\bibinfo {volume} {107}},\ \bibinfo
  {pages} {027001} (\bibinfo {year} {2011})}\BibitemShut {NoStop}%
\bibitem [{\citenamefont {Garcia-Barriocanal}\ \emph
  {et~al.}(2013)\citenamefont {Garcia-Barriocanal}, \citenamefont {Kobrinskii},
  \citenamefont {Leng}, \citenamefont {Kinney}, \citenamefont {Yang},
  \citenamefont {Snyder},\ and\ \citenamefont
  {Goldman}}]{Garcia-Barriocanal2013}%
  \BibitemOpen
  \bibfield  {author} {\bibinfo {author} {\bibfnamefont {J.}~\bibnamefont
  {Garcia-Barriocanal}}, \bibinfo {author} {\bibfnamefont {A.}~\bibnamefont
  {Kobrinskii}}, \bibinfo {author} {\bibfnamefont {X.}~\bibnamefont {Leng}},
  \bibinfo {author} {\bibfnamefont {J.}~\bibnamefont {Kinney}}, \bibinfo
  {author} {\bibfnamefont {B.}~\bibnamefont {Yang}}, \bibinfo {author}
  {\bibfnamefont {S.}~\bibnamefont {Snyder}}, \ and\ \bibinfo {author}
  {\bibfnamefont {A.~M.}\ \bibnamefont {Goldman}},\ }\href {\doibase
  10.1103/PhysRevB.87.024509} {\bibfield  {journal} {\bibinfo  {journal} {Phys.
  Rev. B}\ }\textbf {\bibinfo {volume} {87}},\ \bibinfo {pages} {024509}
  (\bibinfo {year} {2013})}\BibitemShut {NoStop}%
\bibitem [{\citenamefont {Vu{\v{c}}i{\v{c}}evi{\'{c}}}\ \emph
  {et~al.}(2015)\citenamefont {Vu{\v{c}}i{\v{c}}evi{\'{c}}}, \citenamefont
  {Tanaskovi{\'{c}}}, \citenamefont {Rozenberg},\ and\ \citenamefont
  {Dobrosavljevi{\'{c}}}}]{Vucicevic2015}%
  \BibitemOpen
  \bibfield  {author} {\bibinfo {author} {\bibfnamefont {J.}~\bibnamefont
  {Vu{\v{c}}i{\v{c}}evi{\'{c}}}}, \bibinfo {author} {\bibfnamefont
  {D.}~\bibnamefont {Tanaskovi{\'{c}}}}, \bibinfo {author} {\bibfnamefont
  {M.~J.}\ \bibnamefont {Rozenberg}}, \ and\ \bibinfo {author} {\bibfnamefont
  {V.}~\bibnamefont {Dobrosavljevi{\'{c}}}},\ }\href {\doibase
  10.1103/PhysRevLett.114.246402} {\bibfield  {journal} {\bibinfo  {journal}
  {Phys. Rev. Lett.}\ }\textbf {\bibinfo {volume} {114}},\ \bibinfo {pages}
  {246402} (\bibinfo {year} {2015})}\BibitemShut {NoStop}%
\bibitem [{\citenamefont {Terletska}\ \emph {et~al.}(2011)\citenamefont
  {Terletska}, \citenamefont {Vu{\v{c}}i{\v{c}}evi{\'{c}}}, \citenamefont
  {Tanaskovi{\'{c}}},\ and\ \citenamefont
  {Dobrosavljevi{\'{c}}}}]{Terletska2011}%
  \BibitemOpen
  \bibfield  {author} {\bibinfo {author} {\bibfnamefont {H.}~\bibnamefont
  {Terletska}}, \bibinfo {author} {\bibfnamefont {J.}~\bibnamefont
  {Vu{\v{c}}i{\v{c}}evi{\'{c}}}}, \bibinfo {author} {\bibfnamefont
  {D.}~\bibnamefont {Tanaskovi{\'{c}}}}, \ and\ \bibinfo {author}
  {\bibfnamefont {V.}~\bibnamefont {Dobrosavljevi{\'{c}}}},\ }\href {\doibase
  10.1103/PhysRevLett.107.026401} {\bibfield  {journal} {\bibinfo  {journal}
  {Phys. Rev. Lett.}\ }\textbf {\bibinfo {volume} {107}},\ \bibinfo {pages}
  {026401} (\bibinfo {year} {2011})}\BibitemShut {NoStop}%
\bibitem [{\citenamefont {Kawasugi}\ \emph {et~al.}(2009)\citenamefont
  {Kawasugi}, \citenamefont {Yamamoto}, \citenamefont {Tajima}, \citenamefont
  {Fukunaga}, \citenamefont {Tsukagoshi},\ and\ \citenamefont
  {Kato}}]{Kawasugi2009a}%
  \BibitemOpen
  \bibfield  {author} {\bibinfo {author} {\bibfnamefont {Y.}~\bibnamefont
  {Kawasugi}}, \bibinfo {author} {\bibfnamefont {H.}~\bibnamefont {Yamamoto}},
  \bibinfo {author} {\bibfnamefont {N.}~\bibnamefont {Tajima}}, \bibinfo
  {author} {\bibfnamefont {T.}~\bibnamefont {Fukunaga}}, \bibinfo {author}
  {\bibfnamefont {K.}~\bibnamefont {Tsukagoshi}}, \ and\ \bibinfo {author}
  {\bibfnamefont {R.}~\bibnamefont {Kato}},\ }\href {\doibase
  10.1103/PhysRevLett.103.116801} {\bibfield  {journal} {\bibinfo  {journal}
  {Phys. Rev. Lett.}\ }\textbf {\bibinfo {volume} {103}},\ \bibinfo {pages}
  {116801} (\bibinfo {year} {2009})}\BibitemShut {NoStop}%
\bibitem [{\citenamefont {Sordi}\ \emph {et~al.}(2011)\citenamefont {Sordi},
  \citenamefont {Haule},\ and\ \citenamefont {Tremblay}}]{Sordi2011}%
  \BibitemOpen
  \bibfield  {author} {\bibinfo {author} {\bibfnamefont {G.}~\bibnamefont
  {Sordi}}, \bibinfo {author} {\bibfnamefont {K.}~\bibnamefont {Haule}}, \ and\
  \bibinfo {author} {\bibfnamefont {a.-M.~S.}\ \bibnamefont {Tremblay}},\
  }\href {\doibase 10.1103/PhysRevB.84.075161} {\bibfield  {journal} {\bibinfo
  {journal} {Phys. Rev. B}\ }\textbf {\bibinfo {volume} {84}},\ \bibinfo
  {pages} {075161} (\bibinfo {year} {2011})}\BibitemShut {NoStop}%
\bibitem [{\citenamefont {Cooper}\ \emph {et~al.}(1990)\citenamefont {Cooper},
  \citenamefont {Thomas}, \citenamefont {Orenstein}, \citenamefont {Rapkine},
  \citenamefont {Millis}, \citenamefont {Cheong}, \citenamefont {Cooper},\ and\
  \citenamefont {Fisk}}]{Cooper1990}%
  \BibitemOpen
  \bibfield  {author} {\bibinfo {author} {\bibfnamefont {S.~L.}\ \bibnamefont
  {Cooper}}, \bibinfo {author} {\bibfnamefont {G.~A.}\ \bibnamefont {Thomas}},
  \bibinfo {author} {\bibfnamefont {J.}~\bibnamefont {Orenstein}}, \bibinfo
  {author} {\bibfnamefont {D.~H.}\ \bibnamefont {Rapkine}}, \bibinfo {author}
  {\bibfnamefont {A.~J.}\ \bibnamefont {Millis}}, \bibinfo {author}
  {\bibfnamefont {S.-W.}\ \bibnamefont {Cheong}}, \bibinfo {author}
  {\bibfnamefont {A.~S.}\ \bibnamefont {Cooper}}, \ and\ \bibinfo {author}
  {\bibfnamefont {Z.}~\bibnamefont {Fisk}},\ }\href {\doibase
  10.1103/PhysRevB.41.11605} {\bibfield  {journal} {\bibinfo  {journal} {Phys.
  Rev. B}\ }\textbf {\bibinfo {volume} {41}},\ \bibinfo {pages} {11605}
  (\bibinfo {year} {1990})}\BibitemShut {NoStop}%
\bibitem [{\citenamefont {Radisavljevic}\ and\ \citenamefont
  {Kis}(2013)}]{Radisavljevic2013}%
  \BibitemOpen
  \bibfield  {author} {\bibinfo {author} {\bibfnamefont {B.}~\bibnamefont
  {Radisavljevic}}\ and\ \bibinfo {author} {\bibfnamefont {A.}~\bibnamefont
  {Kis}},\ }\href {\doibase 10.1038/nmat3687} {\bibfield  {journal} {\bibinfo
  {journal} {Nat. Mater.}\ }\textbf {\bibinfo {volume} {12}},\ \bibinfo {pages}
  {815} (\bibinfo {year} {2013})}\BibitemShut {NoStop}%
\bibitem [{\citenamefont {Knyazev}\ \emph {et~al.}(2008)\citenamefont
  {Knyazev}, \citenamefont {Omel'yanovskii}, \citenamefont {Pudalov},\ and\
  \citenamefont {Burmistrov}}]{Knyazev2008}%
  \BibitemOpen
  \bibfield  {author} {\bibinfo {author} {\bibfnamefont {D.~A.}\ \bibnamefont
  {Knyazev}}, \bibinfo {author} {\bibfnamefont {O.~E.}\ \bibnamefont
  {Omel'yanovskii}}, \bibinfo {author} {\bibfnamefont {V.~M.}\ \bibnamefont
  {Pudalov}}, \ and\ \bibinfo {author} {\bibfnamefont {I.~S.}\ \bibnamefont
  {Burmistrov}},\ }\href {\doibase 10.1103/PhysRevLett.100.046405} {\bibfield
  {journal} {\bibinfo  {journal} {Phys. Rev. Lett.}\ }\textbf {\bibinfo
  {volume} {100}},\ \bibinfo {pages} {046405} (\bibinfo {year}
  {2008})}\BibitemShut {NoStop}%
\bibitem [{\citenamefont {Suda}\ \emph {et~al.}(2015)\citenamefont {Suda},
  \citenamefont {Kato},\ and\ \citenamefont {Yamamoto}}]{Suda2015}%
  \BibitemOpen
  \bibfield  {author} {\bibinfo {author} {\bibfnamefont {M.}~\bibnamefont
  {Suda}}, \bibinfo {author} {\bibfnamefont {R.}~\bibnamefont {Kato}}, \ and\
  \bibinfo {author} {\bibfnamefont {H.~M.}\ \bibnamefont {Yamamoto}},\ }\href
  {\doibase 10.1126/science.1256783} {\bibfield  {journal} {\bibinfo  {journal}
  {Science}\ }\textbf {\bibinfo {volume} {347}},\ \bibinfo {pages} {743}
  (\bibinfo {year} {2015})}\BibitemShut {NoStop}%
\bibitem [{\citenamefont {Hikami}\ \emph {et~al.}(1980)\citenamefont {Hikami},
  \citenamefont {Larkin},\ and\ \citenamefont {Nagaoka}}]{Hikami1980}%
  \BibitemOpen
  \bibfield  {author} {\bibinfo {author} {\bibfnamefont {S.}~\bibnamefont
  {Hikami}}, \bibinfo {author} {\bibfnamefont {A.~I.}\ \bibnamefont {Larkin}},
  \ and\ \bibinfo {author} {\bibfnamefont {Y.}~\bibnamefont {Nagaoka}},\ }\href
  {\doibase 10.1143/PTP.63.707} {\bibfield  {journal} {\bibinfo  {journal}
  {Prog. Theor. Phys.}\ }\textbf {\bibinfo {volume} {63}},\ \bibinfo {pages}
  {707} (\bibinfo {year} {1980})}\BibitemShut {NoStop}%
\bibitem [{\citenamefont {Limelette}\ \emph
  {et~al.}(2003{\natexlab{a}})\citenamefont {Limelette}, \citenamefont
  {Wzietek}, \citenamefont {Florens}, \citenamefont {Georges}, \citenamefont
  {Costi}, \citenamefont {Pasquier}, \citenamefont {J{\'{e}}rome},
  \citenamefont {M{\'{e}}zi{\`{e}}re},\ and\ \citenamefont
  {Batail}}]{Limelette2003a}%
  \BibitemOpen
  \bibfield  {author} {\bibinfo {author} {\bibfnamefont {P.}~\bibnamefont
  {Limelette}}, \bibinfo {author} {\bibfnamefont {P.}~\bibnamefont {Wzietek}},
  \bibinfo {author} {\bibfnamefont {S.}~\bibnamefont {Florens}}, \bibinfo
  {author} {\bibfnamefont {A.}~\bibnamefont {Georges}}, \bibinfo {author}
  {\bibfnamefont {T.~A.}\ \bibnamefont {Costi}}, \bibinfo {author}
  {\bibfnamefont {C.}~\bibnamefont {Pasquier}}, \bibinfo {author}
  {\bibfnamefont {D.}~\bibnamefont {J{\'{e}}rome}}, \bibinfo {author}
  {\bibfnamefont {C.}~\bibnamefont {M{\'{e}}zi{\`{e}}re}}, \ and\ \bibinfo
  {author} {\bibfnamefont {P.}~\bibnamefont {Batail}},\ }\href {\doibase
  10.1103/PhysRevLett.91.016401} {\bibfield  {journal} {\bibinfo  {journal}
  {Phys. Rev. Lett.}\ }\textbf {\bibinfo {volume} {91}},\ \bibinfo {pages}
  {016401} (\bibinfo {year} {2003}{\natexlab{a}})}\BibitemShut {NoStop}%
\bibitem [{\citenamefont {Emery}\ and\ \citenamefont
  {Kivelson}(1995)}]{Emery1995}%
  \BibitemOpen
  \bibfield  {author} {\bibinfo {author} {\bibfnamefont {V.~J.}\ \bibnamefont
  {Emery}}\ and\ \bibinfo {author} {\bibfnamefont {S.~A.}\ \bibnamefont
  {Kivelson}},\ }\href {\doibase 10.1103/PhysRevLett.74.3253} {\bibfield
  {journal} {\bibinfo  {journal} {Phys. Rev. Lett.}\ }\textbf {\bibinfo
  {volume} {74}},\ \bibinfo {pages} {3253} (\bibinfo {year}
  {1995})}\BibitemShut {NoStop}%
\bibitem [{\citenamefont {Takagi}\ \emph {et~al.}(1992)\citenamefont {Takagi},
  \citenamefont {Batlogg}, \citenamefont {Kao}, \citenamefont {Kwo},
  \citenamefont {Cava}, \citenamefont {Krajewski},\ and\ \citenamefont
  {Peck}}]{Takagi1992}%
  \BibitemOpen
  \bibfield  {author} {\bibinfo {author} {\bibfnamefont {H.}~\bibnamefont
  {Takagi}}, \bibinfo {author} {\bibfnamefont {B.}~\bibnamefont {Batlogg}},
  \bibinfo {author} {\bibfnamefont {H.~L.}\ \bibnamefont {Kao}}, \bibinfo
  {author} {\bibfnamefont {J.}~\bibnamefont {Kwo}}, \bibinfo {author}
  {\bibfnamefont {R.~J.}\ \bibnamefont {Cava}}, \bibinfo {author}
  {\bibfnamefont {J.~J.}\ \bibnamefont {Krajewski}}, \ and\ \bibinfo {author}
  {\bibfnamefont {W.~F.}\ \bibnamefont {Peck}, \bibfnamefont {Jr.}},\ }\href
  {\doibase 10.1103/PhysRevLett.69.2975} {\bibfield  {journal} {\bibinfo
  {journal} {Phys. Rev. Lett.}\ }\textbf {\bibinfo {volume} {69}},\ \bibinfo
  {pages} {2975} (\bibinfo {year} {1992})}\BibitemShut {NoStop}%
\bibitem [{\citenamefont {Si}\ \emph {et~al.}(2016)\citenamefont {Si},
  \citenamefont {Yu},\ and\ \citenamefont {Abrahams}}]{Si2016}%
  \BibitemOpen
  \bibfield  {author} {\bibinfo {author} {\bibfnamefont {Q.}~\bibnamefont
  {Si}}, \bibinfo {author} {\bibfnamefont {R.}~\bibnamefont {Yu}}, \ and\
  \bibinfo {author} {\bibfnamefont {E.}~\bibnamefont {Abrahams}},\ }\href
  {\doibase 10.1038/natrevmats.2016.17} {\bibfield  {journal} {\bibinfo
  {journal} {Nat. Rev. Mater.}\ }\textbf {\bibinfo {volume} {1}},\ \bibinfo
  {pages} {16017} (\bibinfo {year} {2016})}\BibitemShut {NoStop}%
\bibitem [{\citenamefont {Vu{\v{c}}i{\v{c}}evi{\'{c}}}\ \emph
  {et~al.}(2013)\citenamefont {Vu{\v{c}}i{\v{c}}evi{\'{c}}}, \citenamefont
  {Terletska}, \citenamefont {Tanaskovi{\'{c}}},\ and\ \citenamefont
  {Dobrosavljevi{\'{c}}}}]{Vucicevic2013}%
  \BibitemOpen
  \bibfield  {author} {\bibinfo {author} {\bibfnamefont {J.}~\bibnamefont
  {Vu{\v{c}}i{\v{c}}evi{\'{c}}}}, \bibinfo {author} {\bibfnamefont
  {H.}~\bibnamefont {Terletska}}, \bibinfo {author} {\bibfnamefont
  {D.}~\bibnamefont {Tanaskovi{\'{c}}}}, \ and\ \bibinfo {author}
  {\bibfnamefont {V.}~\bibnamefont {Dobrosavljevi{\'{c}}}},\ }\href {\doibase
  10.1103/PhysRevB.88.075143} {\bibfield  {journal} {\bibinfo  {journal} {Phys.
  Rev. B}\ }\textbf {\bibinfo {volume} {88}},\ \bibinfo {pages} {075143}
  (\bibinfo {year} {2013})}\BibitemShut {NoStop}%
\bibitem [{\citenamefont {Komatsu}\ \emph {et~al.}(1996)\citenamefont
  {Komatsu}, \citenamefont {Matsukawa}, \citenamefont {Inoue},\ and\
  \citenamefont {Saito}}]{Komatsu1996}%
  \BibitemOpen
  \bibfield  {author} {\bibinfo {author} {\bibfnamefont {T.}~\bibnamefont
  {Komatsu}}, \bibinfo {author} {\bibfnamefont {N.}~\bibnamefont {Matsukawa}},
  \bibinfo {author} {\bibfnamefont {T.}~\bibnamefont {Inoue}}, \ and\ \bibinfo
  {author} {\bibfnamefont {G.}~\bibnamefont {Saito}},\ }\href {\doibase
  10.1143/JPSJ.65.1340} {\bibfield  {journal} {\bibinfo  {journal} {J. Phys.
  Soc. Jpn.}\ }\textbf {\bibinfo {volume} {65}},\ \bibinfo {pages} {1340}
  (\bibinfo {year} {1996})}\BibitemShut {NoStop}%
\bibitem [{\citenamefont {Phillips}(2012)}]{Phillips2012}%
  \BibitemOpen
  \bibfield  {author} {\bibinfo {author} {\bibfnamefont {P.}~\bibnamefont
  {Phillips}},\ }\href@noop {} {\emph {\bibinfo {title} {{Advancded solid state
  physics}}}},\ \bibinfo {edition} {2nd}\ ed.\ (\bibinfo  {publisher}
  {Cambridge University Press},\ \bibinfo {year} {2012})\BibitemShut {NoStop}%
\bibitem [{\citenamefont {Gantmakher}\ and\ \citenamefont
  {Dolgopolov}(2010)}]{Gantmakher2010}%
  \BibitemOpen
  \bibfield  {author} {\bibinfo {author} {\bibfnamefont {V.~F.}\ \bibnamefont
  {Gantmakher}}\ and\ \bibinfo {author} {\bibfnamefont {V.~T.}\ \bibnamefont
  {Dolgopolov}},\ }\href {\doibase 10.3367/UFNe.0180.201001a.0003} {\bibfield
  {journal} {\bibinfo  {journal} {Physics-Uspekhi}\ }\textbf {\bibinfo {volume}
  {53}},\ \bibinfo {pages} {57} (\bibinfo {year} {2010})}\BibitemShut {NoStop}%
\bibitem [{\citenamefont {Sondhi}\ \emph {et~al.}(1997)\citenamefont {Sondhi},
  \citenamefont {Girvin}, \citenamefont {Carini},\ and\ \citenamefont
  {Shahar}}]{Sondhi1997}%
  \BibitemOpen
  \bibfield  {author} {\bibinfo {author} {\bibfnamefont {S.~L.}\ \bibnamefont
  {Sondhi}}, \bibinfo {author} {\bibfnamefont {S.~M.}\ \bibnamefont {Girvin}},
  \bibinfo {author} {\bibfnamefont {J.~P.}\ \bibnamefont {Carini}}, \ and\
  \bibinfo {author} {\bibfnamefont {D.}~\bibnamefont {Shahar}},\ }\href
  {\doibase 10.1103/RevModPhys.69.315} {\bibfield  {journal} {\bibinfo
  {journal} {Rev. Mod. Phys.}\ }\textbf {\bibinfo {volume} {69}},\ \bibinfo
  {pages} {315} (\bibinfo {year} {1997})}\BibitemShut {NoStop}%
\bibitem [{\citenamefont {Dobrosavljevi{\'{c}}}\ \emph
  {et~al.}(1997)\citenamefont {Dobrosavljevi{\'{c}}}, \citenamefont {Abrahams},
  \citenamefont {Miranda},\ and\ \citenamefont
  {Chakravarty}}]{Dobrosavljevic1997}%
  \BibitemOpen
  \bibfield  {author} {\bibinfo {author} {\bibfnamefont {V.}~\bibnamefont
  {Dobrosavljevi{\'{c}}}}, \bibinfo {author} {\bibfnamefont {E.}~\bibnamefont
  {Abrahams}}, \bibinfo {author} {\bibfnamefont {E.}~\bibnamefont {Miranda}}, \
  and\ \bibinfo {author} {\bibfnamefont {S.}~\bibnamefont {Chakravarty}},\
  }\href {\doibase 10.1103/PhysRevLett.79.455} {\bibfield  {journal} {\bibinfo
  {journal} {Phys. Rev. Lett.}\ }\textbf {\bibinfo {volume} {79}},\ \bibinfo
  {pages} {455} (\bibinfo {year} {1997})}\BibitemShut {NoStop}%
\bibitem [{\citenamefont {Simeoni}\ \emph {et~al.}(2010)\citenamefont
  {Simeoni}, \citenamefont {Bryk}, \citenamefont {Gorelli}, \citenamefont
  {Krisch}, \citenamefont {Ruocco}, \citenamefont {Santoro},\ and\
  \citenamefont {Scopigno}}]{Simeoni2010}%
  \BibitemOpen
  \bibfield  {author} {\bibinfo {author} {\bibfnamefont {G.~G.}\ \bibnamefont
  {Simeoni}}, \bibinfo {author} {\bibfnamefont {T.}~\bibnamefont {Bryk}},
  \bibinfo {author} {\bibfnamefont {F.~A.}\ \bibnamefont {Gorelli}}, \bibinfo
  {author} {\bibfnamefont {M.}~\bibnamefont {Krisch}}, \bibinfo {author}
  {\bibfnamefont {G.}~\bibnamefont {Ruocco}}, \bibinfo {author} {\bibfnamefont
  {M.}~\bibnamefont {Santoro}}, \ and\ \bibinfo {author} {\bibfnamefont
  {T.}~\bibnamefont {Scopigno}},\ }\href {\doibase 10.1038/nphys1683}
  {\bibfield  {journal} {\bibinfo  {journal} {Nat. Phys.}\ }\textbf {\bibinfo
  {volume} {6}},\ \bibinfo {pages} {503} (\bibinfo {year} {2010})}\BibitemShut
  {NoStop}%
\bibitem [{\citenamefont {Abrahams}\ \emph {et~al.}(2001)\citenamefont
  {Abrahams}, \citenamefont {Kravchenko},\ and\ \citenamefont
  {Sarachik}}]{Abrahams2001}%
  \BibitemOpen
  \bibfield  {author} {\bibinfo {author} {\bibfnamefont {E.}~\bibnamefont
  {Abrahams}}, \bibinfo {author} {\bibfnamefont {S.~V.}\ \bibnamefont
  {Kravchenko}}, \ and\ \bibinfo {author} {\bibfnamefont {M.~P.}\ \bibnamefont
  {Sarachik}},\ }\href {\doibase 10.1103/RevModPhys.73.251} {\bibfield
  {journal} {\bibinfo  {journal} {Rev. Mod. Phys.}\ }\textbf {\bibinfo {volume}
  {73}},\ \bibinfo {pages} {251} (\bibinfo {year} {2001})}\BibitemShut
  {NoStop}%
\bibitem [{\citenamefont {Dobrosavljevi{\'{c}}}(2012)}]{Dobrosavljevic2012}%
  \BibitemOpen
  \bibfield  {author} {\bibinfo {author} {\bibfnamefont {V.}~\bibnamefont
  {Dobrosavljevi{\'{c}}}},\ }in\ \href {http://arxiv.org/abs/1112.6166} {\emph
  {\bibinfo {booktitle} {Conductor-insulator quantum phase transition}}},\
  \bibinfo {editor} {edited by\ \bibinfo {editor} {\bibfnamefont
  {V.}~\bibnamefont {Dobrosavljevi{\'{c}}}}, \bibinfo {editor} {\bibfnamefont
  {N.}~\bibnamefont {Trivedi}}, \ and\ \bibinfo {editor} {\bibfnamefont
  {J.~M.}\ \bibnamefont {{Valles, Jr.}}}}\ (\bibinfo  {publisher} {Oxford
  University Press},\ \bibinfo {year} {2012})\ Chap.~\bibinfo {chapter}
  {1}\BibitemShut {NoStop}%
\bibitem [{\citenamefont {Abdel-Jawad}\ \emph {et~al.}(2015)\citenamefont
  {Abdel-Jawad}, \citenamefont {Kato}, \citenamefont {Watanabe}, \citenamefont
  {Tajima},\ and\ \citenamefont {Ishii}}]{Abdel-Jawad2015}%
  \BibitemOpen
  \bibfield  {author} {\bibinfo {author} {\bibfnamefont {M.}~\bibnamefont
  {Abdel-Jawad}}, \bibinfo {author} {\bibfnamefont {R.}~\bibnamefont {Kato}},
  \bibinfo {author} {\bibfnamefont {I.}~\bibnamefont {Watanabe}}, \bibinfo
  {author} {\bibfnamefont {N.}~\bibnamefont {Tajima}}, \ and\ \bibinfo {author}
  {\bibfnamefont {Y.}~\bibnamefont {Ishii}},\ }\href {\doibase
  10.1103/PhysRevLett.114.106401} {\bibfield  {journal} {\bibinfo  {journal}
  {Phys. Rev. Lett.}\ }\textbf {\bibinfo {volume} {114}},\ \bibinfo {pages}
  {106401} (\bibinfo {year} {2015})}\BibitemShut {NoStop}%
\bibitem [{\citenamefont {Gunnarsson}\ \emph {et~al.}(2003)\citenamefont
  {Gunnarsson}, \citenamefont {Calandra},\ and\ \citenamefont
  {Han}}]{Gunnarsson2003}%
  \BibitemOpen
  \bibfield  {author} {\bibinfo {author} {\bibfnamefont {O.}~\bibnamefont
  {Gunnarsson}}, \bibinfo {author} {\bibfnamefont {M.}~\bibnamefont
  {Calandra}}, \ and\ \bibinfo {author} {\bibfnamefont {J.~E.}\ \bibnamefont
  {Han}},\ }\href {\doibase 10.1103/RevModPhys.75.1085} {\bibfield  {journal}
  {\bibinfo  {journal} {Rev. Mod. Phys.}\ }\textbf {\bibinfo {volume} {75}},\
  \bibinfo {pages} {1085} (\bibinfo {year} {2003})}\BibitemShut {NoStop}%
\bibitem [{\citenamefont {Phillips}\ and\ \citenamefont
  {Chamon}(2005)}]{Phillips2005}%
  \BibitemOpen
  \bibfield  {author} {\bibinfo {author} {\bibfnamefont {P.}~\bibnamefont
  {Phillips}}\ and\ \bibinfo {author} {\bibfnamefont {C.}~\bibnamefont
  {Chamon}},\ }\href {\doibase 10.1103/PhysRevLett.95.107002} {\bibfield
  {journal} {\bibinfo  {journal} {Phys. Rev. Lett.}\ }\textbf {\bibinfo
  {volume} {95}},\ \bibinfo {pages} {107002} (\bibinfo {year}
  {2005})}\BibitemShut {NoStop}%
\bibitem [{\citenamefont {Varma}\ \emph {et~al.}(1989)\citenamefont {Varma},
  \citenamefont {Littlewood}, \citenamefont {Schmitt-Rink}, \citenamefont
  {Abrahams},\ and\ \citenamefont {Ruckenstein}}]{Varma1989}%
  \BibitemOpen
  \bibfield  {author} {\bibinfo {author} {\bibfnamefont {C.~M.}\ \bibnamefont
  {Varma}}, \bibinfo {author} {\bibfnamefont {P.~B.}\ \bibnamefont
  {Littlewood}}, \bibinfo {author} {\bibfnamefont {S.}~\bibnamefont
  {Schmitt-Rink}}, \bibinfo {author} {\bibfnamefont {E.}~\bibnamefont
  {Abrahams}}, \ and\ \bibinfo {author} {\bibfnamefont {A.~E.}\ \bibnamefont
  {Ruckenstein}},\ }\href {\doibase 10.1103/PhysRevLett.63.1996} {\bibfield
  {journal} {\bibinfo  {journal} {Phys. Rev. Lett.}\ }\textbf {\bibinfo
  {volume} {63}},\ \bibinfo {pages} {1996} (\bibinfo {year}
  {1989})}\BibitemShut {NoStop}%
\bibitem [{\citenamefont {Hartnoll}(2014)}]{Hartnoll2014}%
  \BibitemOpen
  \bibfield  {author} {\bibinfo {author} {\bibfnamefont {S.~A.}\ \bibnamefont
  {Hartnoll}},\ }\href {\doibase 10.1038/NPHYS3174} {\bibfield  {journal}
  {\bibinfo  {journal} {Nat. Phys.}\ }\textbf {\bibinfo {volume} {11}},\
  \bibinfo {pages} {54} (\bibinfo {year} {2014})}\BibitemShut {NoStop}%
\bibitem [{\citenamefont {Davison}\ \emph {et~al.}(2014)\citenamefont
  {Davison}, \citenamefont {Schalm},\ and\ \citenamefont
  {Zaanen}}]{Davison2014}%
  \BibitemOpen
  \bibfield  {author} {\bibinfo {author} {\bibfnamefont {R.~A.}\ \bibnamefont
  {Davison}}, \bibinfo {author} {\bibfnamefont {K.}~\bibnamefont {Schalm}}, \
  and\ \bibinfo {author} {\bibfnamefont {J.}~\bibnamefont {Zaanen}},\ }\href
  {\doibase 10.1103/PhysRevB.89.245116} {\bibfield  {journal} {\bibinfo
  {journal} {Phys. Rev. B}\ }\textbf {\bibinfo {volume} {89}},\ \bibinfo
  {pages} {245116} (\bibinfo {year} {2014})}\BibitemShut {NoStop}%
\bibitem [{\citenamefont {Kawasugi}\ \emph {et~al.}(2008)\citenamefont
  {Kawasugi}, \citenamefont {Yamamoto}, \citenamefont {Hosoda}, \citenamefont
  {Tajima}, \citenamefont {Fukunaga}, \citenamefont {Tsukagoshi},\ and\
  \citenamefont {Kato}}]{Kawasugi2008}%
  \BibitemOpen
  \bibfield  {author} {\bibinfo {author} {\bibfnamefont {Y.}~\bibnamefont
  {Kawasugi}}, \bibinfo {author} {\bibfnamefont {H.~M.}\ \bibnamefont
  {Yamamoto}}, \bibinfo {author} {\bibfnamefont {M.}~\bibnamefont {Hosoda}},
  \bibinfo {author} {\bibfnamefont {N.}~\bibnamefont {Tajima}}, \bibinfo
  {author} {\bibfnamefont {T.}~\bibnamefont {Fukunaga}}, \bibinfo {author}
  {\bibfnamefont {K.}~\bibnamefont {Tsukagoshi}}, \ and\ \bibinfo {author}
  {\bibfnamefont {R.}~\bibnamefont {Kato}},\ }\href {\doibase
  10.1063/1.2949316} {\bibfield  {journal} {\bibinfo  {journal} {Appl. Phys.
  Lett.}\ }\textbf {\bibinfo {volume} {92}},\ \bibinfo {pages} {243508}
  (\bibinfo {year} {2008})}\BibitemShut {NoStop}%
\bibitem [{\citenamefont {Schlich}\ and\ \citenamefont
  {Spolenak}(2013)}]{Schlich2013}%
  \BibitemOpen
  \bibfield  {author} {\bibinfo {author} {\bibfnamefont {F.~F.}\ \bibnamefont
  {Schlich}}\ and\ \bibinfo {author} {\bibfnamefont {R.}~\bibnamefont
  {Spolenak}},\ }\href {\doibase 10.1063/1.4833537} {\bibfield  {journal}
  {\bibinfo  {journal} {Appl. Phys. Lett.}\ }\textbf {\bibinfo {volume}
  {103}},\ \bibinfo {pages} {213112} (\bibinfo {year} {2013})}\BibitemShut
  {NoStop}%
\bibitem [{\citenamefont {Blake}\ \emph {et~al.}(2007)\citenamefont {Blake},
  \citenamefont {Hill}, \citenamefont {Castro~Neto}, \citenamefont {Novoselov},
  \citenamefont {Jiang}, \citenamefont {Yang}, \citenamefont {Booth},\ and\
  \citenamefont {Geim}}]{Blake2007}%
  \BibitemOpen
  \bibfield  {author} {\bibinfo {author} {\bibfnamefont {P.}~\bibnamefont
  {Blake}}, \bibinfo {author} {\bibfnamefont {E.~W.}\ \bibnamefont {Hill}},
  \bibinfo {author} {\bibfnamefont {A.~H.}\ \bibnamefont {Castro~Neto}},
  \bibinfo {author} {\bibfnamefont {K.~S.}\ \bibnamefont {Novoselov}}, \bibinfo
  {author} {\bibfnamefont {D.}~\bibnamefont {Jiang}}, \bibinfo {author}
  {\bibfnamefont {R.}~\bibnamefont {Yang}}, \bibinfo {author} {\bibfnamefont
  {T.~J.}\ \bibnamefont {Booth}}, \ and\ \bibinfo {author} {\bibfnamefont
  {A.~K.}\ \bibnamefont {Geim}},\ }\href {\doibase 10.1063/1.2768624}
  {\bibfield  {journal} {\bibinfo  {journal} {Appl. Phys. Lett.}\ }\textbf
  {\bibinfo {volume} {91}},\ \bibinfo {pages} {063124} (\bibinfo {year}
  {2007})}\BibitemShut {NoStop}%
\bibitem [{\citenamefont {Nakagawa}\ \emph {et~al.}(2006)\citenamefont
  {Nakagawa}, \citenamefont {Hwang},\ and\ \citenamefont
  {Muller}}]{Nakagawa2006}%
  \BibitemOpen
  \bibfield  {author} {\bibinfo {author} {\bibfnamefont {N.}~\bibnamefont
  {Nakagawa}}, \bibinfo {author} {\bibfnamefont {H.~Y.}\ \bibnamefont {Hwang}},
  \ and\ \bibinfo {author} {\bibfnamefont {D.~A.}\ \bibnamefont {Muller}},\
  }\href {\doibase 10.1038/nmat1569} {\bibfield  {journal} {\bibinfo  {journal}
  {Nat. Mater.}\ }\textbf {\bibinfo {volume} {5}},\ \bibinfo {pages} {204}
  (\bibinfo {year} {2006})}\BibitemShut {NoStop}%
\bibitem [{\citenamefont {Yokota}\ \emph {et~al.}(2011)\citenamefont {Yokota},
  \citenamefont {Takai},\ and\ \citenamefont {Enoki}}]{YokotaK2011}%
  \BibitemOpen
  \bibfield  {author} {\bibinfo {author} {\bibfnamefont {K.}~\bibnamefont
  {Yokota}}, \bibinfo {author} {\bibfnamefont {K.}~\bibnamefont {Takai}}, \
  and\ \bibinfo {author} {\bibfnamefont {T.}~\bibnamefont {Enoki}},\ }\href
  {\doibase 10.1021/nl201607t} {\bibfield  {journal} {\bibinfo  {journal} {Nano
  Lett.}\ }\textbf {\bibinfo {volume} {11}},\ \bibinfo {pages} {3669} (\bibinfo
  {year} {2011})}\BibitemShut {NoStop}%
\bibitem [{\citenamefont {Liu}\ \emph {et~al.}(2011)\citenamefont {Liu},
  \citenamefont {Bol},\ and\ \citenamefont {Haensch}}]{Liu2011e}%
  \BibitemOpen
  \bibfield  {author} {\bibinfo {author} {\bibfnamefont {Z.}~\bibnamefont
  {Liu}}, \bibinfo {author} {\bibfnamefont {A.~A.}\ \bibnamefont {Bol}}, \ and\
  \bibinfo {author} {\bibfnamefont {W.}~\bibnamefont {Haensch}},\ }\href
  {\doibase 10.1021/nl1033842} {\bibfield  {journal} {\bibinfo  {journal} {Nano
  Lett.}\ }\textbf {\bibinfo {volume} {11}},\ \bibinfo {pages} {523} (\bibinfo
  {year} {2011})}\BibitemShut {NoStop}%
\bibitem [{\citenamefont {Tomi{\'{c}}}\ \emph {et~al.}(2013)\citenamefont
  {Tomi{\'{c}}}, \citenamefont {Pinteri{\'{c}}}, \citenamefont {Ivek},
  \citenamefont {Sedlmeier}, \citenamefont {Beyer}, \citenamefont {Wu},
  \citenamefont {Schlueter}, \citenamefont {Schweitzer},\ and\ \citenamefont
  {Dressel}}]{Tomic2013}%
  \BibitemOpen
  \bibfield  {author} {\bibinfo {author} {\bibfnamefont {S.}~\bibnamefont
  {Tomi{\'{c}}}}, \bibinfo {author} {\bibfnamefont {M.}~\bibnamefont
  {Pinteri{\'{c}}}}, \bibinfo {author} {\bibfnamefont {T.}~\bibnamefont
  {Ivek}}, \bibinfo {author} {\bibfnamefont {K.}~\bibnamefont {Sedlmeier}},
  \bibinfo {author} {\bibfnamefont {R.}~\bibnamefont {Beyer}}, \bibinfo
  {author} {\bibfnamefont {D.}~\bibnamefont {Wu}}, \bibinfo {author}
  {\bibfnamefont {J.~A.}\ \bibnamefont {Schlueter}}, \bibinfo {author}
  {\bibfnamefont {D.}~\bibnamefont {Schweitzer}}, \ and\ \bibinfo {author}
  {\bibfnamefont {M.}~\bibnamefont {Dressel}},\ }\href {\doibase
  10.1088/0953-8984/25/43/436004} {\bibfield  {journal} {\bibinfo  {journal}
  {J. Phys. Condens. Matter}\ }\textbf {\bibinfo {volume} {25}},\ \bibinfo
  {pages} {436004} (\bibinfo {year} {2013})}\BibitemShut {NoStop}%
\bibitem [{\citenamefont {Ito}\ \emph {et~al.}(1996)\citenamefont {Ito},
  \citenamefont {Ishiguro}, \citenamefont {Kubota},\ and\ \citenamefont
  {Saito}}]{Ito1996}%
  \BibitemOpen
  \bibfield  {author} {\bibinfo {author} {\bibfnamefont {H.}~\bibnamefont
  {Ito}}, \bibinfo {author} {\bibfnamefont {T.}~\bibnamefont {Ishiguro}},
  \bibinfo {author} {\bibfnamefont {M.}~\bibnamefont {Kubota}}, \ and\ \bibinfo
  {author} {\bibfnamefont {G.}~\bibnamefont {Saito}},\ }\href {\doibase
  10.1143/JPSJ.65.2987} {\bibfield  {journal} {\bibinfo  {journal} {J. Phys.
  Soc. Jpn.}\ }\textbf {\bibinfo {volume} {65}},\ \bibinfo {pages} {2987}
  (\bibinfo {year} {1996})}\BibitemShut {NoStop}%
\bibitem [{\citenamefont {Els{\"{a}}sser}\ \emph {et~al.}(2012)\citenamefont
  {Els{\"{a}}sser}, \citenamefont {Wu}, \citenamefont {Dressel},\ and\
  \citenamefont {Schlueter}}]{Elsasser2012}%
  \BibitemOpen
  \bibfield  {author} {\bibinfo {author} {\bibfnamefont {S.}~\bibnamefont
  {Els{\"{a}}sser}}, \bibinfo {author} {\bibfnamefont {D.}~\bibnamefont {Wu}},
  \bibinfo {author} {\bibfnamefont {M.}~\bibnamefont {Dressel}}, \ and\
  \bibinfo {author} {\bibfnamefont {J.~A.}\ \bibnamefont {Schlueter}},\ }\href
  {\doibase 10.1103/PhysRevB.86.155150} {\bibfield  {journal} {\bibinfo
  {journal} {Phys. Rev. B}\ }\textbf {\bibinfo {volume} {86}},\ \bibinfo
  {pages} {155150} (\bibinfo {year} {2012})}\BibitemShut {NoStop}%
\bibitem [{\citenamefont {Williams}\ \emph {et~al.}(1990)\citenamefont
  {Williams}, \citenamefont {Kini}, \citenamefont {Wang}, \citenamefont
  {Carlson}, \citenamefont {Geiser}, \citenamefont {Montgomery}, \citenamefont
  {Pyrka}, \citenamefont {Watkins},\ and\ \citenamefont
  {Kommers}}]{Williams1990}%
  \BibitemOpen
  \bibfield  {author} {\bibinfo {author} {\bibfnamefont {J.~M.}\ \bibnamefont
  {Williams}}, \bibinfo {author} {\bibfnamefont {A.~M.}\ \bibnamefont {Kini}},
  \bibinfo {author} {\bibfnamefont {H.~H.}\ \bibnamefont {Wang}}, \bibinfo
  {author} {\bibfnamefont {K.~D.}\ \bibnamefont {Carlson}}, \bibinfo {author}
  {\bibfnamefont {U.}~\bibnamefont {Geiser}}, \bibinfo {author} {\bibfnamefont
  {L.~K.}\ \bibnamefont {Montgomery}}, \bibinfo {author} {\bibfnamefont
  {G.~J.}\ \bibnamefont {Pyrka}}, \bibinfo {author} {\bibfnamefont {D.~M.}\
  \bibnamefont {Watkins}}, \ and\ \bibinfo {author} {\bibfnamefont {J.~M.}\
  \bibnamefont {Kommers}},\ }\href {\doibase 10.1021/ic00343a003} {\bibfield
  {journal} {\bibinfo  {journal} {Inorg. Chem.}\ }\textbf {\bibinfo {volume}
  {29}},\ \bibinfo {pages} {3272} (\bibinfo {year} {1990})}\BibitemShut
  {NoStop}%
\bibitem [{\citenamefont {Kagawa}\ \emph {et~al.}(2009)\citenamefont {Kagawa},
  \citenamefont {Miyagawa},\ and\ \citenamefont {Kanoda}}]{Kagawa2009}%
  \BibitemOpen
  \bibfield  {author} {\bibinfo {author} {\bibfnamefont {F.}~\bibnamefont
  {Kagawa}}, \bibinfo {author} {\bibfnamefont {K.}~\bibnamefont {Miyagawa}}, \
  and\ \bibinfo {author} {\bibfnamefont {K.}~\bibnamefont {Kanoda}},\ }\href
  {\doibase 10.1038/nphys1428} {\bibfield  {journal} {\bibinfo  {journal} {Nat.
  Phys.}\ }\textbf {\bibinfo {volume} {5}},\ \bibinfo {pages} {880} (\bibinfo
  {year} {2009})}\BibitemShut {NoStop}%
\bibitem [{\citenamefont {Kanoda}(2006)}]{Kanoda2006}%
  \BibitemOpen
  \bibfield  {author} {\bibinfo {author} {\bibfnamefont {K.}~\bibnamefont
  {Kanoda}},\ }\href {\doibase 10.1143/JPSJ.75.051007} {\bibfield  {journal}
  {\bibinfo  {journal} {J. Phys. Soc. Jpn.}\ }\textbf {\bibinfo {volume}
  {75}},\ \bibinfo {pages} {051007} (\bibinfo {year} {2006})}\BibitemShut
  {NoStop}%
\bibitem [{\citenamefont {Lefebvre}\ \emph {et~al.}(2000)\citenamefont
  {Lefebvre}, \citenamefont {Wzietek}, \citenamefont {Brown}, \citenamefont
  {Bourbonnais}, \citenamefont {J{\'{e}}rome}, \citenamefont
  {M{\'{e}}zi{\`{e}}re}, \citenamefont {Fourmigu{\'{e}}},\ and\ \citenamefont
  {Batail}}]{Lefebvre2000}%
  \BibitemOpen
  \bibfield  {author} {\bibinfo {author} {\bibfnamefont {S.}~\bibnamefont
  {Lefebvre}}, \bibinfo {author} {\bibfnamefont {P.}~\bibnamefont {Wzietek}},
  \bibinfo {author} {\bibfnamefont {S.}~\bibnamefont {Brown}}, \bibinfo
  {author} {\bibfnamefont {C.}~\bibnamefont {Bourbonnais}}, \bibinfo {author}
  {\bibfnamefont {D.}~\bibnamefont {J{\'{e}}rome}}, \bibinfo {author}
  {\bibfnamefont {C.}~\bibnamefont {M{\'{e}}zi{\`{e}}re}}, \bibinfo {author}
  {\bibfnamefont {M.}~\bibnamefont {Fourmigu{\'{e}}}}, \ and\ \bibinfo {author}
  {\bibfnamefont {P.}~\bibnamefont {Batail}},\ }\href
  {http://www.ncbi.nlm.nih.gov/pubmed/11136011} {\bibfield  {journal} {\bibinfo
   {journal} {Phys. Rev. Lett.}\ }\textbf {\bibinfo {volume} {85}},\ \bibinfo
  {pages} {5420} (\bibinfo {year} {2000})}\BibitemShut {NoStop}%
\bibitem [{\citenamefont {Yamamoto}\ \emph {et~al.}(2013)\citenamefont
  {Yamamoto}, \citenamefont {Nakano}, \citenamefont {Suda}, \citenamefont
  {Iwasa}, \citenamefont {Kawasaki},\ and\ \citenamefont
  {Kato}}]{Yamamoto2013}%
  \BibitemOpen
  \bibfield  {author} {\bibinfo {author} {\bibfnamefont {H.~M.}\ \bibnamefont
  {Yamamoto}}, \bibinfo {author} {\bibfnamefont {M.}~\bibnamefont {Nakano}},
  \bibinfo {author} {\bibfnamefont {M.}~\bibnamefont {Suda}}, \bibinfo {author}
  {\bibfnamefont {Y.}~\bibnamefont {Iwasa}}, \bibinfo {author} {\bibfnamefont
  {M.}~\bibnamefont {Kawasaki}}, \ and\ \bibinfo {author} {\bibfnamefont
  {R.}~\bibnamefont {Kato}},\ }\href {\doibase 10.1038/ncomms3379} {\bibfield
  {journal} {\bibinfo  {journal} {Nat. Commun.}\ }\textbf {\bibinfo {volume}
  {4}},\ \bibinfo {pages} {2379} (\bibinfo {year} {2013})}\BibitemShut
  {NoStop}%
\bibitem [{\citenamefont {Suda}\ \emph {et~al.}(2014)\citenamefont {Suda},
  \citenamefont {Kawasugi}, \citenamefont {Minari}, \citenamefont {Tsukagoshi},
  \citenamefont {Kato},\ and\ \citenamefont {Yamamoto}}]{Suda2014}%
  \BibitemOpen
  \bibfield  {author} {\bibinfo {author} {\bibfnamefont {M.}~\bibnamefont
  {Suda}}, \bibinfo {author} {\bibfnamefont {Y.}~\bibnamefont {Kawasugi}},
  \bibinfo {author} {\bibfnamefont {T.}~\bibnamefont {Minari}}, \bibinfo
  {author} {\bibfnamefont {K.}~\bibnamefont {Tsukagoshi}}, \bibinfo {author}
  {\bibfnamefont {R.}~\bibnamefont {Kato}}, \ and\ \bibinfo {author}
  {\bibfnamefont {H.~M.}\ \bibnamefont {Yamamoto}},\ }\href {\doibase
  10.1002/adma.201305797} {\bibfield  {journal} {\bibinfo  {journal} {Adv.
  Mater.}\ }\textbf {\bibinfo {volume} {26}},\ \bibinfo {pages} {3490}
  (\bibinfo {year} {2014})}\BibitemShut {NoStop}%
\bibitem [{\citenamefont {Hussey}\ \emph {et~al.}(2004)\citenamefont {Hussey},
  \citenamefont {Takenaka},\ and\ \citenamefont {Takagi}}]{Hussey2004}%
  \BibitemOpen
  \bibfield  {author} {\bibinfo {author} {\bibfnamefont {N.~E.}\ \bibnamefont
  {Hussey}}, \bibinfo {author} {\bibfnamefont {K.}~\bibnamefont {Takenaka}}, \
  and\ \bibinfo {author} {\bibfnamefont {H.}~\bibnamefont {Takagi}},\ }\href
  {\doibase 10.1080/14786430410001716944} {\bibfield  {journal} {\bibinfo
  {journal} {Phil. Mag.}\ }\textbf {\bibinfo {volume} {84}},\ \bibinfo {pages}
  {2847} (\bibinfo {year} {2004})}\BibitemShut {NoStop}%
\bibitem [{\citenamefont {Sordi}\ \emph {et~al.}(2012)\citenamefont {Sordi},
  \citenamefont {S{\'{e}}mon}, \citenamefont {Haule},\ and\ \citenamefont
  {Tremblay}}]{Sordi2012}%
  \BibitemOpen
  \bibfield  {author} {\bibinfo {author} {\bibfnamefont {G.}~\bibnamefont
  {Sordi}}, \bibinfo {author} {\bibfnamefont {P.}~\bibnamefont {S{\'{e}}mon}},
  \bibinfo {author} {\bibfnamefont {K.}~\bibnamefont {Haule}}, \ and\ \bibinfo
  {author} {\bibfnamefont {A.~M.~S.}\ \bibnamefont {Tremblay}},\ }\href
  {\doibase 10.1103/PhysRevLett.108.216401} {\bibfield  {journal} {\bibinfo
  {journal} {Phys. Rev. Lett.}\ }\textbf {\bibinfo {volume} {108}},\ \bibinfo
  {pages} {216401} (\bibinfo {year} {2012})}\BibitemShut {NoStop}%
\bibitem [{\citenamefont {Pradhan}\ \emph {et~al.}(2015)\citenamefont
  {Pradhan}, \citenamefont {McCreary}, \citenamefont {Rhodes}, \citenamefont
  {Lu}, \citenamefont {Feng}, \citenamefont {Manousakis}, \citenamefont
  {Smirnov}, \citenamefont {Namburu}, \citenamefont {Dubey}, \citenamefont
  {{Hight Walker}}, \citenamefont {Terrones}, \citenamefont {Terrones},
  \citenamefont {Dobrosavljevi{\'{c}}},\ and\ \citenamefont
  {Balicas}}]{Pradhan2015}%
  \BibitemOpen
  \bibfield  {author} {\bibinfo {author} {\bibfnamefont {N.~R.}\ \bibnamefont
  {Pradhan}}, \bibinfo {author} {\bibfnamefont {A.}~\bibnamefont {McCreary}},
  \bibinfo {author} {\bibfnamefont {D.}~\bibnamefont {Rhodes}}, \bibinfo
  {author} {\bibfnamefont {Z.}~\bibnamefont {Lu}}, \bibinfo {author}
  {\bibfnamefont {S.}~\bibnamefont {Feng}}, \bibinfo {author} {\bibfnamefont
  {E.}~\bibnamefont {Manousakis}}, \bibinfo {author} {\bibfnamefont
  {D.}~\bibnamefont {Smirnov}}, \bibinfo {author} {\bibfnamefont
  {R.}~\bibnamefont {Namburu}}, \bibinfo {author} {\bibfnamefont
  {M.}~\bibnamefont {Dubey}}, \bibinfo {author} {\bibfnamefont {A.~R.}\
  \bibnamefont {{Hight Walker}}}, \bibinfo {author} {\bibfnamefont
  {H.}~\bibnamefont {Terrones}}, \bibinfo {author} {\bibfnamefont
  {M.}~\bibnamefont {Terrones}}, \bibinfo {author} {\bibfnamefont
  {V.}~\bibnamefont {Dobrosavljevi{\'{c}}}}, \ and\ \bibinfo {author}
  {\bibfnamefont {L.}~\bibnamefont {Balicas}},\ }\href {\doibase
  10.1021/acs.nanolett.5b04100} {\bibfield  {journal} {\bibinfo  {journal}
  {Nano Lett.}\ }\textbf {\bibinfo {volume} {15}},\ \bibinfo {pages} {8377}
  (\bibinfo {year} {2015})}\BibitemShut {NoStop}%
\bibitem [{\citenamefont {Limelette}\ \emph
  {et~al.}(2003{\natexlab{b}})\citenamefont {Limelette}, \citenamefont
  {Georges}, \citenamefont {J{\'{e}}rome}, \citenamefont {Wzietek},
  \citenamefont {Metcalf},\ and\ \citenamefont {Honig}}]{Limelette2003}%
  \BibitemOpen
  \bibfield  {author} {\bibinfo {author} {\bibfnamefont {P.}~\bibnamefont
  {Limelette}}, \bibinfo {author} {\bibfnamefont {A.}~\bibnamefont {Georges}},
  \bibinfo {author} {\bibfnamefont {D.}~\bibnamefont {J{\'{e}}rome}}, \bibinfo
  {author} {\bibfnamefont {P.}~\bibnamefont {Wzietek}}, \bibinfo {author}
  {\bibfnamefont {P.}~\bibnamefont {Metcalf}}, \ and\ \bibinfo {author}
  {\bibfnamefont {J.~M.}\ \bibnamefont {Honig}},\ }\href {\doibase
  10.1126/science.1088386} {\bibfield  {journal} {\bibinfo  {journal}
  {Science}\ }\textbf {\bibinfo {volume} {302}},\ \bibinfo {pages} {89}
  (\bibinfo {year} {2003}{\natexlab{b}})}\BibitemShut {NoStop}%
\bibitem [{\citenamefont {Pelissetto}\ and\ \citenamefont
  {Vicari}(2002)}]{Pelissetto2002}%
  \BibitemOpen
  \bibfield  {author} {\bibinfo {author} {\bibfnamefont {A.}~\bibnamefont
  {Pelissetto}}\ and\ \bibinfo {author} {\bibfnamefont {E.}~\bibnamefont
  {Vicari}},\ }\href {\doibase 10.1016/S0370-1573(02)00219-3} {\bibfield
  {journal} {\bibinfo  {journal} {Phys. Rep.}\ }\textbf {\bibinfo {volume}
  {368}},\ \bibinfo {pages} {549} (\bibinfo {year} {2002})}\BibitemShut
  {NoStop}%
\bibitem [{\citenamefont {Simons}(1997)}]{Simons1997}%
  \BibitemOpen
  \bibfield  {author} {\bibinfo {author} {\bibfnamefont {B.}~\bibnamefont
  {Simons}},\ }\href {http://www.tcm.phy.cam.ac.uk/~bds10/phase.html} {\emph
  {\bibinfo {title} {{Phase transitions and collective phenomena}}}}\ (\bibinfo
   {publisher} {Cambridge University Press},\ \bibinfo {year} {1997})\ \bibinfo
  {note} {{A}vailable at
  \url{http://www.tcm.phy.cam.ac.uk/~bds10/phase.html}}\BibitemShut {NoStop}%
\bibitem [{\citenamefont {Georges}(2004)}]{Georges2004}%
  \BibitemOpen
  \bibfield  {author} {\bibinfo {author} {\bibfnamefont {A.}~\bibnamefont
  {Georges}},\ }\href {\doibase 10.1063/1.1800733} {\bibfield  {journal}
  {\bibinfo  {journal} {AIP Conf. Proc.}\ }\textbf {\bibinfo {volume} {715}},\
  \bibinfo {pages} {3} (\bibinfo {year} {2004})}\BibitemShut {NoStop}%
\end{thebibliography}

\end{document}